\documentclass[aps,pre,superscriptaddress]{revtex4}
\usepackage[dvips]{graphics}
\usepackage{graphicx}
\usepackage{amsfonts}
\usepackage{amssymb}
\usepackage{amsmath}
\usepackage{subfigure}

\begin{document}

\title{Stability of Quantized Vortices in a Bose-Einstein condensate confined in an Optical Lattice}
\author{K.J.H. Law}
\affiliation{Department of Mathematics and Statistics, University of
Massachusetts, Amherst MA 01003-4515}
\author{L. Qiao}
\affiliation{Department of Chemical Engineering, 
Princeton University, Princeton, NJ 08540}
\author{P.G. Kevrekidis}
\affiliation{Department of Mathematics and Statistics, University of
Massachusetts, Amherst MA 01003-4515}
\author{I.G. Kevrekidis}
\affiliation{PACM and 
Department of Chemical Engineering, Princeton University, Princeton, NJ 
08540}
\begin{abstract}
We investigate the existence and especially the linear stability of 
single and multiple-charge quantized vortex states of nonlinear
Schr{\"o}dinger equations in the presence of a
periodic and a parabolic potential in two spatial dimensions.
The study is motivated by the examination of pancake-shaped
Bose-Einstein condensates in the presence of magnetic and optical
confinement.  A two-parameter space of the condensate's chemical potential 
versus the periodic potential's strength is scanned for both single- and 
double-quantized  vortex states located at 
a local minimum or a local maximum 
of the lattice. Triply charged vortices are also briefly discussed. 
Single-charged vortices are found to be stable for cosinusoidal
potentials and unstable for sinusoidal ones above a critical strength.
Higher charge vortices are more unstable for both types of potentials
and their dynamical evolution leads to breakup into single-charged vortices.
\end{abstract}

\maketitle

\section{Introduction}

In the past decade, the study of coherent structures with topological
charge has become increasingly popular in Hamiltonian systems. This has
been especially so due to the emergence of both nonlinear optical,
as well as atomic physics settings where such studies were experimentally
relevant. Recent reviews both in the framework of optical fibers
and waveguides \cite{desyatnikov}, as well as in that of Bose-Einstein
condensates (BECs) \cite{fetter,us} have given overviews of this 
rapidly growing area of research; see also \cite{Pismen}. 
It is particularly interesting
that both of these directions are associated with the prototypical
dispersive equation that supports such structures, namely the
nonlinear Schr{\"o}dinger equation \cite{sulem}.

More specifically, in the context of dilute alkali vapors forming
BECs at extremely low temperatures, the experimental realization
of one as well as a few vortices \cite{dalib1,cornell1}, followed by the
realization of very robust vortex lattices \cite{kett1} has had
a profound influence on the field. More recently, experimental
efforts have turned to higher-charged vortices, illustrating not
only how vortices of topological charge $S=2$ and
even $S=4$ can be imprinted \cite{kett2}, 
but also the dynamical instability of such waves \cite{kett3}.

These experimental works have caused an intense theoretical
interest in the existence and stability of vortices, especially
under the case of parabolic confinement which is customary in
BECs due to magnetic trapping. While vortices of topological
charge $S=1$ were found to be stable in a parabolic potential,
higher topological charges were found to be unstable. The work
of \cite{pu} illustrated the potential instability (for appropriate
values of the number of atoms i.e., the spatially integrated
square modulus of the solution) of $S=2$ and $S=3$ waveforms.
Later, the work of \cite{kawa} shed light into the case of
$S=4$ and the role of negative energy eigenmodes in the instability.
The studies of \cite{motto} and \cite{ueda} examined the various 
scenaria of split-up of higher charge vortices during dynamical
evolution simulations for repulsive and attractive interactions
(i.e., defocusing and focusing nonlinear Schr{\"o}dinger equations)
respectively. The work of \cite{carr} examined such vortices riding
on the background of not just the ground state, but also of higher,
ring-like, excited states of the system. In the context of
an optical lattice, the authors of  \cite{borissak}
tried to construct both higher charge vortices, and super-vortices
for attractive condensates in optical lattices, while in the repulsive
setting the so-called gap vortices have been predicted \cite{ostr1} and
systematically created \cite{ostr2}.
More recently, these notions have also been examined
from a more mathematically rigorous point of view, using tools such
as the Evans function to compute the eigenvalue spectrum \cite{kollar},
or weak perturbations from the linear limit, to obtain a full count
on the potentially unstable eigenvalues \cite{todd}. 

One essential limitation of almost all of the above studies regarding
the existence and stability of the vortices has been the radial nature
of the symmetry of the parabolically confined system. In that radially
symmetric setting, the authors of the above theoretical references
proceeded to convert the problem to an effectively one-dimensional
one along the radial direction, which it is straightforward to solve
by a shooting method or otherwise. Then, when the so-called Bogolyubov-de
Gennes (or linear stability) analysis was performed around the ensuing
radial excitations, the radial symmetry was again crucially involved,
as the angular dependence of the perturbation modes was decomposed 
into Fourier modes, resulting into a one-parameter infinity of
radial problems (indexed by the wavenumber of the angular mode); 
for the mathematical details of this analysis see e.g., \cite{pegowar}. 
As discussed in \cite{pegowar}, subsequent examination of only
the lower modes in the resulting quasi-1d stability problem is sufficient
to extract conclusions on the stability of the full 2d radial state.
It should be mentioned in passing that the more recent works of
\cite{lundh1,lundh2} considered the stability of multiply quantized
vortices in three-dimensional BECs.

This above radial 
approach, however, does not allow us to examine the stability 
of states in the presence of {\it both} a magnetic trap and an
optical lattice. In the latter setting, the combined potential 
breaks the radial symmetry and defies any one-dimensional approach
towards the identification of the solutions or of their stability.
In view of that, while numerous works examine the existence and
linear stability of states in the presence of the parabolic trap,
references that address the properties of the vortices in combined
potentials are scant; in fact, we are only aware of the
direct integration results of \cite{pgk} which illustrate the
disparity of the effects of the optical lattice 
phase on vortices of $S=1$,
and of the entirely theoretical work of \cite{bhatta} that calculated
the normal modes of the system based on a Lagrangian approach.

Our purpose in the present work is to examine systematically
both the existence and the stability of both single-charged
as well as multiply charged vortices in the presence of the combined
parabolic and periodic potentials. 
The considerable numerical
load that is required to perform the relevant computations
(and which, we believe, is what prevented an earlier study of
this topic) is circumvented with two separate variations of Newton's
method which both allow us to perform equally efficient fixed 
point iteration that converges on numerically exact solutions of the system. 
The variations differ in the method of handling the linear system
in each Newton iteration, which is far beyond the capacity of a
standard direct linear solvers. One scheme implements an 
iterative linear solver in each Newton step, and the other implements 
a standard sparse banded matrix solver.
We subsequently
critically use the sparse nature of the relevant stability matrices
to invoke a sparse Arnoldi iterative algorithm for approximating
the system's eigenvalues, which, in turn, allows us to determine
the linear stability (i.e., to computationally perform the Bogolyubov-de Gennes
analysis) of the resulting vortices. This permits us to develop
two-parameter diagrams (as a function of the chemical potential,
and of the strength of the periodic potential) for the existence
and stability of solutions with $S=1$ and $S=2$; we also briefly
touch upon solutions with $S=3$ that can also be obtained by
the above means. Finally, we examine the dynamical evolution
of these solutions when they are unstable. 

Our presentation will be structured as follows. In the next
section, we give an overview of our theoretical setup. In
section 3, we develop our numerical results for the vortices
of different topological charge, while in section 4, we summarize
our findings and present some interesting directions for future
research.


\section{Theoretical Setup}

We assume the well known dimensionless 
setting for a confined BEC in harmonic oscillator units \cite{book1,book2}, 
given as the following:

 \begin{equation}
iu_t = -\frac{1}{2}\nabla^2 u+g|u|^2u+V(x,y)u
\label{eq1a}
\end{equation}

 \begin{eqnarray*}
V_M  &=& \frac{1}{2} \Omega^2 (x^2+y^2) 
\label{eq1b}
\\
V_{OL} &=& V_0 [\cos^2(k_x x+\phi) + \cos^2(k_y y+\phi)]
\label{eq1c}
\\
V(x,y) &=& V_M+ V_{OL}.
\label{2}
\end{eqnarray*}
where $\nabla^2$ is the two dimensional Laplacian operator 
$\nabla^2=\partial_{xx}+\partial_{yy}$, $V_0$ denotes the
strength of the optical lattice and $\Omega$ characterizes the
effective
tightness of the magnetic trap, while $\phi$ is an arbitrary 
phase shift. 
In particular, we will focus on the case of the, so-called, 
pancake-shaped (i.e., quasi-two-dimensional) BECs where 
$\Omega \ll 1$.
By assuming a standing wave solution of the form $u(r,t)=e^{-i\mu t}w(r)$, 
we are able to locate stationary states by solving the 
following equation with a fixed point algorithm 
(such as a Newton method):

\begin{equation}
F(w) \equiv \mu w+\frac{1}{2}\nabla^2 w-g|w|^2w-V(r)w =0 .
\label{eqn2}
\end{equation}
  
In the Newton iteration, we have the following update scheme 
(where we define $w=(w_r,w_i)^T$ in the equivalent 
case of the real formulation and $J$ represents the Jacobian of $F$ with
respect to $w$):
\begin{eqnarray*}
-J(w) \Delta w &=& F(w)
\\
w &=& w+\Delta w,
\end{eqnarray*}
where 

\begin{eqnarray*}
F(w) &=& Lw+N(w)\\
L &=& \frac{1}{2}\nabla^2\\
N(w) &=& (\mu-V-g |w|^2)w\\
J(w) &=& L+\frac{\partial N}{\partial w}
\end{eqnarray*}

For $V_0=0$, in the linear limit, $g\rightarrow 0$, 
we recognize the two-dimensional Quantum Harmonic Oscillator (QHO), 
for which the eigenfunctions are explicitly known in the form :
\begin{eqnarray*}
w_0=e^{-r^2 \Omega/2}H_{n,m}(x,y)\\
\mu=\Omega(n+m+1)
\end{eqnarray*}
where $H_{n,m}$ is the product of the $nth$ in x and $mth$ in y 
Hermite polynomials. 
The advantage of this underlying linear problem is that it allows
us to generate the structure of a single charge vortex in 2d as a 
linear combination of 
$|1,0\rangle=H_{1,0}e^{-\Omega r^2/2}$ and  
$|0,1\rangle=H_{0,1}e^{-\Omega r^2/2}$,
where $H_{1,0}=x$  and $H_{0,1}=y$,  according to 
$w(x,y)=|1,0\rangle+i|0,1\rangle$. 
Similarly, a linear, doubly charged vortex-type state can be constructed
as $w(x,y)=|2,0\rangle-|0,2\rangle+2 i |1,1\rangle$ and 
such constructions can also
be produced for higher charges, as well. These can be used as good
initial guesses for the fixed point iteration, towards producing
a solution of the full nonlinear problem.

In what follows, we perform a two-parameter continuation in $\mu$, which 
corresponds to the chemical potential of the BEC (and is related to
its number of atoms), and in $V_0$ which parametrizes the strength of
the OL potential. 
A uniform space mesh with $\Delta x=\Delta y=0.15$ is used to
prevent contamination of spurious modes which may
occur when the spatial scale of the unstable bound state in the
vortex core competes with the grid size. 
Solutions were found with both a standard 
banded solver for each Newton iteration 
as well as a Newton-Krylov GMRES scheme with the MATLAB script 
{\bf nsoli} \cite{kelley}. 
The total necessary storage space was minimized with the Krylov method, 
by approximating the directional derivative in the direction of 
subsequent Krylov vectors and iteratively solving the linear system
in each Newton step.  
For a system of size $267 \times 267$ (about $7 \times 10^4$ nodes) 
no more than 400 MB of RAM 
was used and a solution (within $10^{-8}$) 
for a single charge vortex with 
$V_0=0$ and $\mu=0.45$ took 80 seconds to compute. 
The same solution using a standard sparse banded solver 
took 60 seconds and used 500 MB of RAM. 
A 64-bit AMD processor was used for the computations.



Once the solution of the nonlinear problem has been identified to a 
prescribed accuracy (typically $10^{-7}$-$10^{-8}$), we solve the 
linearized eigenvalue problem of the Bogolyubov-de Gennes equations
\cite{book1,book2} to determine the stability of each solution. 
To do so, we introduce a complex-valued perturbation as follows:
  \begin{equation}
u(x,t)=\exp(-i \mu t) [ w(x) + \delta ( W_1(x) \exp(i \omega t) + W_2(x) 
\exp(-i \omega^* t) ) ] ,
\end{equation}
where $^*$ denotes complex conjugation. Then, substituting $u$ into 
(\ref{eq1a}), 
ignoring terms of order $\delta^2$, 
collecting terms of $O(\delta)$, and then separating the ones 
with like phase ($i \omega$ and $-i \omega^*$ ) we obtain the following 
eigensystem for $\omega$:

\begin{center}
\begin{math}
\bordermatrix{& \cr &\mathcal{L}_1 &\mathcal{L}_2 \cr
 &-\mathcal{L}_2^* &-\mathcal{L}_1^* \cr}
\bordermatrix{& \cr & W_1 \cr & W_2^* \cr}=\omega \bordermatrix{& \cr & W_1 \cr & W_2^* \cr},
\end{math}
\end{center}
where we define
\begin{eqnarray*}
\mathcal{L}_1 &=& \mu+\frac{1}{2}\nabla^2-V-2g|w|^2\\
\mathcal{L}_2 &=& -w^2.
\end{eqnarray*}

In this eigenvalue problem for 
$\omega$, we see from the expansion above that Im$(\omega)<0 \rightarrow$ 
Re$(i\omega)>0$, and so the eigenfrequencies that satisfy this condition are 
those which lead to instability of our stationary wave solution. We will 
henceforth refer to the eigenvalues $\lambda=i \omega$ for clarity. 
Returning to our discretized space mesh and the 
finite difference approximation to the Laplacian in 
order to calculate the values for $\omega$, we notice that 
the system is sparse and so 
it is well suited for the use of a sparse Arnoldi iterative algorithm for 
approximating the eigenvalues. Such a method is implemented by ARPACK 
that is invoked by the MATLAB function \textbf{eigs}, which we use for our 
calculations. The Hamiltonian nature of the 
system guarantees that the eigenvalues come in quadruples 
$\pm\lambda$, $\pm\lambda^*$. 
So we are only concerned with $\lambda$ such that Re$(\lambda)$ is 
non-zero. 

We will follow the above procedure, scanning the parameter space for  
quantized vortex states in the presence of the potential $V$,
for both the cases where the vortex 
is centered at a local maximum ($\phi=0$) and a minimum ($\phi=\pi/2$)
at the bottom of the parabolic trap. Finally, once the solution is
identified and the eigenvalues of linearization around it are obtained,
we confirm our stability results by  
dynamically evolving these solutions over time. Such numerical results
for vortices with $S=1$, $S=2$ and $S=3$ are presented below.

\section{Numerical Results}

\subsection{$S=1$ Vortices}

\begin{figure}
\includegraphics[width=60mm]{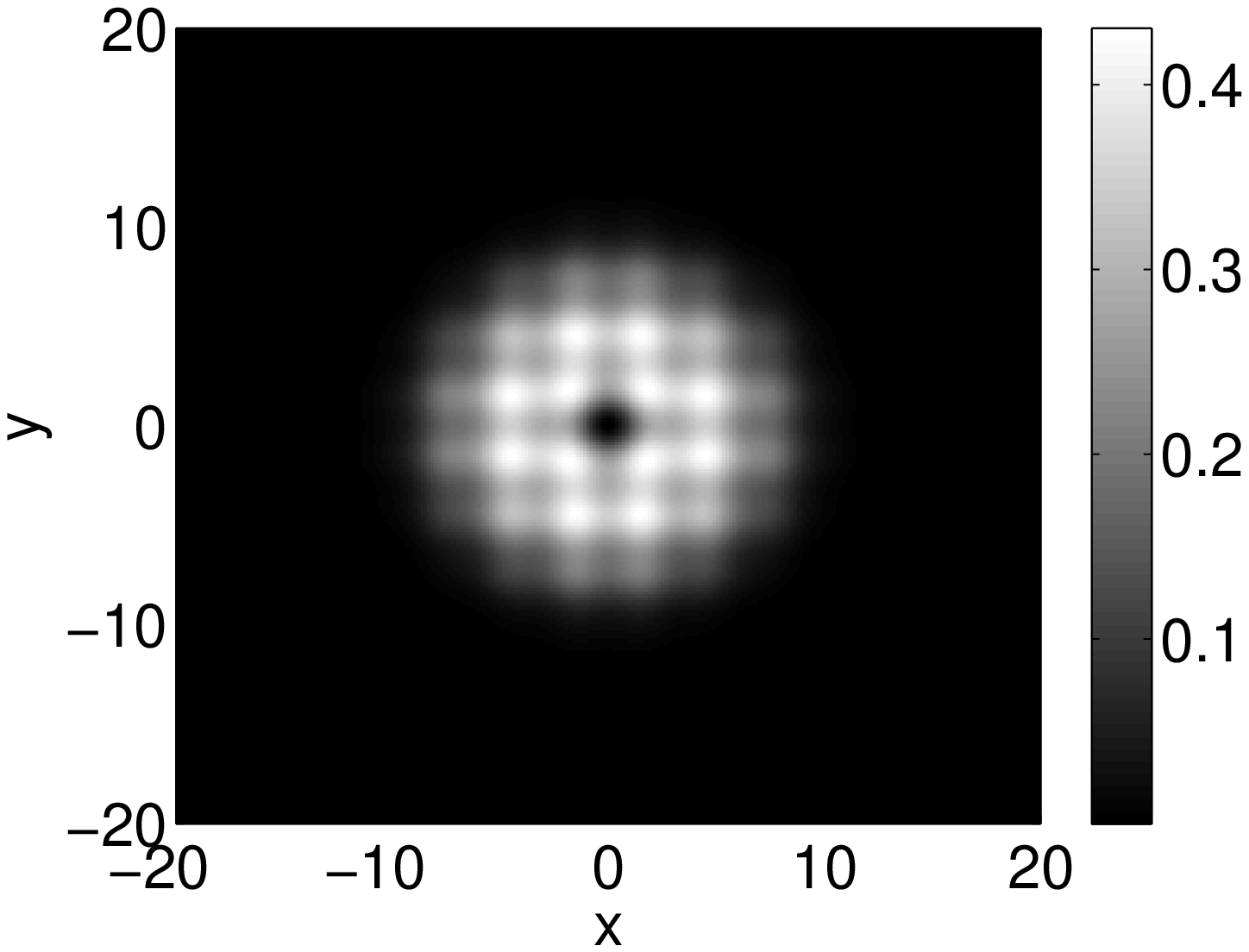}
\includegraphics[width=60mm]{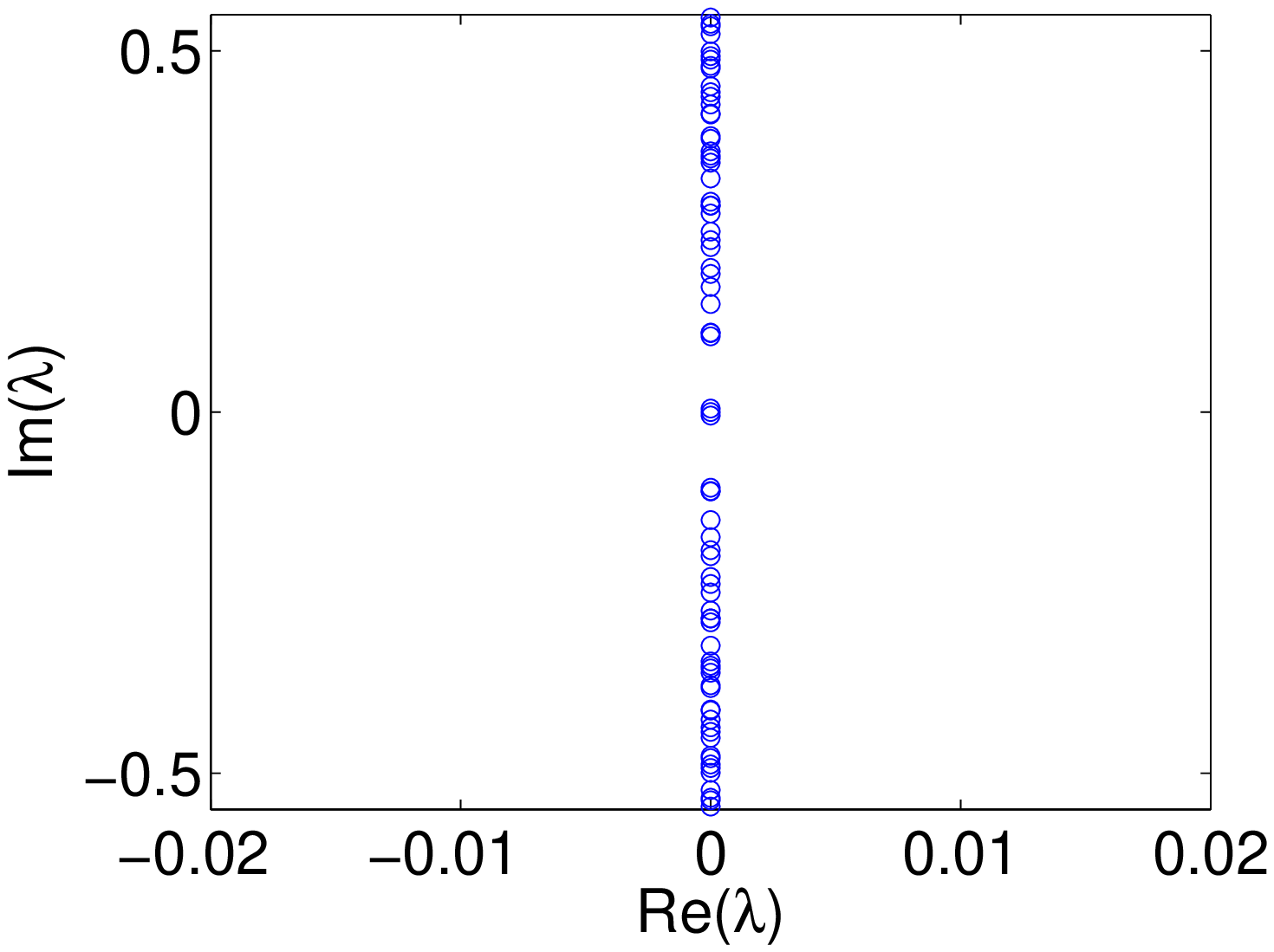}
\includegraphics[width=60mm]{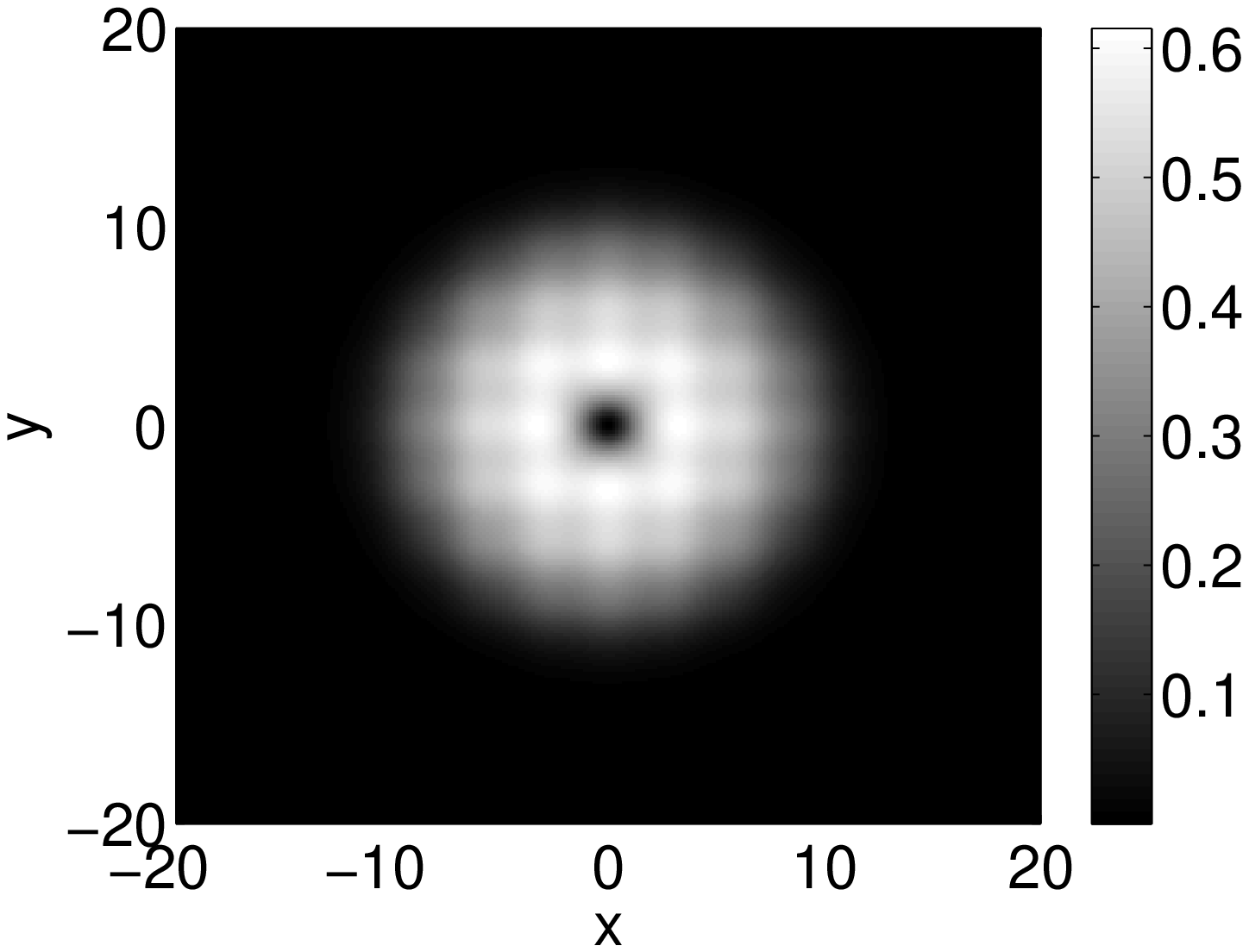}
\includegraphics[width=60mm]{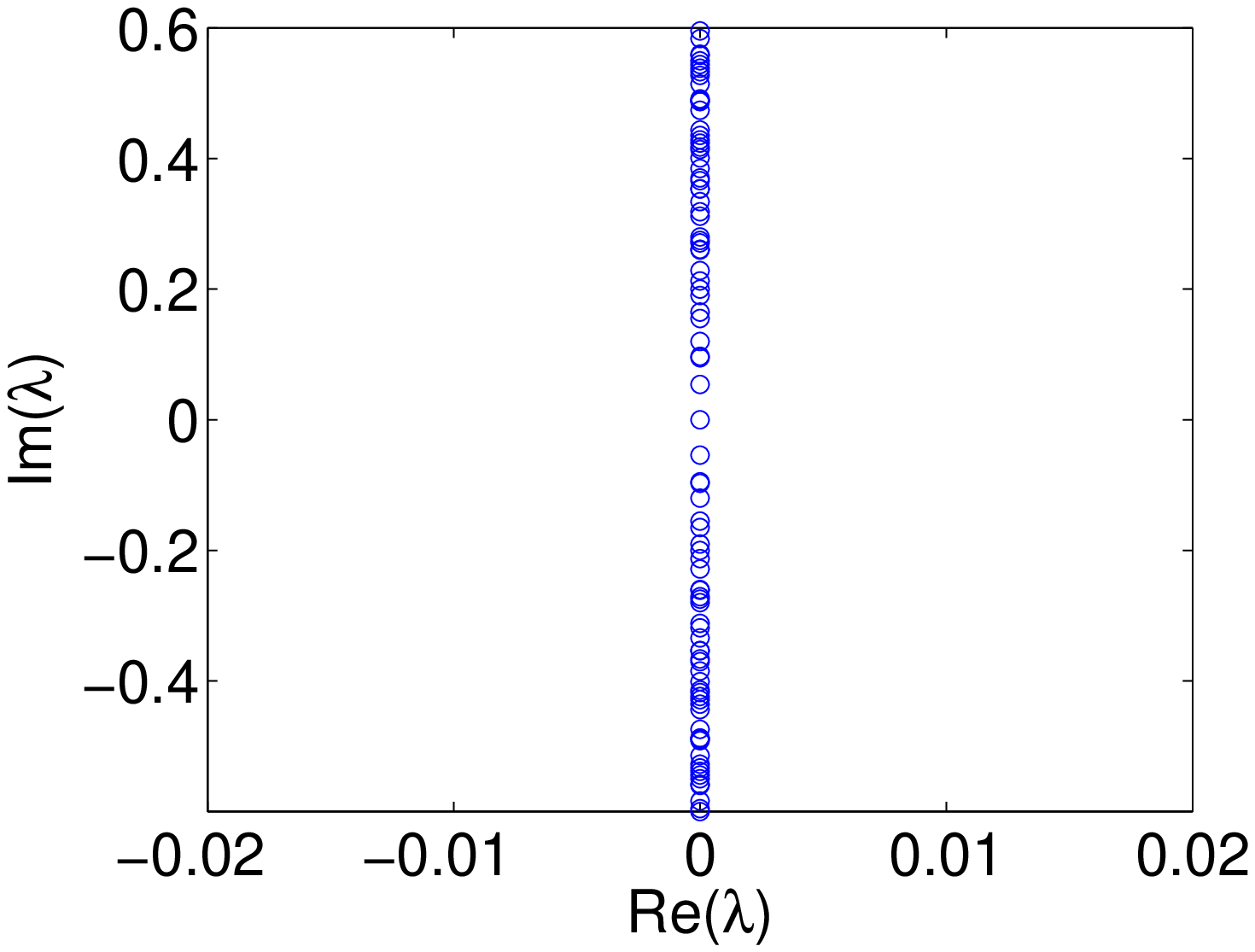}
\includegraphics[width=60mm]{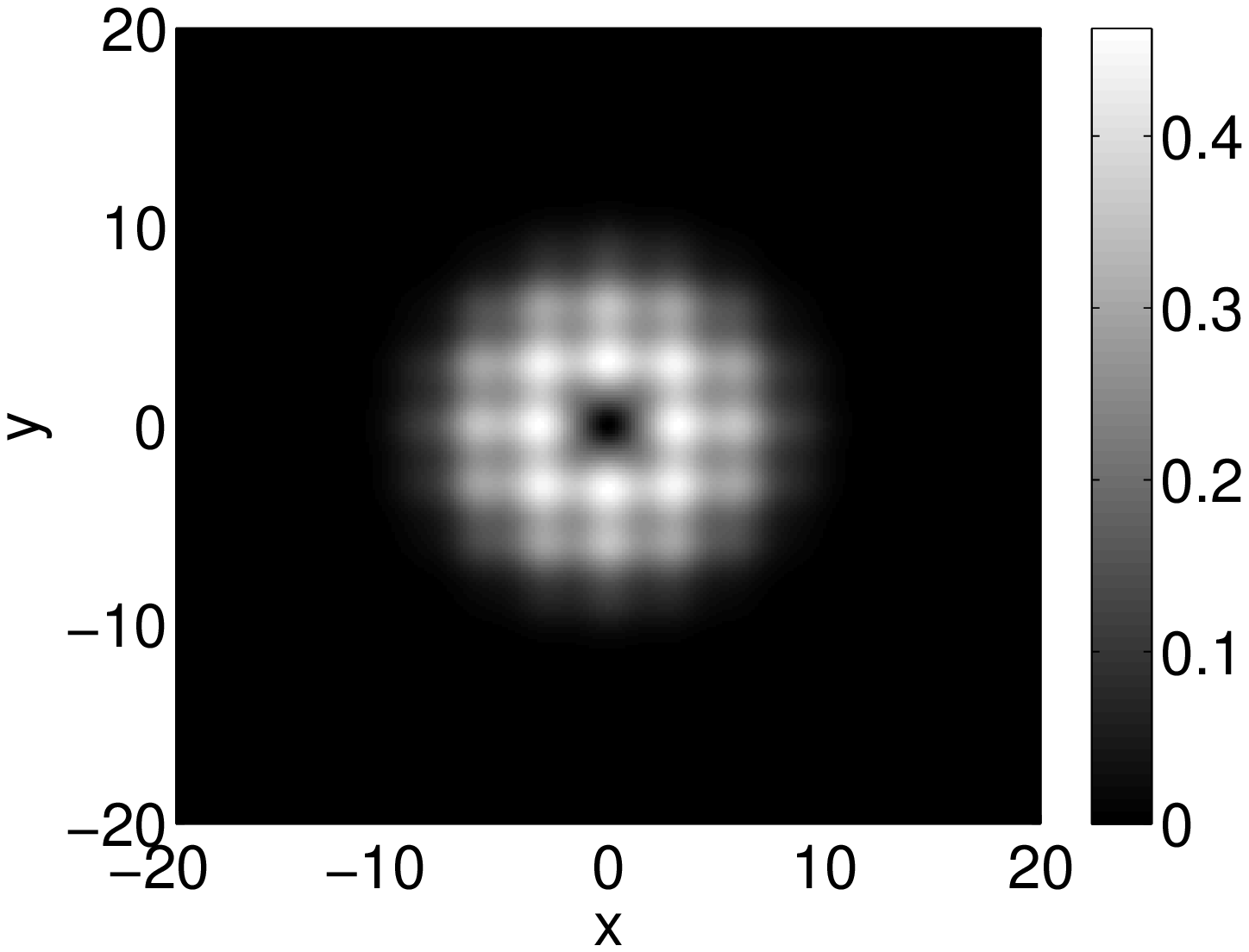}
\includegraphics[width=60mm]{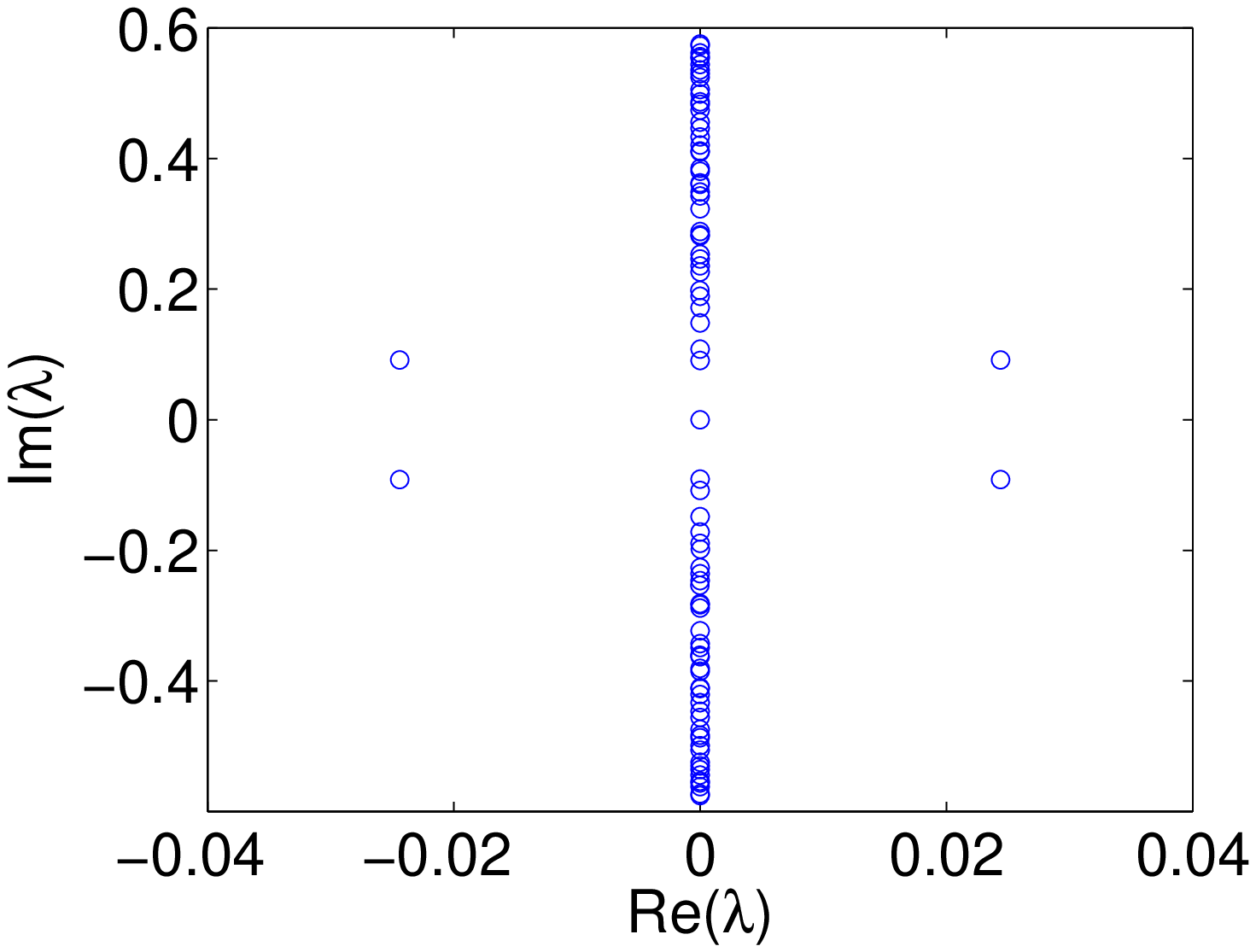}
\caption{(Color Online) Top: profile (left) and spectrum 
(right) for a stable vortex 
with $\phi=0$ at $(V_0,\mu)=(0.3,0.8)$.
Middle: stable vortex with $\phi=\pi/2$ at the point $(0.1,0.8)$ 
in our parameter plane. Bottom: same for the unstable case at the 
intersection of the slices from the previous image, $(0.3,0.8)$.} 
\label{fig1}
\end{figure}

We start by investigating the stability of the single quantized 
vortex state in the cases of $\phi=0$ and $\phi=\pi/2$ over a range of 
chemical potentials, $\mu$ between $0.2$ (the QHO linear limit for such 
a state with $V_0=0$), and $1.2$. For $V_0=0$, our results confirm previous 
ones \cite{pu} in that the single quantized vortex is stable in the 
presence of a pure parabolic potential over all values of $\mu$.  Perhaps 
more interestingly, we found the $\phi=0$ case to be stable over all $V_0$ as 
well, in agreement with the dynamical observations of \cite{pgk}, where
such a solution was observed to remain at its location, when 
initialized in the center of such a potential. An example density profile 
and spectrum are presented at the top panel of figure \ref{fig1}.

\begin{figure}
\includegraphics[width=60mm]{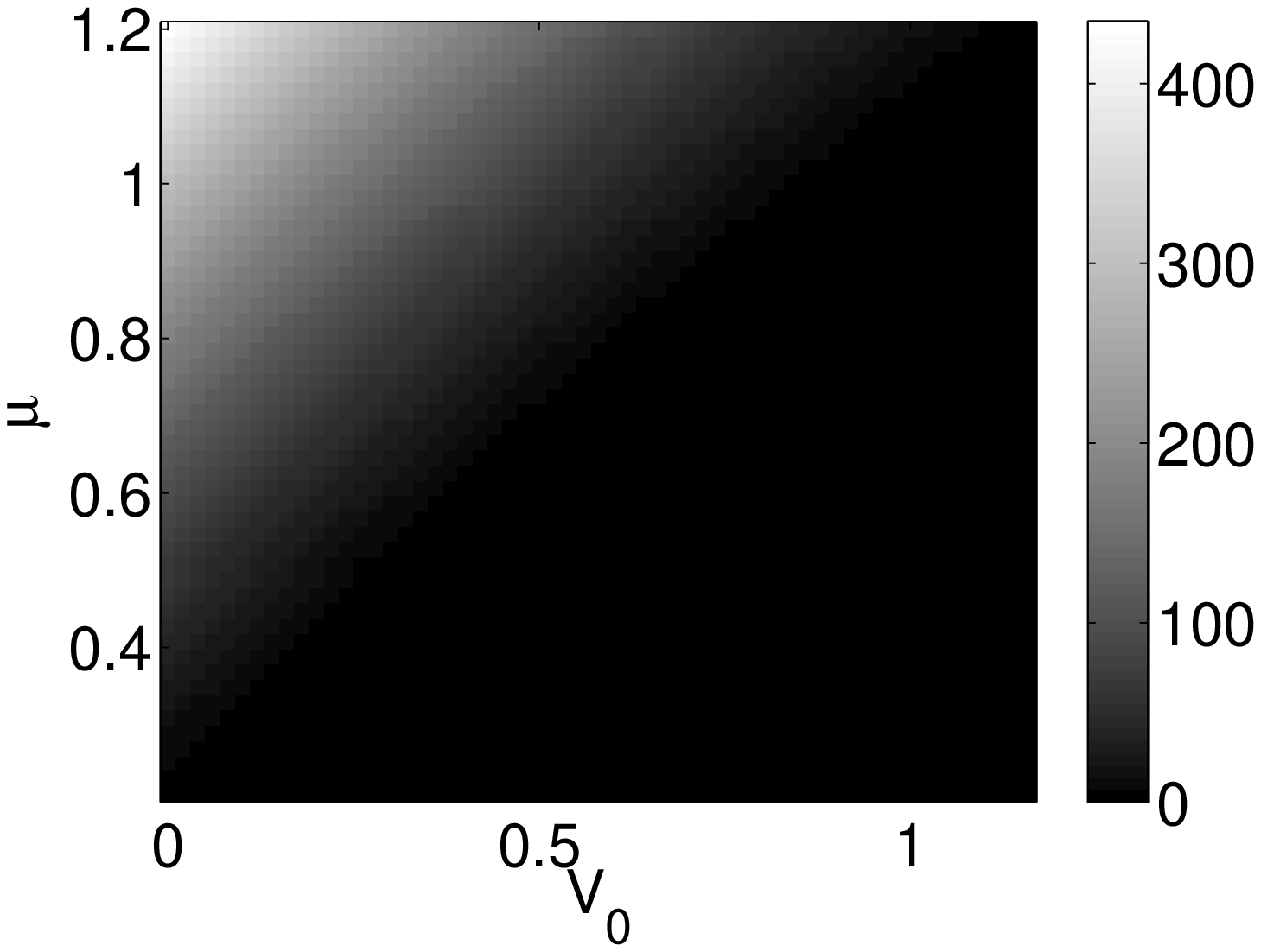}
\includegraphics[width=60mm]{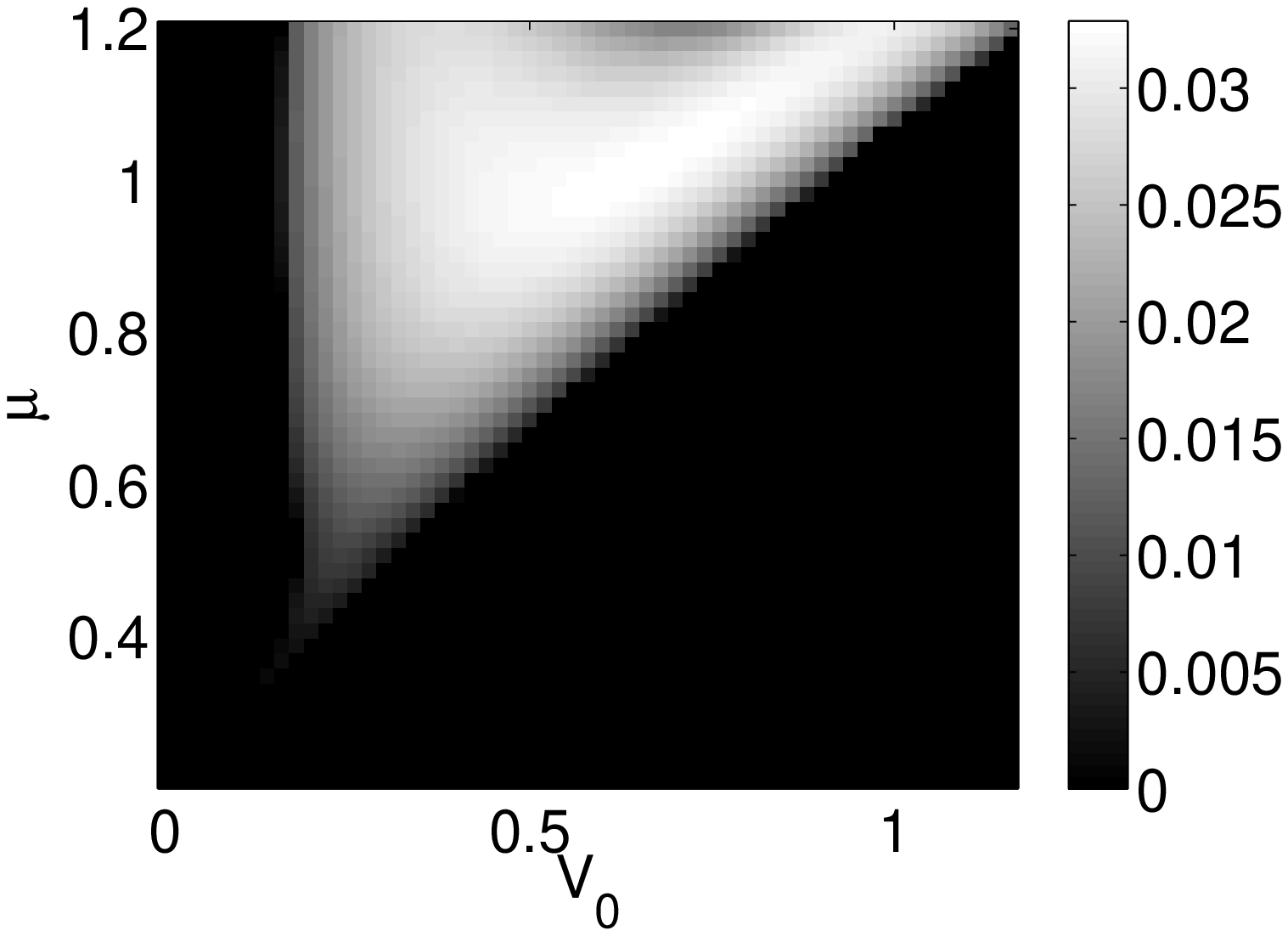}
\includegraphics[width=60mm]{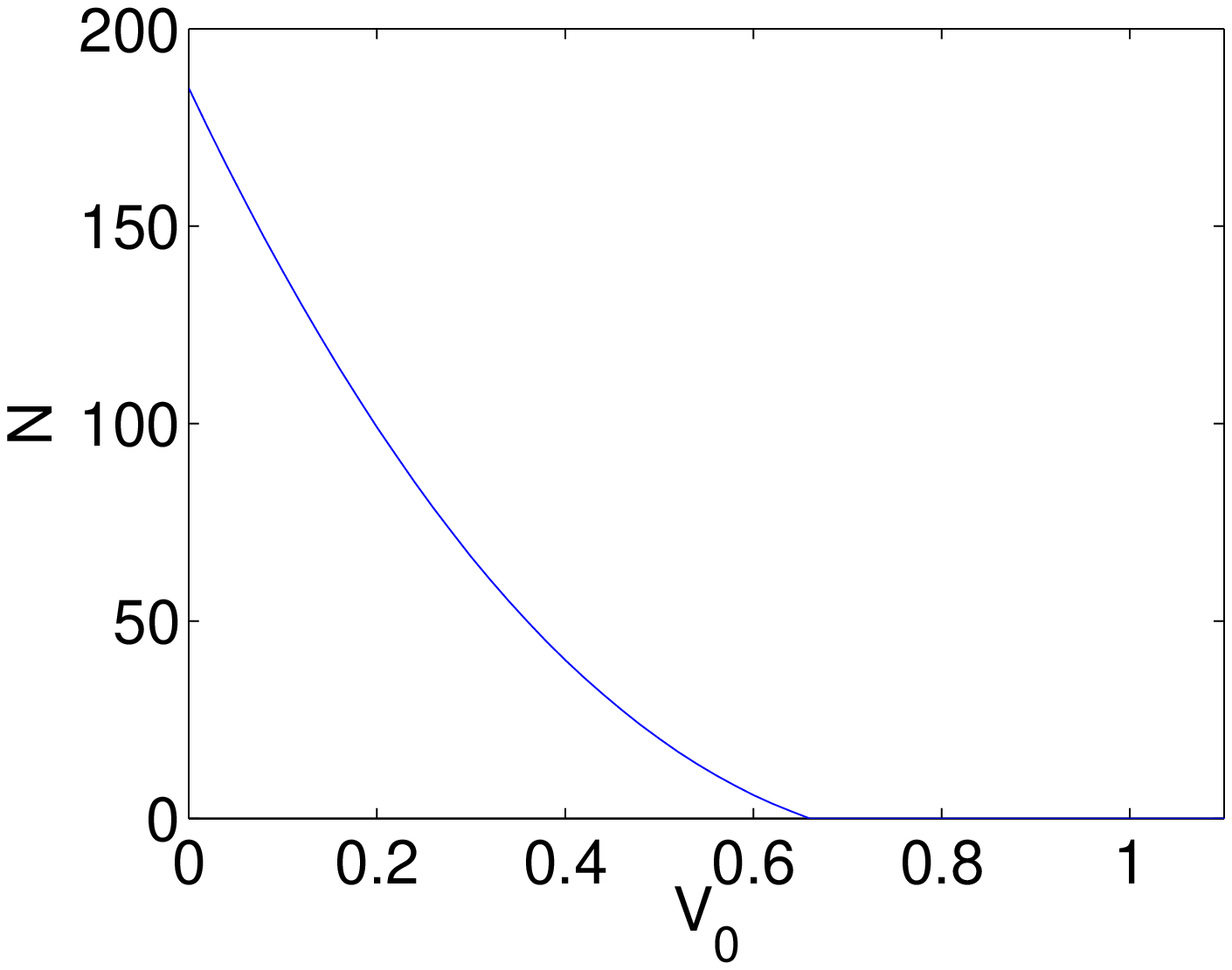}
\includegraphics[width=60mm]{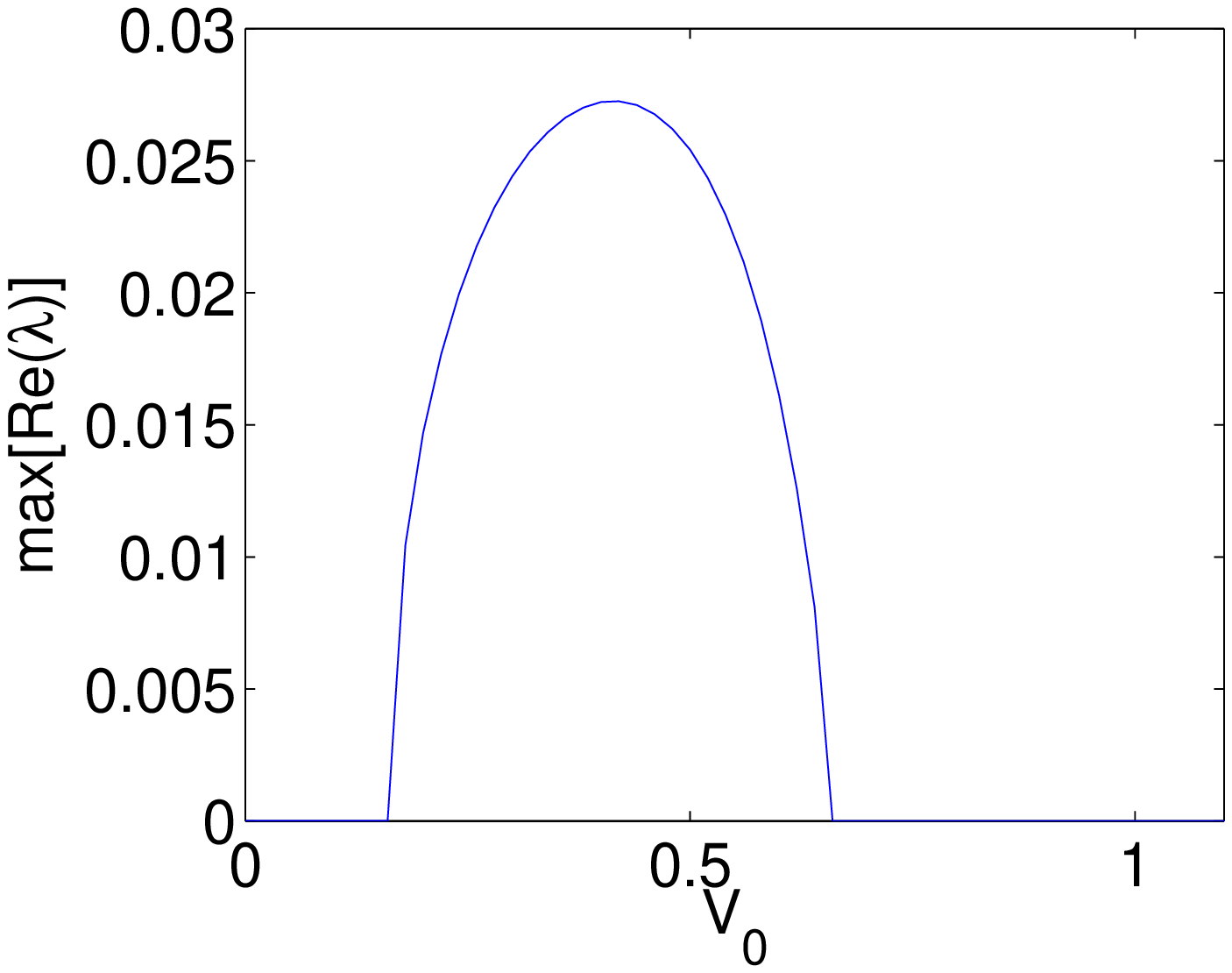}
\includegraphics[width=60mm]{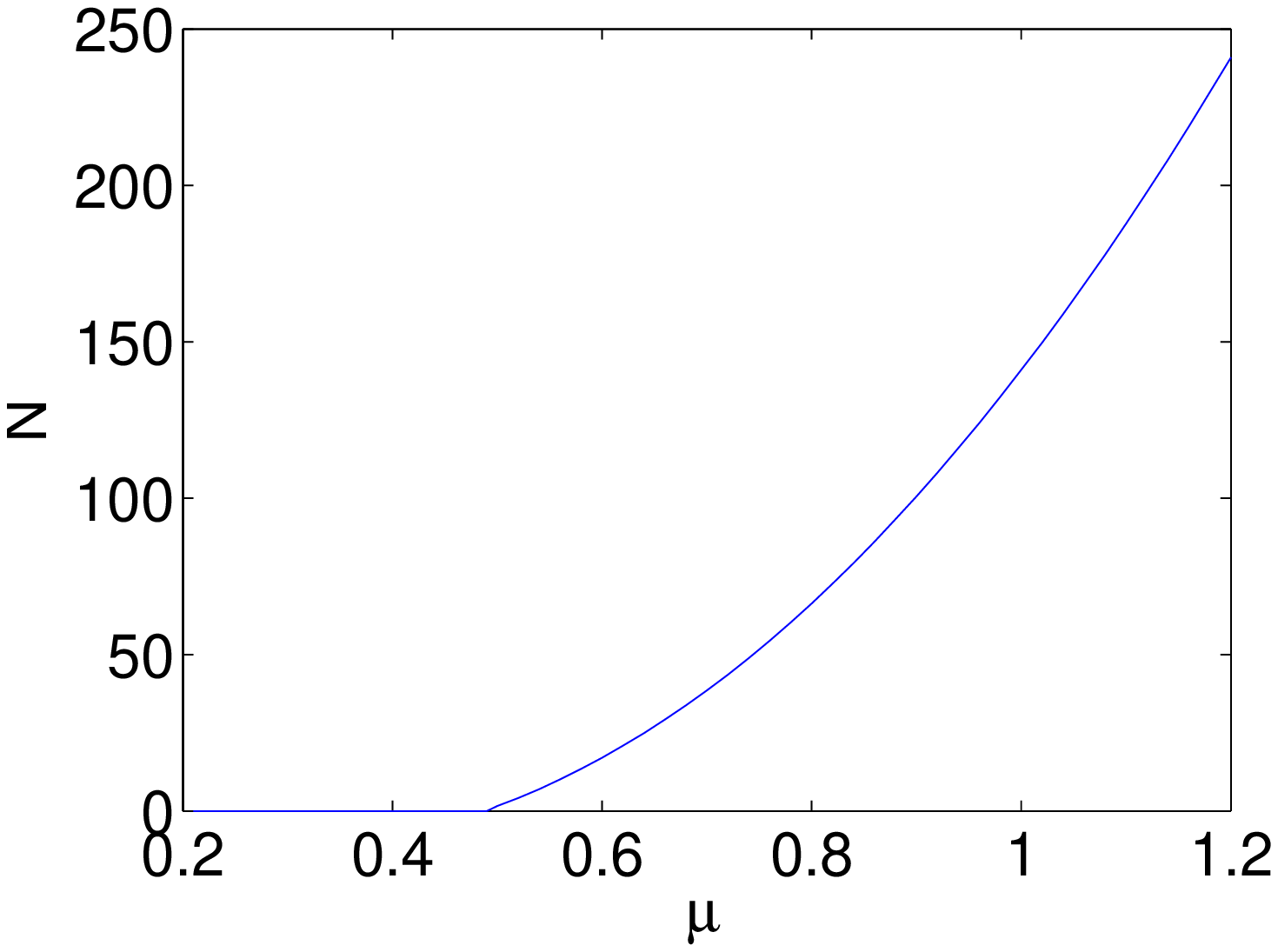}
\includegraphics[width=60mm]{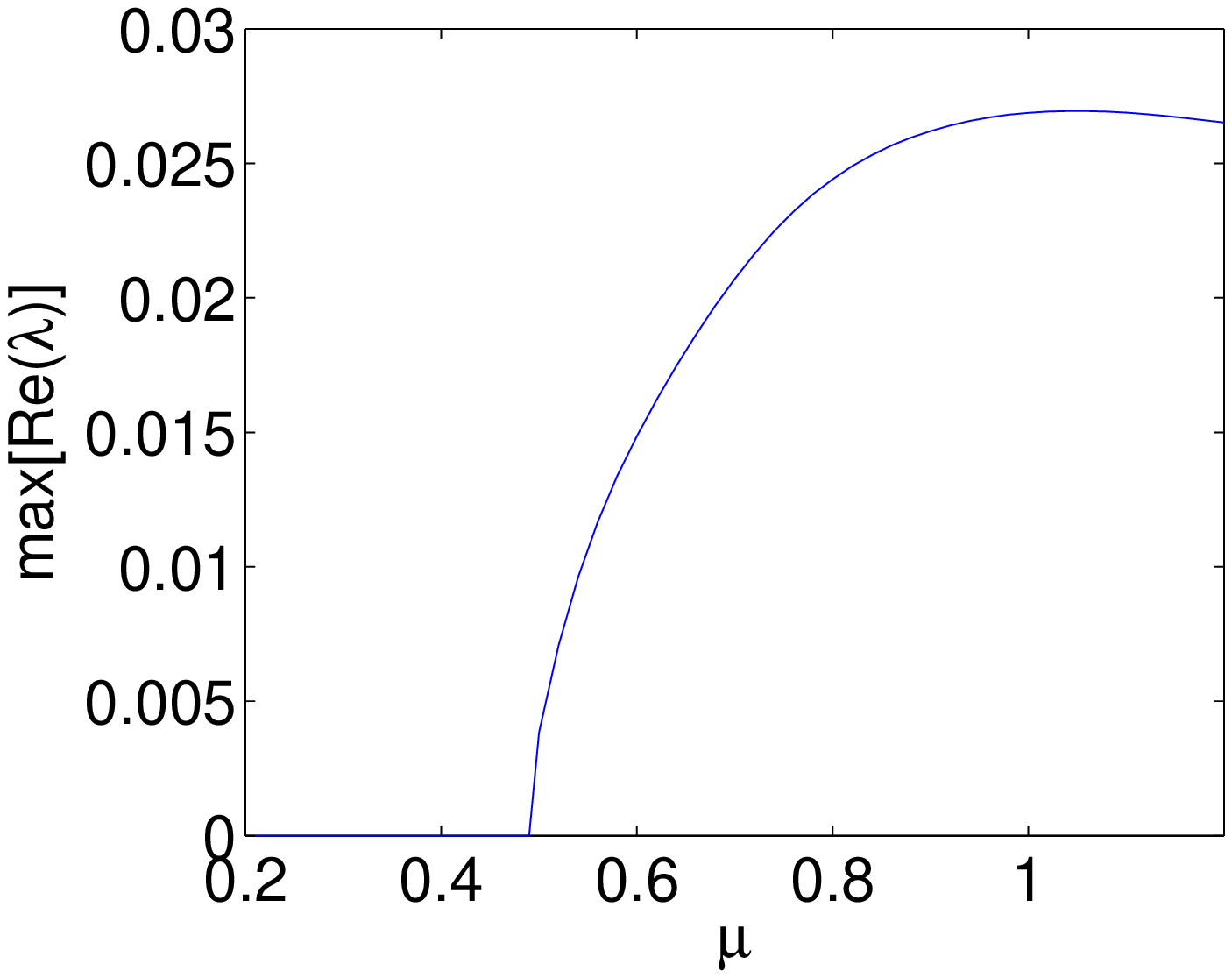}
\caption{(Color Online) Top: the number of particles, $N$, is shown as a contour plot
in the two-parameter plane of  $(V_0,\mu)$ in the left panel, while the 
maximum real part of $\lambda$ is shown in a similar way in the right
panel for a single charged vortex with $\phi=\pi/2$. 
Middle: slices of $N(V_0)$ (left) and 
${\rm max}[{\rm Re}(\lambda(V_0))]$ (right) for $\mu=0.8$. 
Bottom: slices of $N(\mu)$ (left) and 
${\rm max}[{\rm Re}(\lambda(\mu))]$ (right) for $V_0=0.3$.}
\label{fig2}
\end{figure}

On the other hand, when the vortex of single charge is located at a 
local minimum 
($\phi=\pi/2$),  instability sets in for a 
sufficiently modulated trap (i.e., for $V_0 \approx 0.2$) and the solutions 
remain unstable essentially throughout their interval of existence.
Notice that for each value of $\mu$, the solution can only sustain
a certain level of modulation from the optical lattice, before
it finally degenerates into the zero solution (beyond a certain critical,
$\mu$-dependent OL strength).  Typical examples of the subcritical
and supercritical profiles and their corresponding spectra for 
the relevant $S=1$ mode are shown in Fig. \ref{fig1}.
It is evident that in this case, beyond the critical $V_0$, a mode
becomes unstable through an oscillatory instability associated
with an eigenvalue quartet. Both the existence regime (top left
panel) and the stability regimes (top right panel) of such waveforms
are illustrated in Fig. \ref{fig2} (the figure also contains relevant
one-parameter cross-sections of the number of atoms and of the eigenvalue of 
the most unstable mode of the linearization).


\begin{figure}
\includegraphics[width=100mm]{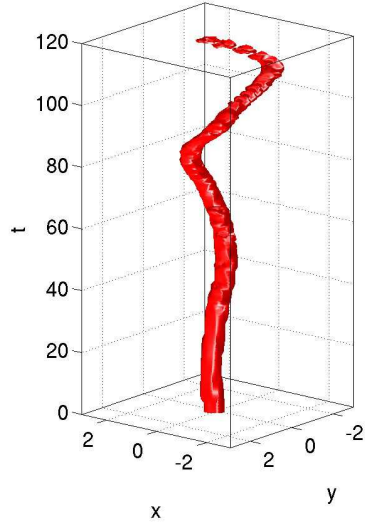}
\caption{(Color Online) The dynamics of the unstable mode when $\phi=\pi/2$ for 
parameter values $(0.3,0.8)$. We show an isosurface of the vorticity, 
or curl of the velocity 
field ($\nabla \times \{ \frac{-i}{|u|^2} 
[  u^* \nabla u - u \nabla u^* ] \}$) over time. This clearly illustrates
the winding out of the vortex towards the periphery of the BEC cloud.}
\label{fig3}
\end{figure}

\begin{figure}
\includegraphics[width=50mm]{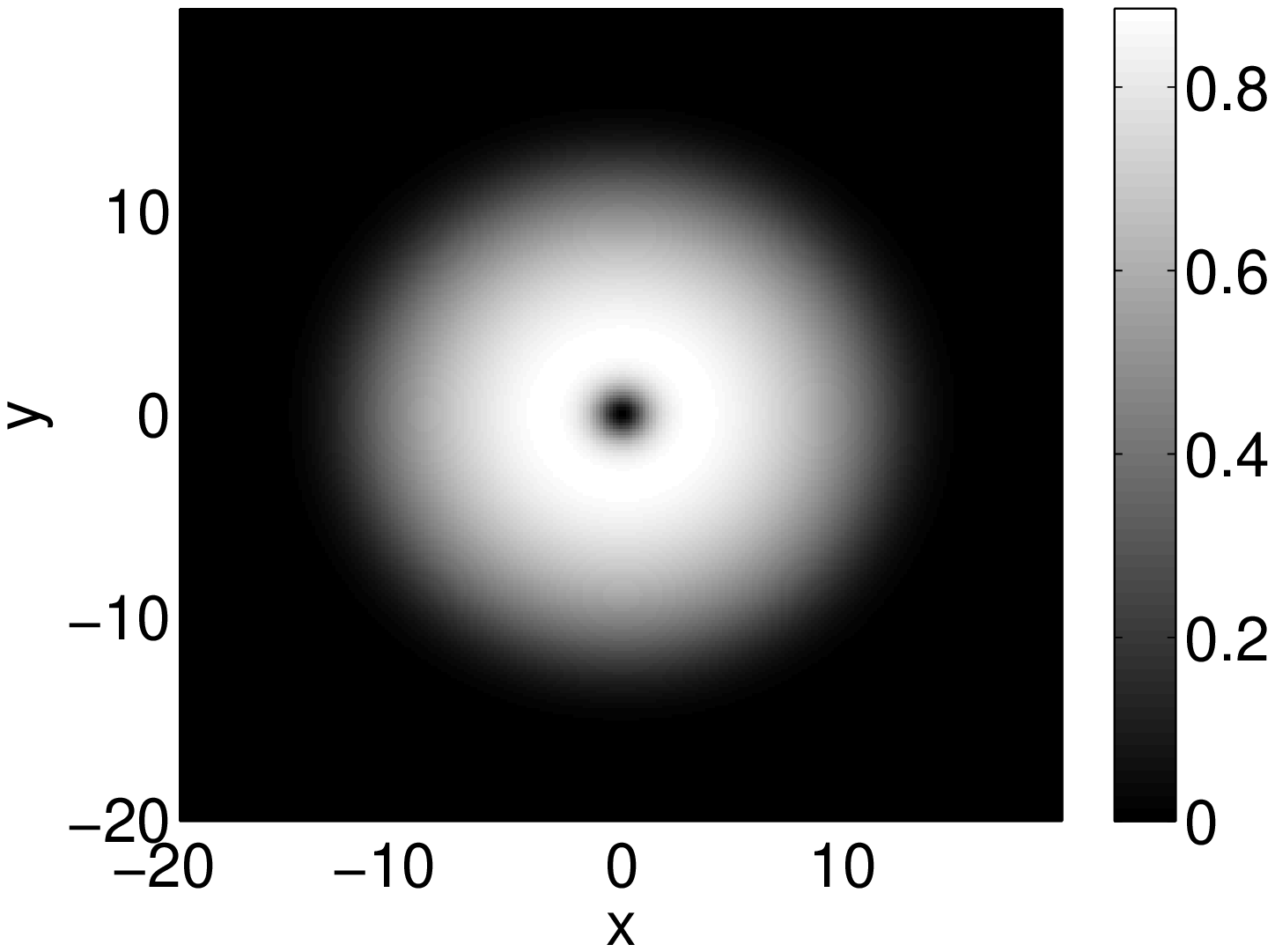}
\includegraphics[width=50mm]{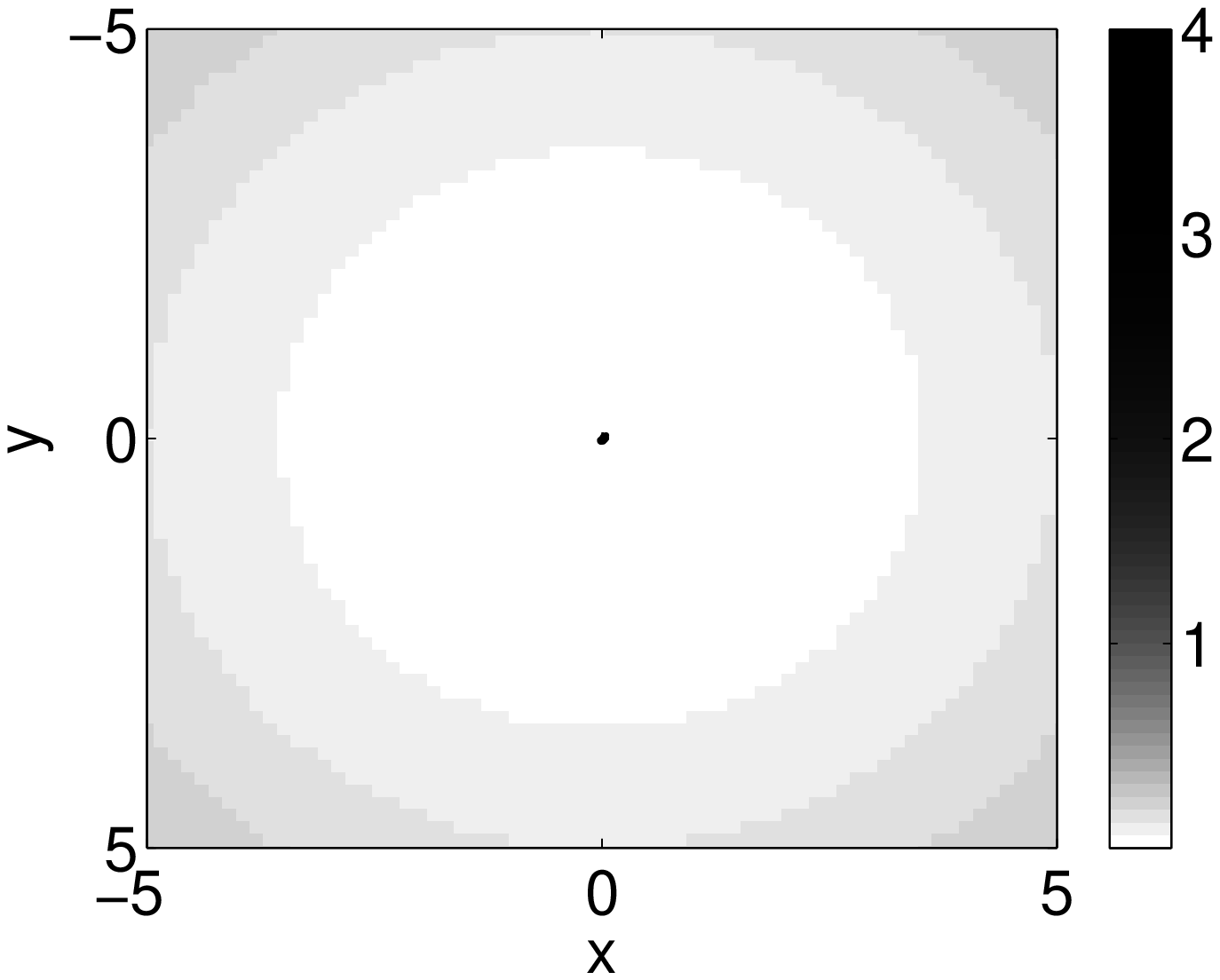}
\includegraphics[width=50mm]{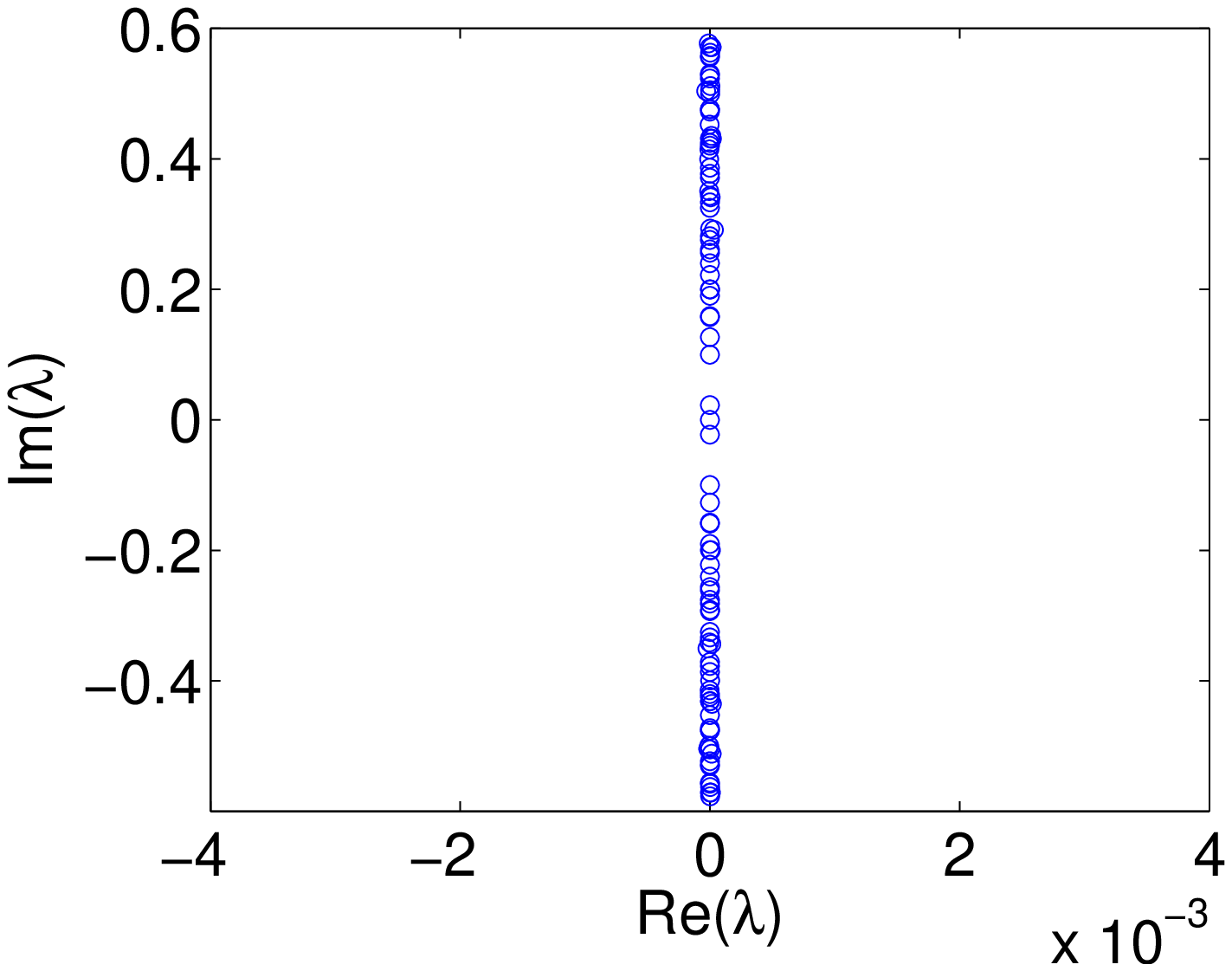}
\includegraphics[width=50mm]{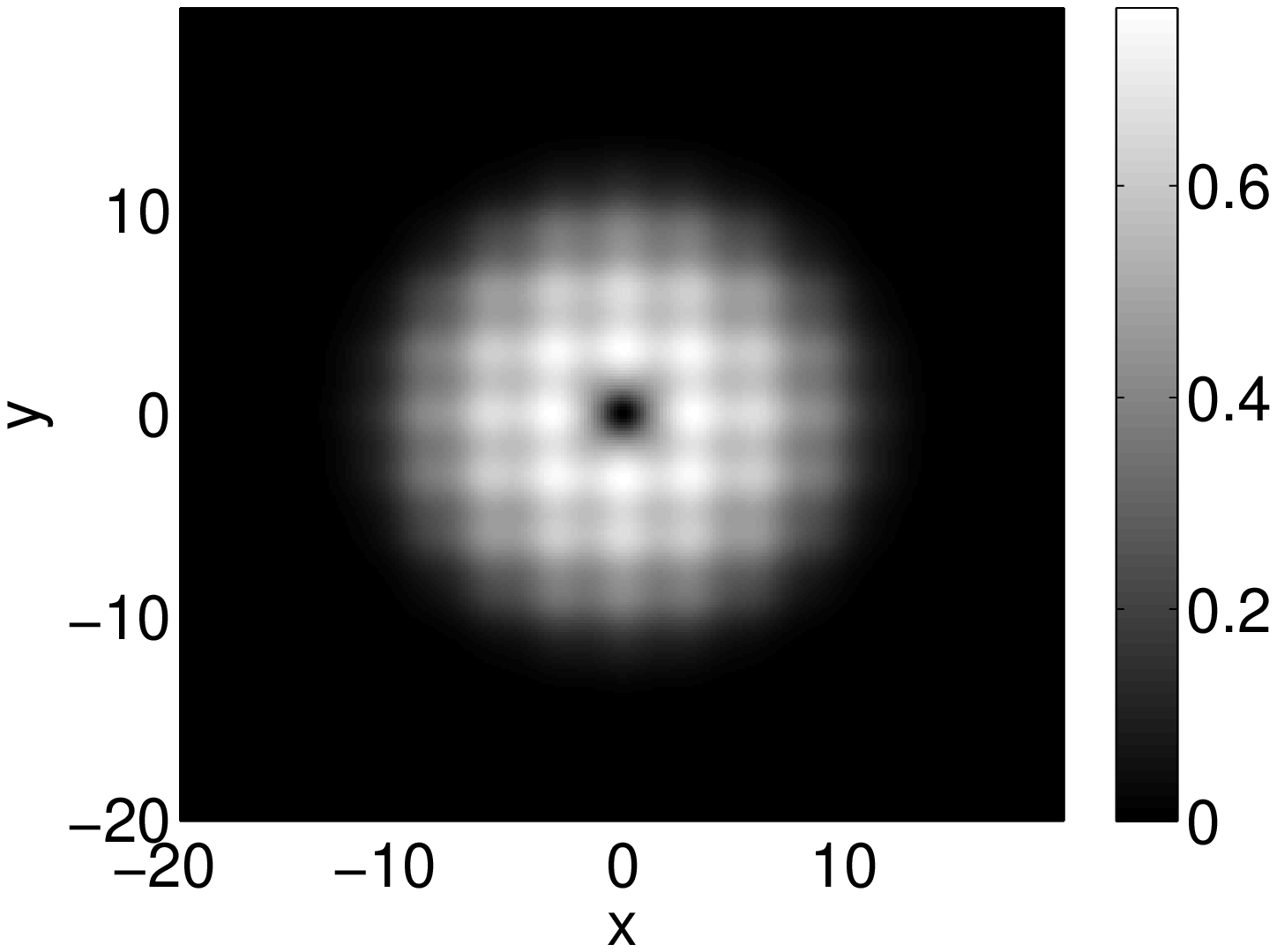}
\includegraphics[width=50mm]{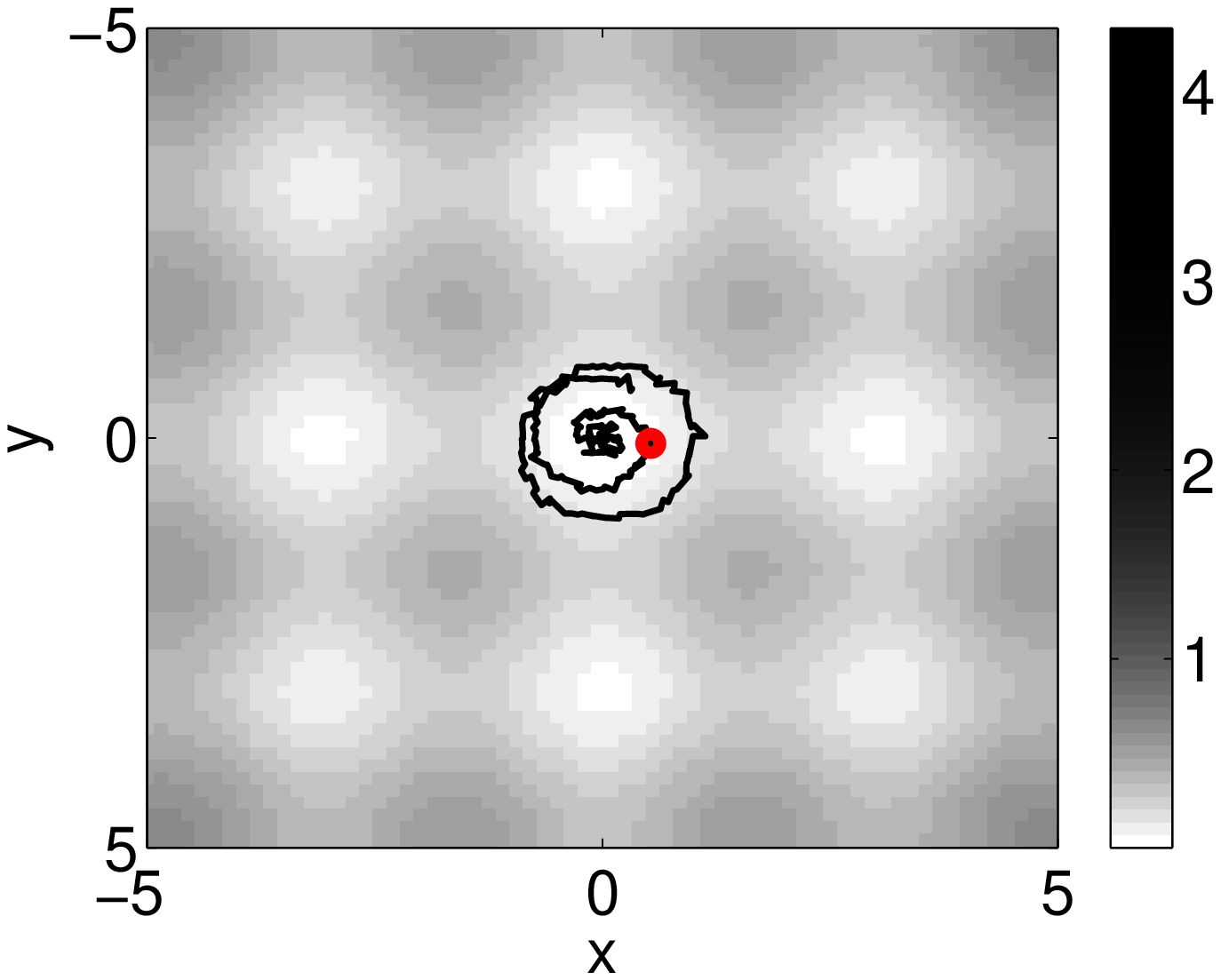}
\includegraphics[width=50mm]{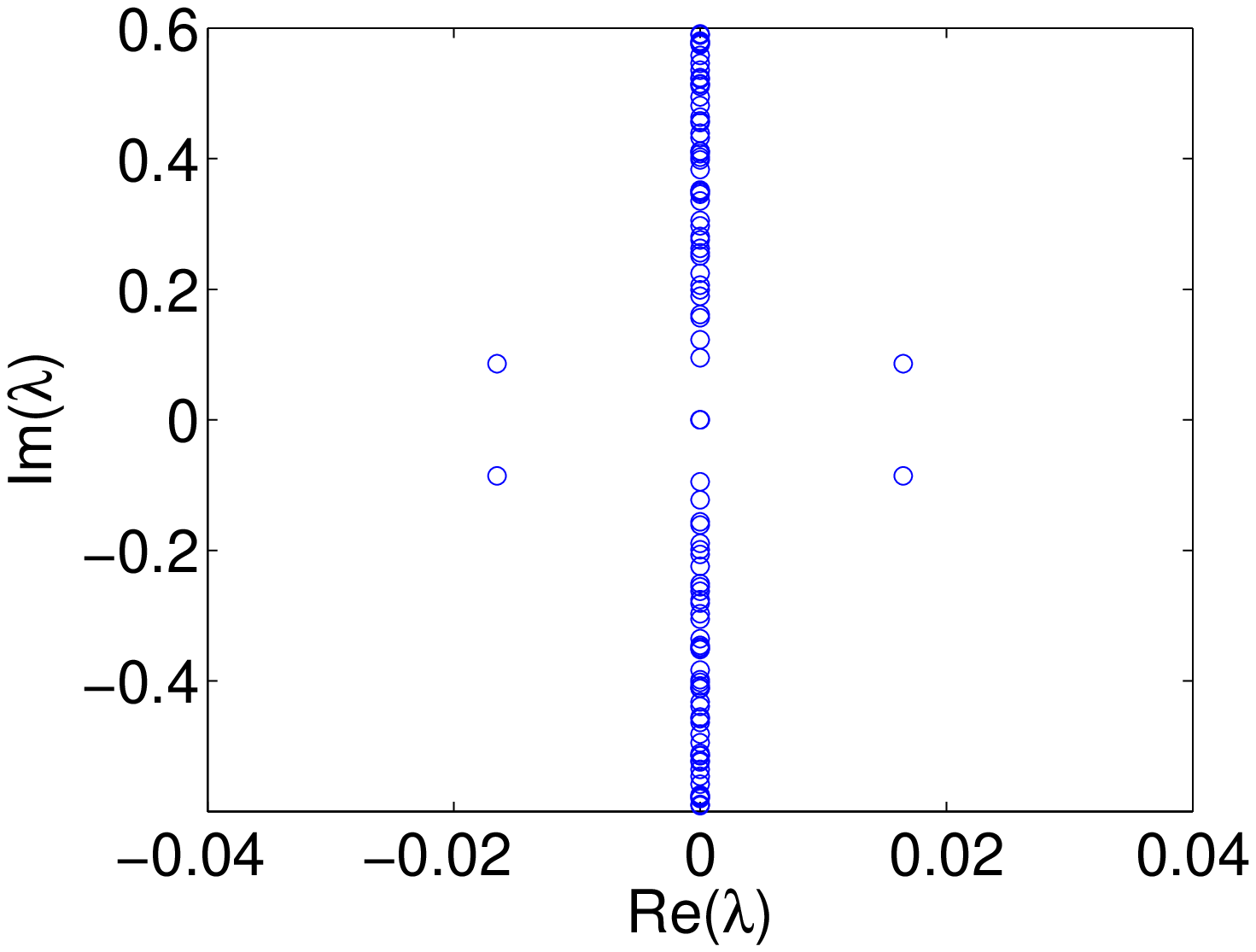}
\includegraphics[width=50mm]{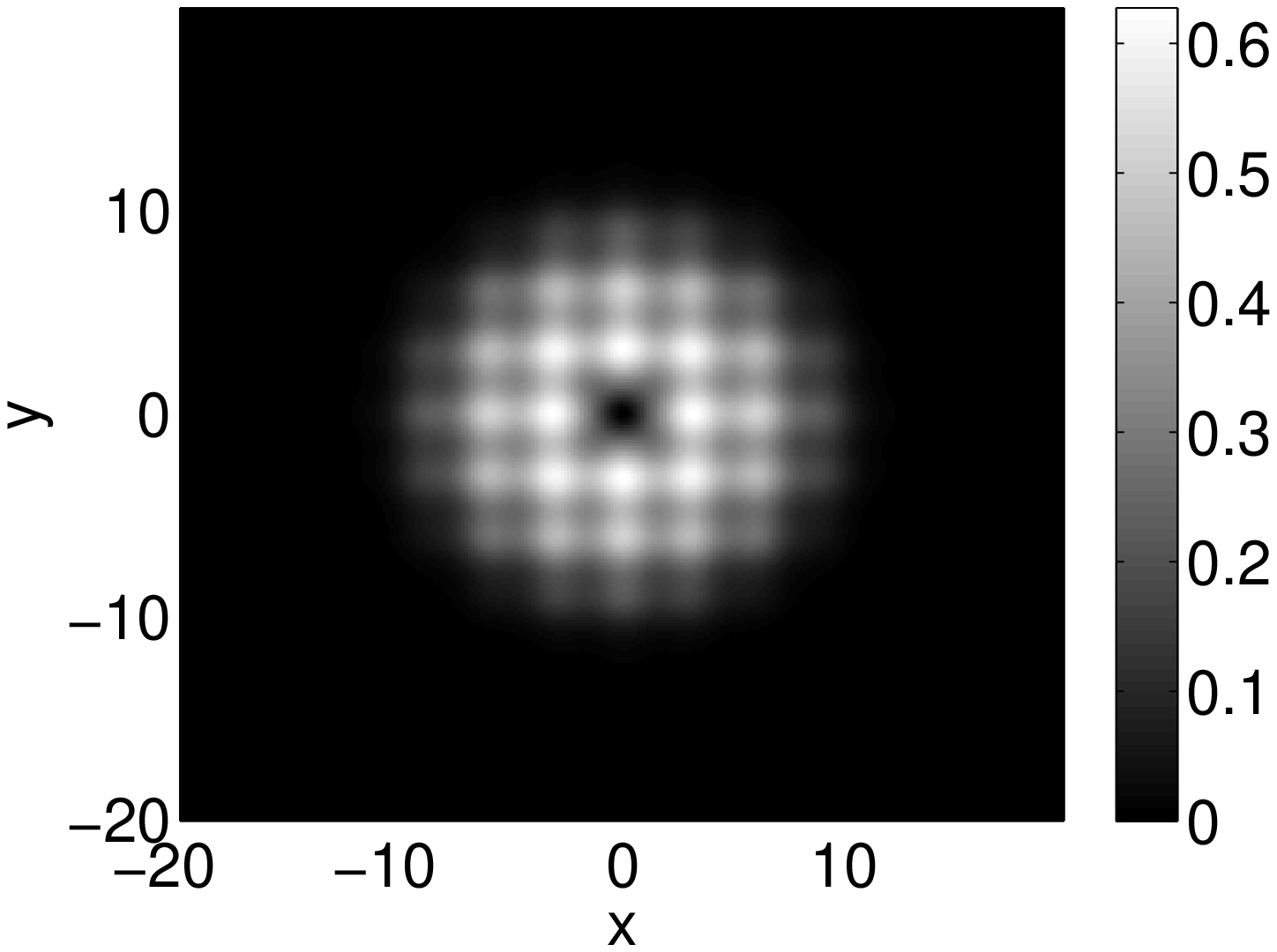}
\includegraphics[width=50mm]{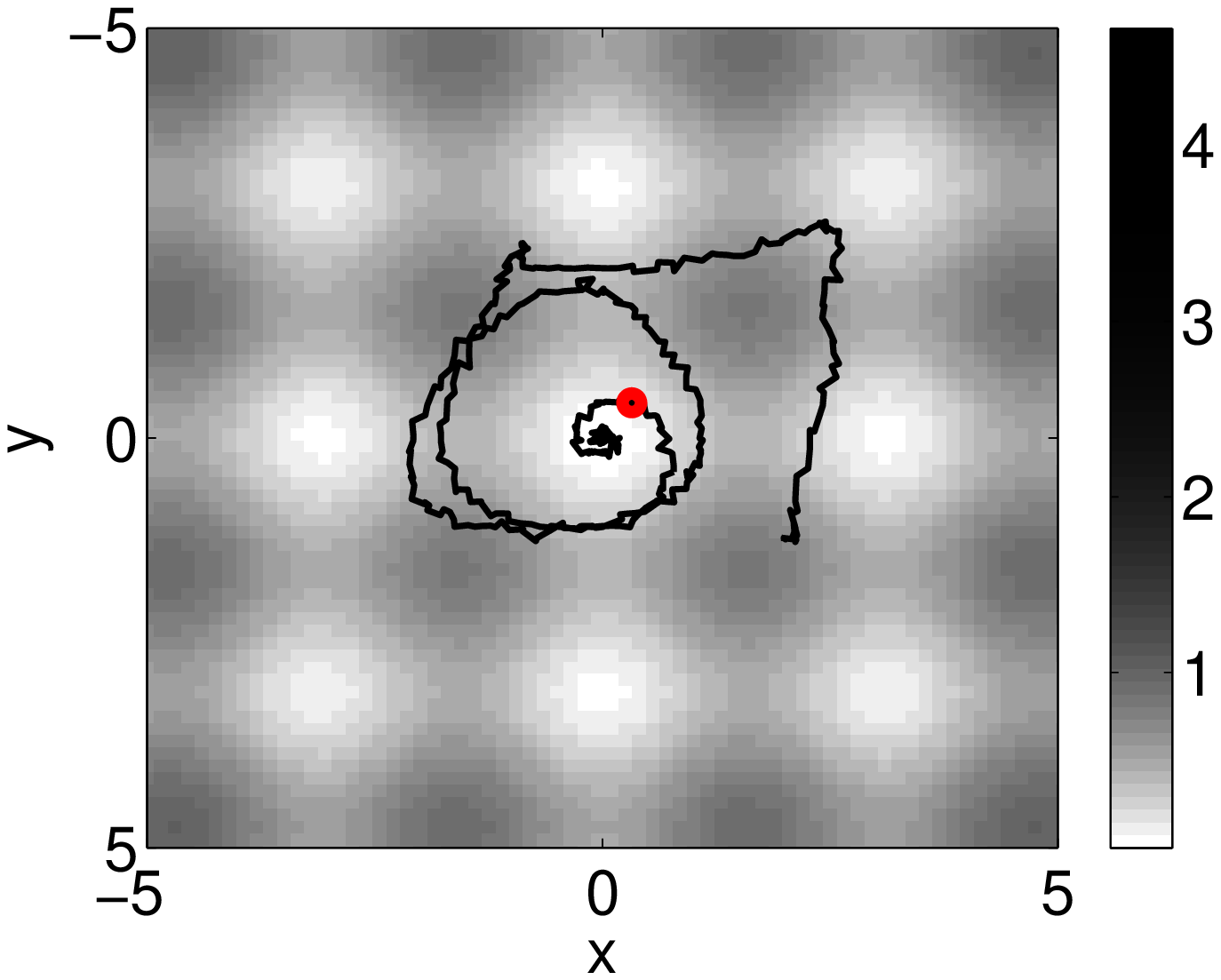}
\includegraphics[width=50mm]{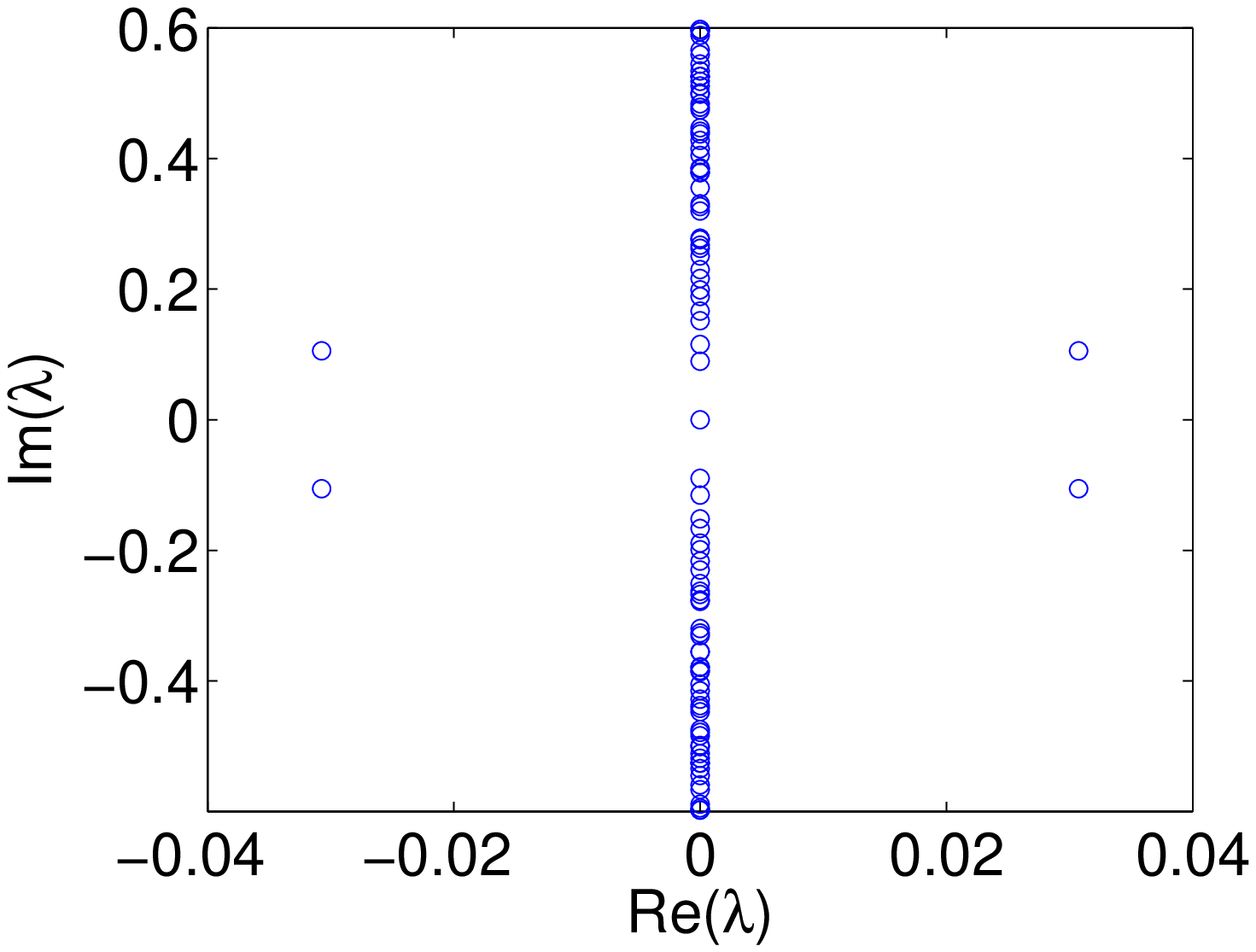}
\caption{(Color Online) The vortex trajectory for 450 time units for
some example charge 1 solutions are given above, center, with
the respective stationary solutions, left, and spectra, right.
All solutions are with $\mu=1$ and are overlayed
on top of the external potential
in order to clarify the setting. 
The top row is for $V_0=0$, middle is $V_0=0.2$, and 
the bottom row is for $V_0=0.4$.  The red dot represents
$\textbf{p}(t^*)$ (see \ref{vtx_traj}, \ref{tcrit}). The 
value of $t^*$ is $300$ for the middle case and $134$
for the bottom.}
\label{more_dyno_1}
\end{figure}

\begin{figure}
\includegraphics[width=100mm]{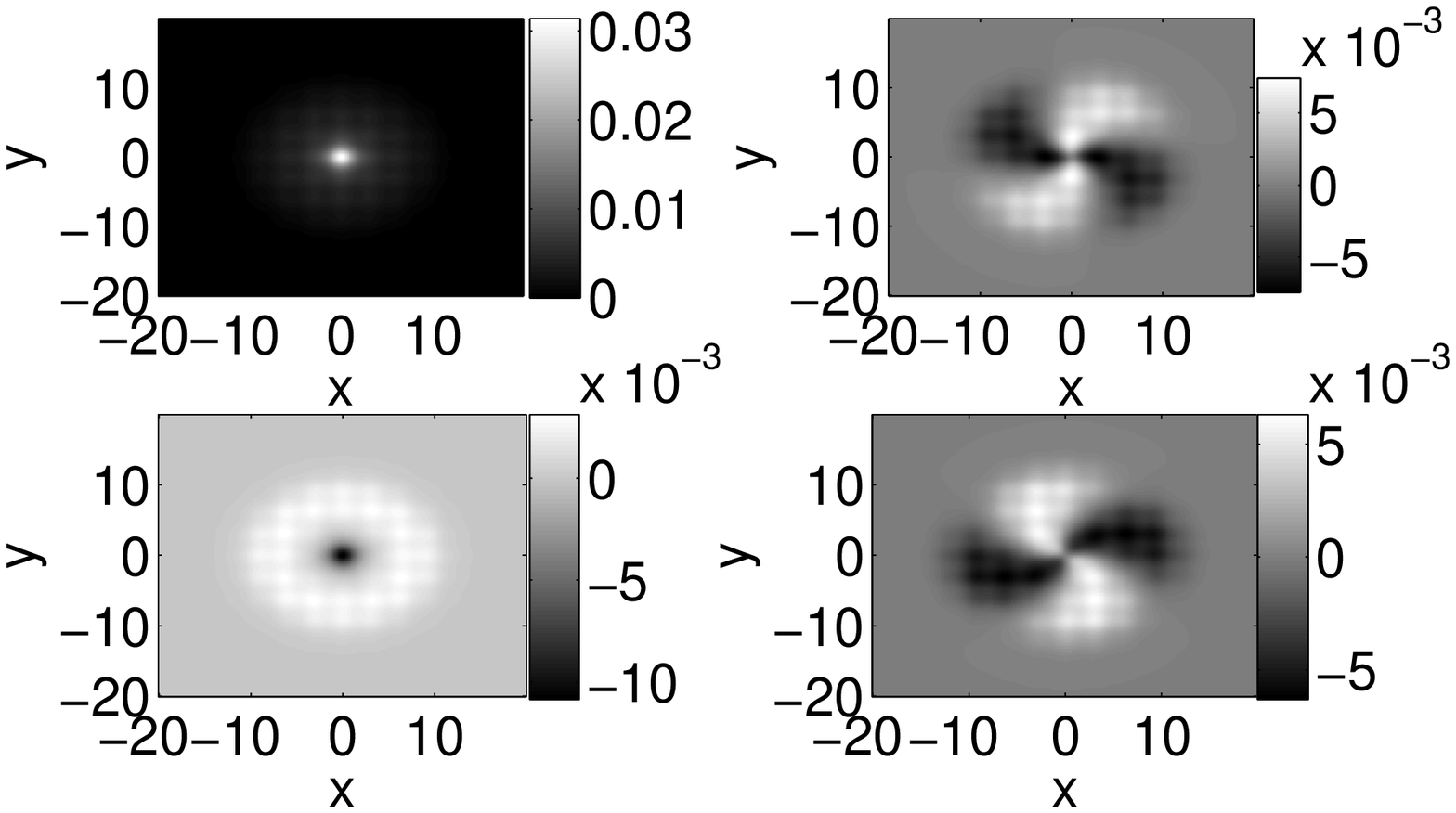}
\caption{(Color Online). The real (top row) and imaginary (bottom row) parts 
of the linearization eigenvectors $W_1$ (left column) and $W_2^*$ 
(right column) corresponding to the unstable eigenvalue of 
the charge 1 solution with $\phi=\pi/2$ for parameter values $(V_0,\mu)=(0.4,1)$ as in the evolution from the bottom row of Figure \ref{more_dyno_1}.  Notice the majority of the support of $W_1$ is real valued and located at the vortex 
core. This accounts for the dominant feature of the dynamics of the perturbed solution, namely, the displacement from it's original location.  The apparent 
vorticity of the secondary mode, $W_2^*$, then contributes to the circular path.}
\label{more_dyno_1_2}
\end{figure}

Our stability calculations above are corroborated by
direct numerical experiments involving the dynamical evolution
of the vortices. We perturb the stationary state at unstable 
parameter values (for $\phi=\pi/2$) in the direction $\tilde{u}$ of
its unstable eigenmode. 
We introduce the initial condition $u+\tilde{u}$ into 
our integrator, in this case a  Runge-Kutta, fourth order scheme.
Then, we integrate with the modest time step of $dt=10^{-2}$ afforded 
(stability-wise) by this method. 
The instability is clearly present and the vortex spirals out from 
the center of the condensate within the first 100 time units, as is 
shown in figure \ref{fig3}. We note in passing that this result is consonant
with the earlier findings of \cite{pgk}.

It is interesting enough to reiterate
that for $\phi=\pi/2$, the instability
growth rate 
is zero for small $V_0$, while for larger
$V_0$, it becomes nonzero, in fact increasing initially as a function
of $V_0$ as seen in Fig. \ref{fig2} 
(before it eventually decreases back to zero along with the solution,
for very strong lattices). This indicates that the vortices
should spiral outward faster for larger $V_0$ (with the exception
of the case of very strong lattices), which is perhaps counter-intuitive.
In addition to the vorticity dynamics illustrated in figure \ref{fig3},
this fact is confirmed by observing
the trajectory $\textbf{p}(t)$, defined below, of the initially stationary 
vortex slightly perturbed by a random noise of amplitude $\pm 0.05\%$ of
the maximum amplitude of the solution, 
for different values of $V_0$ in the runs of Fig. \ref{more_dyno_1}.
In these runs, the location of the vortex center is obtained by  
\begin{equation}
\textbf{p}(t) = \frac{1}{||v||_2^2} ( \int x |v|^2 dA ,  \int y |v|^2 dA ), 
\label{vtx_traj}
\end{equation}
where v(x,y,t) is the vorticity,   
\begin{equation}
v (x,y,t) = \nabla \times \{- \frac{i}{|u|^2} [  u^* \nabla u - u \nabla u^* ] \}.
\label{vtcty}
\end{equation}
A related quantitative diagnostic is defined as
\begin{equation}
t^*={\rm min}\{t;||\textbf{p}(t)||_2>0.5\},
\label{tcrit}
\end{equation}
illustrating the time at which the vortex arrives at a distance
of $0.5$ from its initial central location 
($||\cdot||_2$ is the $L_2$ norm of Euclidean distance).
In the absence of external potential, the 
perturbed vortex remains at the center of the harmonic trap for the 
entire simulation time of 450 units (see top row of Fig. \ref{more_dyno_1}). 
For a slightly larger value of $V_0=0.2$ it becomes
slightly unstable due to a small magnitude 
eigenvalue quartet 
(with ${\rm max (Re} (\lambda))=0.0165$) which emerges.
See the middle row of Fig. \ref{more_dyno_1}, and notice the (red)
dot in the center image presenting the dynamics.  This 
corresponds to 
$\textbf{p}(t^*)$, where $t^*=300$, 
which represents the first time the vortex leaves the radius of $0.5$ 
from its original position.
Notice that subsequently the perturbed vortex remains within the local 
minimum of the lattice for the duration of 450 time units.  
As $V_0$
is increased further
to the value of $V_0=0.4$, depicted at the 
bottom row of the Fig. (\ref{more_dyno_1}), it moves away from
the initial location faster, having a
critical time of $t^*=134$. This is in line with 
the fact that this case corresponds to a larger
growth rate with ${\rm max (Re} (\lambda))=0.0307$.
We should point out that this outward from the center motion
of the vortex core may, in fact, be expected on the basis
of the unstable eigenmodes of the linearization analysis 
shown in Fig. \ref{more_dyno_1_2}. There, it can be seen that
the dominant instability eigenmode has a maximum (real) amplitude at
the center and hence, when amplified, it will lead to the displacement
of the vortex core from the center of the trap.  The vorticity of 
the secondary mode will then contribute to the circular trajectory.
 
\begin{figure}
\includegraphics[width=100mm]{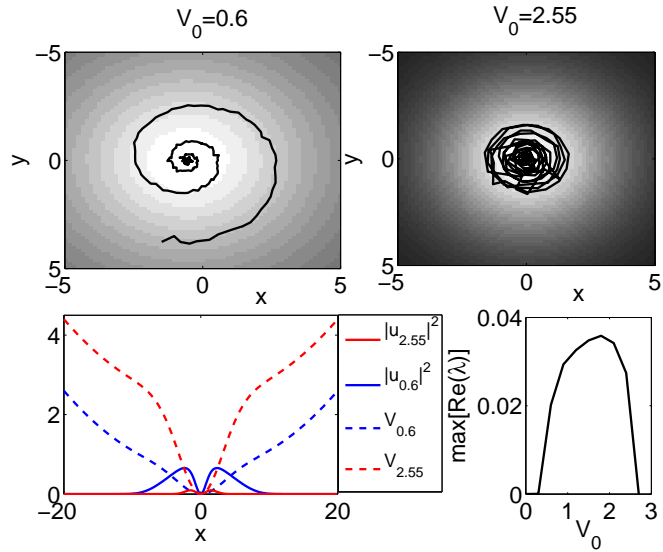}
\caption{(Color Online) The above results are for the Gaussian dimple 
potential given in equation (\ref{dimple}) with $\sigma=2\sqrt{10}$.  
The top row displays the (perturbed) vortex trajectory for $350$
time units of two solutions with fixed $\mu=0.99$. 
The left panel is for the value $V_0=0.6$, and the vortex is 
able to escape the confines of the dimple, while in the right panel, 
where $V_0=2.55$, the vortex remains in the deep dimple which has 
the approximate form of a harmonic potential itself, of frequency 
$2\sqrt{V_0}/\sigma$. The bottom left panel shows the potential 
(dotted) and the density proifile
(solid) for $V_0=0.6$ (blue) and $V_0=2.55$ (red). The 
corresponding instability growth rate branch is given in the 
bottom right panel. It is interesting that, while the growth rates in
the two cases are comparable, the parabolic nature of the deep dimple 
($V_0=2.55$) prevents the spiraling out of the vortex during 
dynamical evolution.}
\label{prl_comp}
\end{figure}

It is interesting to compare these results with the work of
\cite{prl_npp} where the effect of sound emission and corresponding spiraling
of the vortex away from the center of a magnetic trap with
an additional {\it radial} dimple potential is considered.
In the latter case, it was observed that for small amplitude of
the dimple potential, the sound emission and vortex spiraling
were facilitated, while for larger amplitudes of the dimple, the
vortex was trapped within the dimple. While we point out that
the case of \cite{prl_npp} appears to have a reverse dependence 
of the evolution dynamics
on the strength of the additional potential (to the magnetic one),
we should also point out that there are fundamental differences between
the two cases such as for instance the radial form of the dimple
potential in comparison with the anisotropic spatial dependence of the 
periodic potential considered herein. In fact, we have performed stability
computations in the context of the trap of \cite{prl_npp}, 

\begin{equation}
V=V_M+V_0[1-\exp(\frac{2(x^2+y^2)}{\sigma^2})].
\label{dimple}
\end{equation}

We found a similar instability and dependence of its growth rate on $V_0$ 
(the strength of the dimple amplitude) as in our optical
lattice results above (see the bottom right panel of Fig. \ref{prl_comp}); 
our numerical experiments on the evolution of the instability, 
on the other hand, confirm the findings of \cite{prl_npp} (see the 
top row of Fig. \ref{prl_comp}).

\subsection{$S=2$ Vortices}

\begin{figure}
\includegraphics[width=60mm]{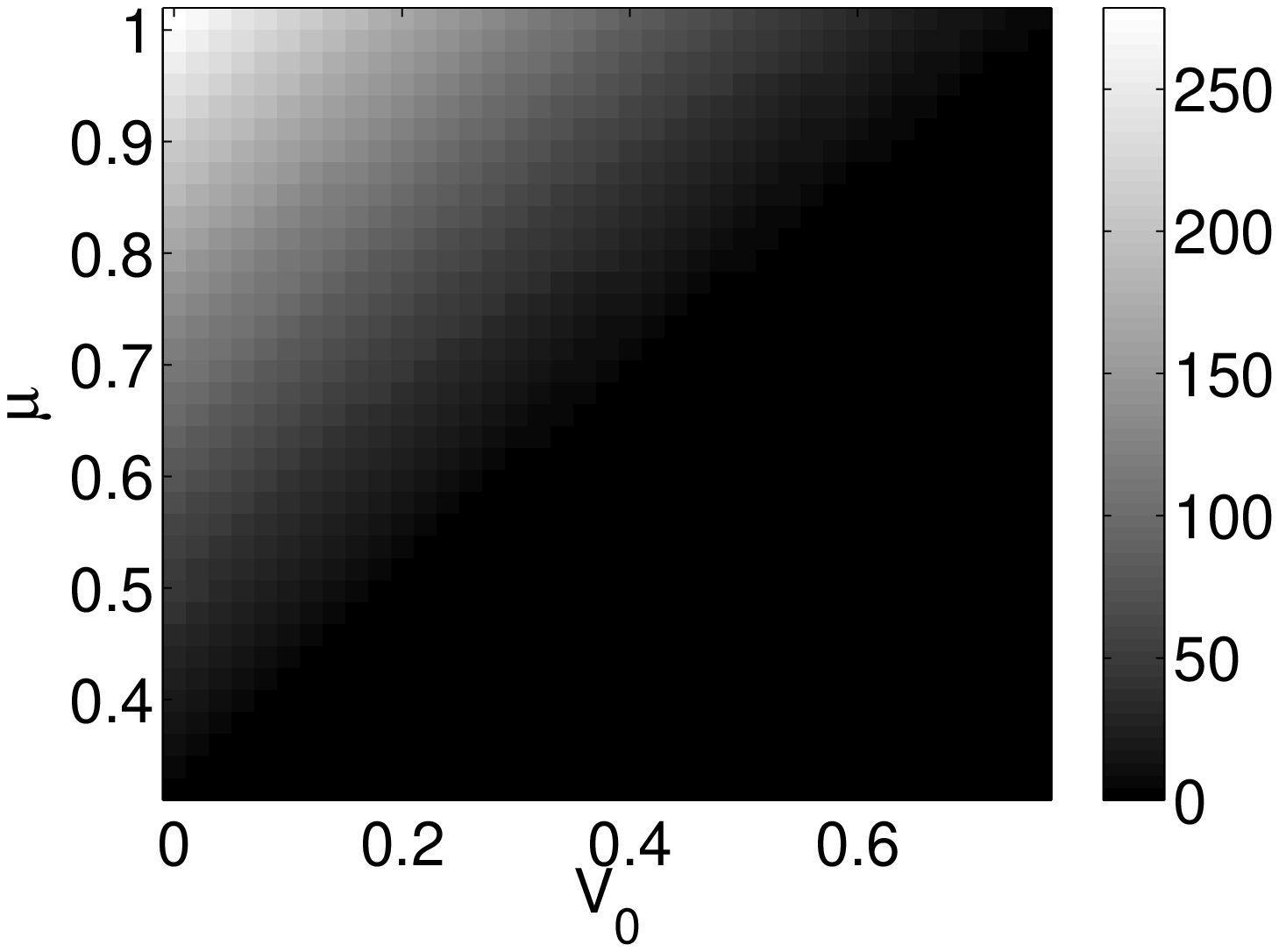}
\includegraphics[width=60mm]{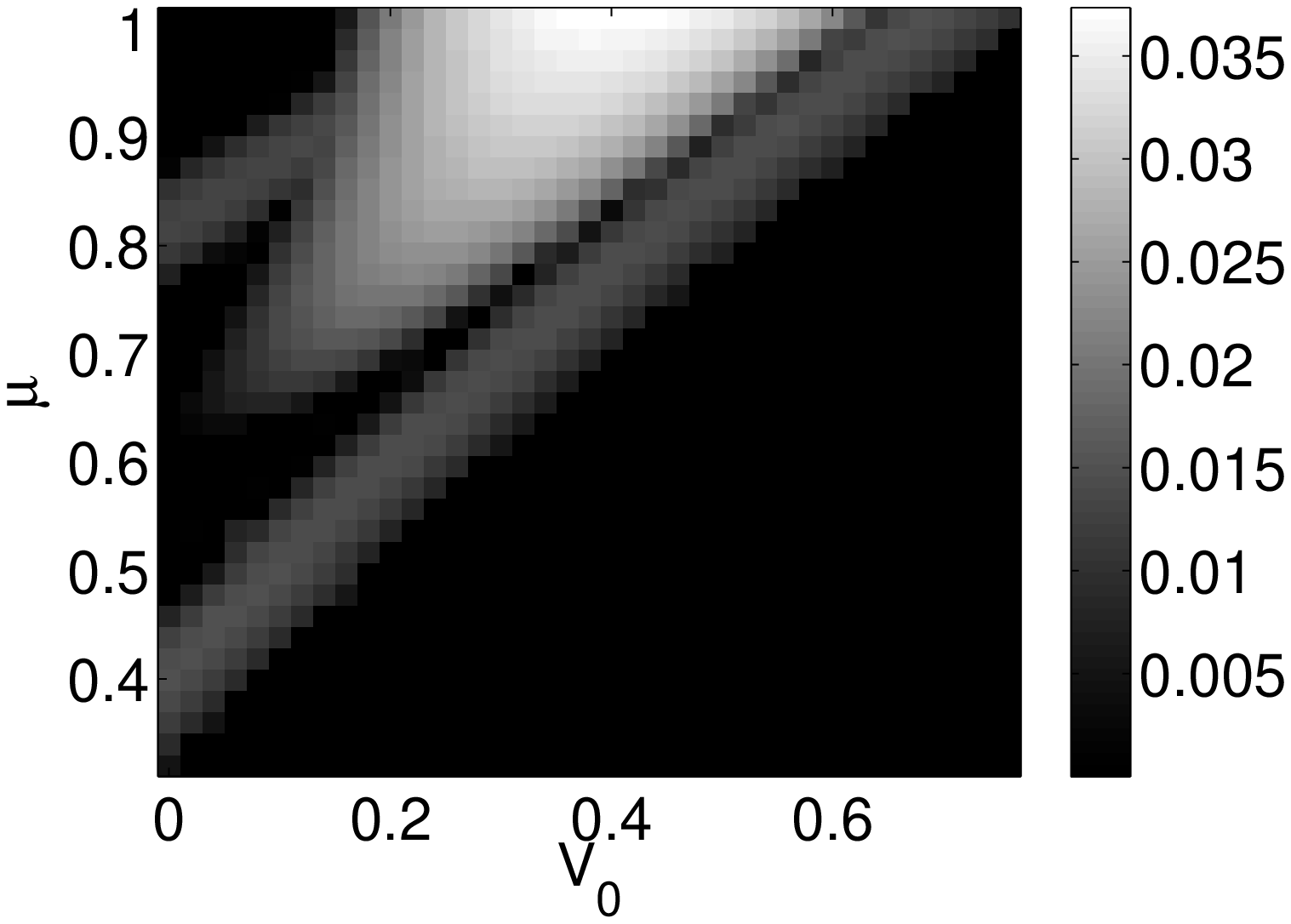}
\includegraphics[width=60mm]{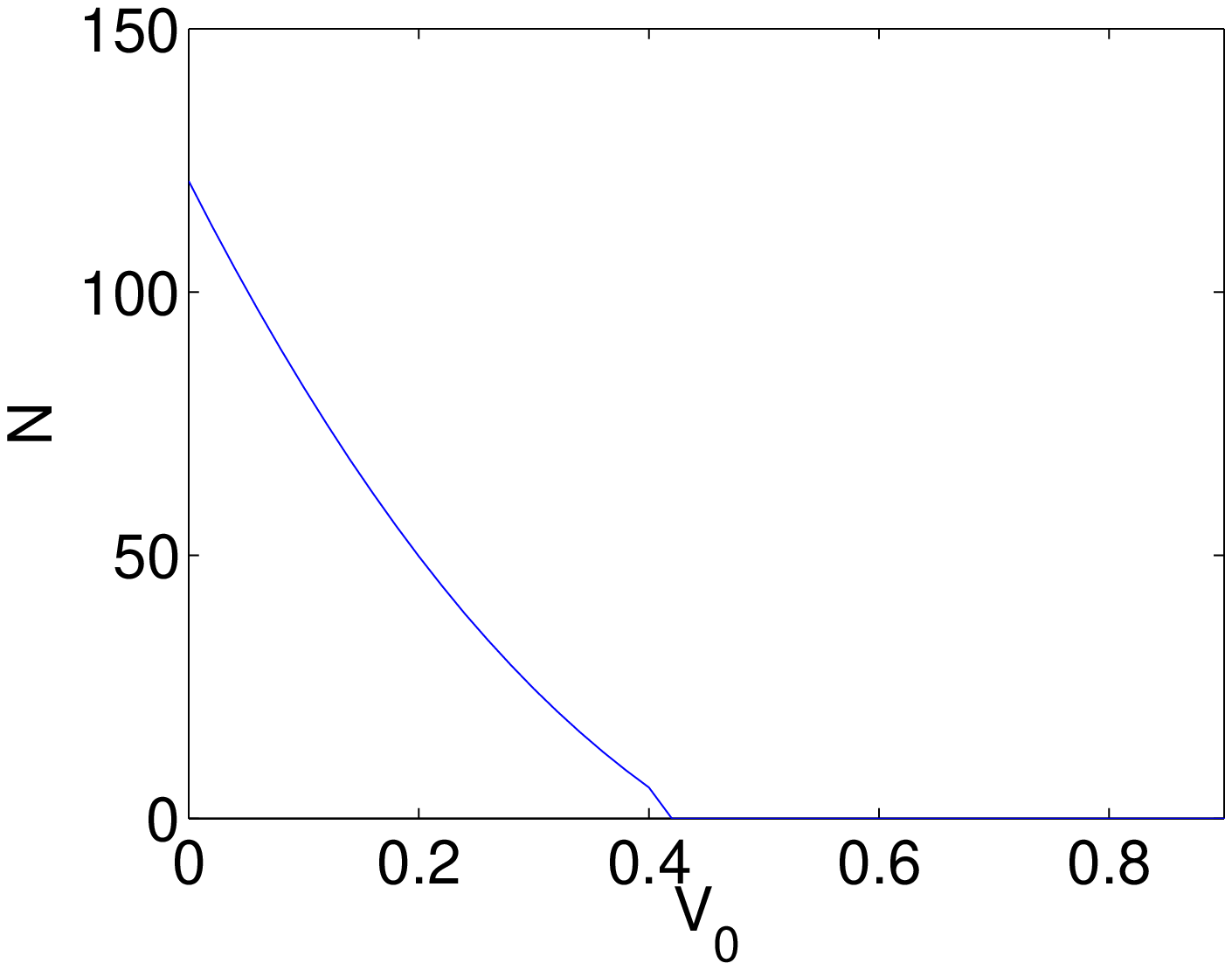}
\includegraphics[width=60mm]{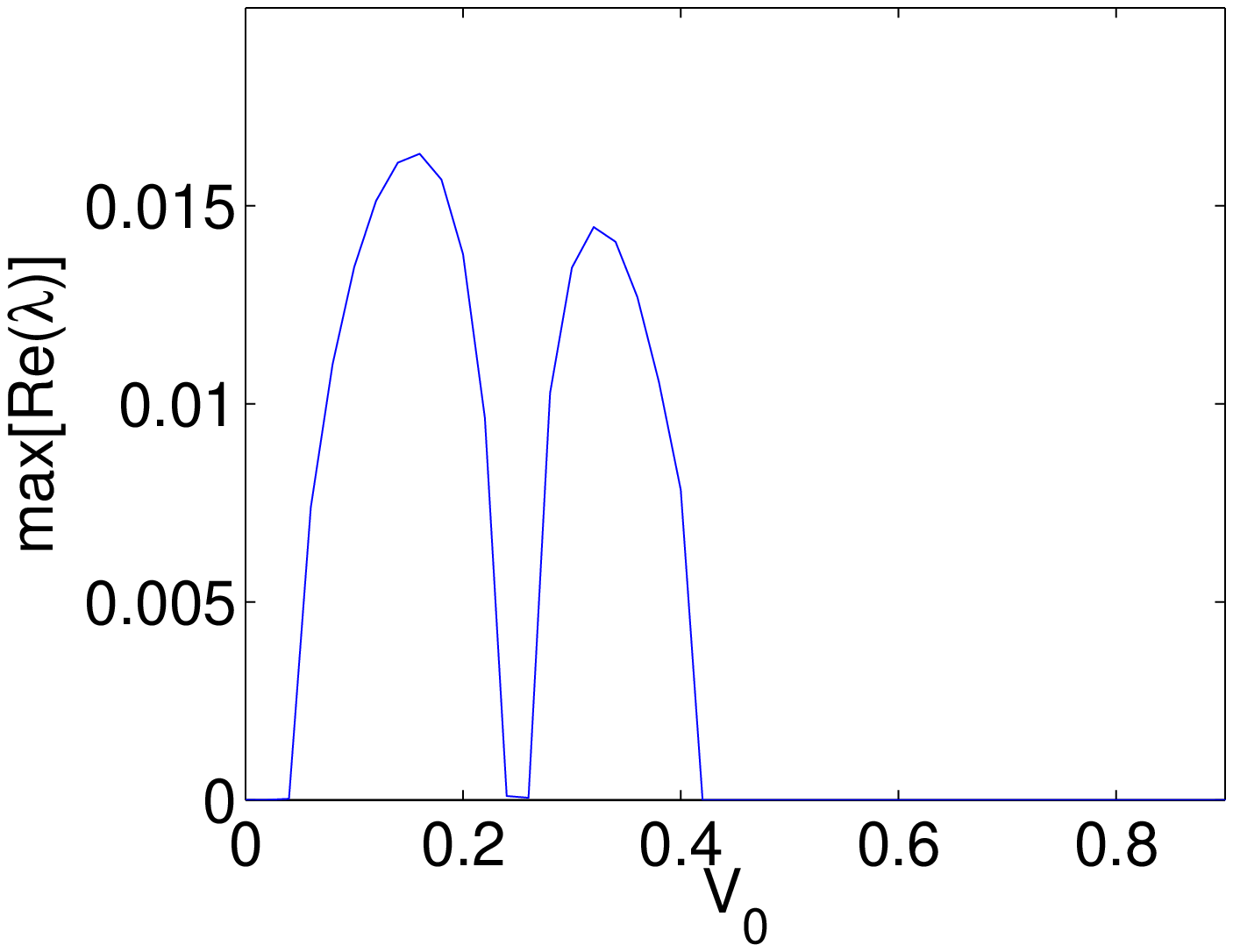}
\includegraphics[width=60mm]{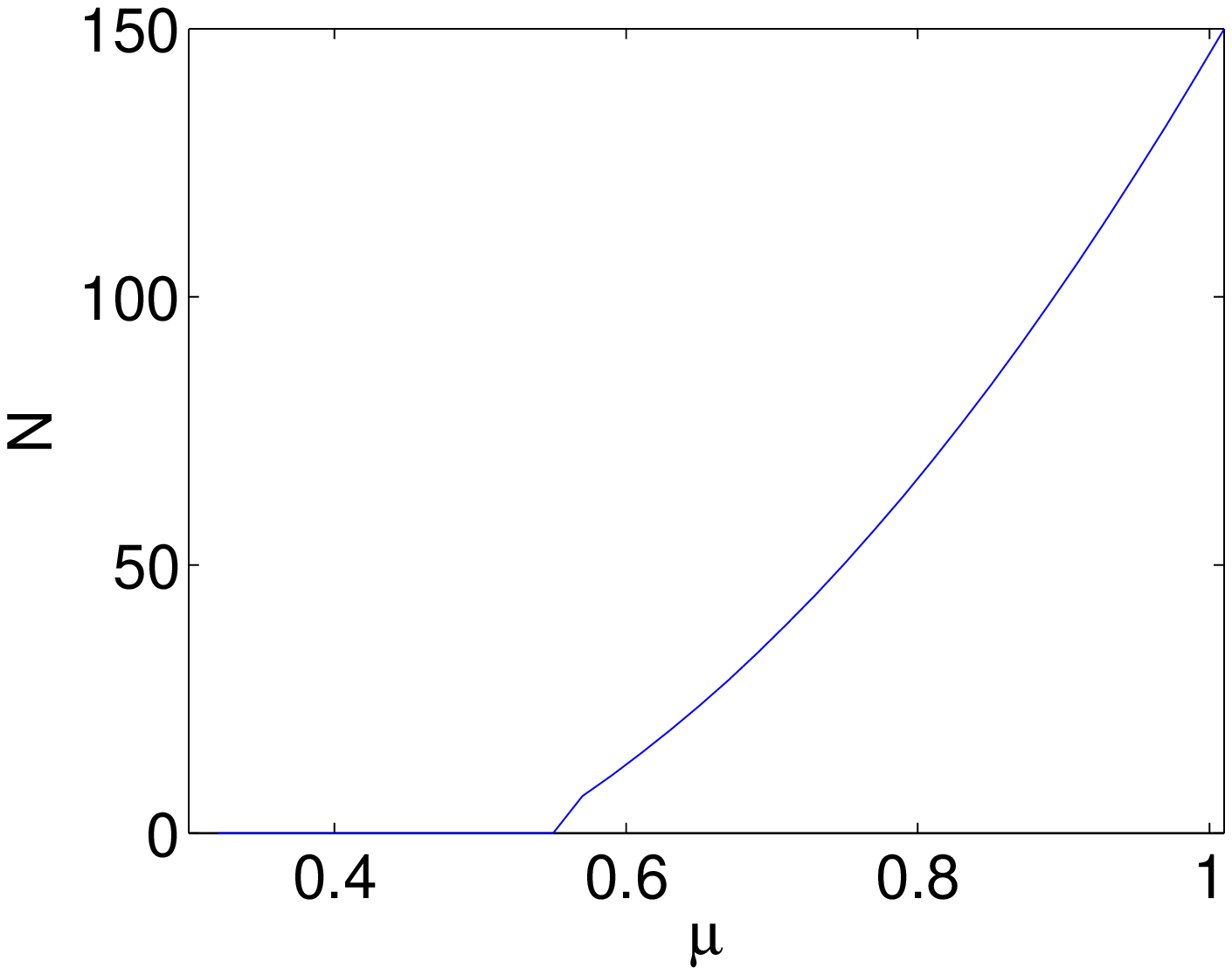}
\includegraphics[width=60mm]{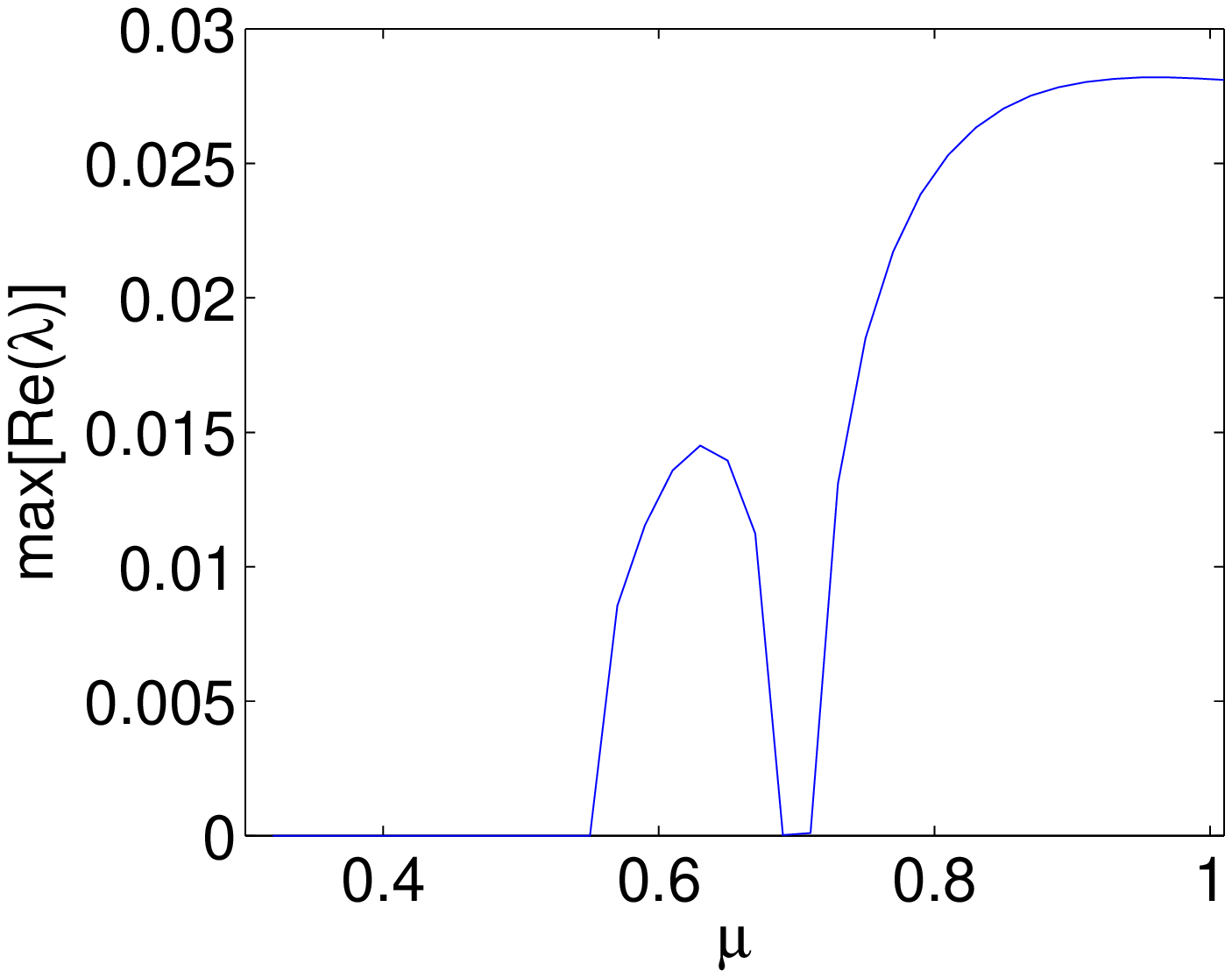}
\caption{(Color Online) The corresponding images to Fig. \ref{fig2} 
for a double charge vortex with 
$\phi=\pi/2$ in the same layout as for the single charge case in 
Fig. \ref{fig2}. $V_0$ slices are for $\mu=0.71$ and $\mu$ slices are for 
$V_0=0.24$.}
\label{fig4}
\end{figure}

We now turn to the case of the doubly quantized vortex.
Again, when $V_0=0$, we observe the solutions disappear 
at the linear QHO limit of $\mu=0.3$.  
Our results in the  $V_0=0$ continuation over $\mu$ are 
consistent with those of \cite{pu}. In particular,
there are windows of stability along this branch 
for  $\mu$ in the intervals $(0.45,0.75)$ and $(0.85,1)$.
Then, as we fill out the surface of solutions in $V_0$ we
observe in Fig. \ref{fig4} a similar behavior of $N$ (as
for the $S=1$ vortices), while 
the stability intervals appear to narrow as $V_0$ increases.
Furthermore the stability landscapes 
appear to be quite similar for $\phi=0$ and $\phi=\pi/2$, as can be 
observed in the comparison of the 
upper right corners of figures \ref{fig4} and \ref{fig7}.
While the stability intervals quickly narrow as $V_0$ increases,
it is worthwhile to note that stable solutions with a considerable
amount of modulation by the optical lattice can still be identified.

\begin{figure}
\includegraphics[width=60mm]{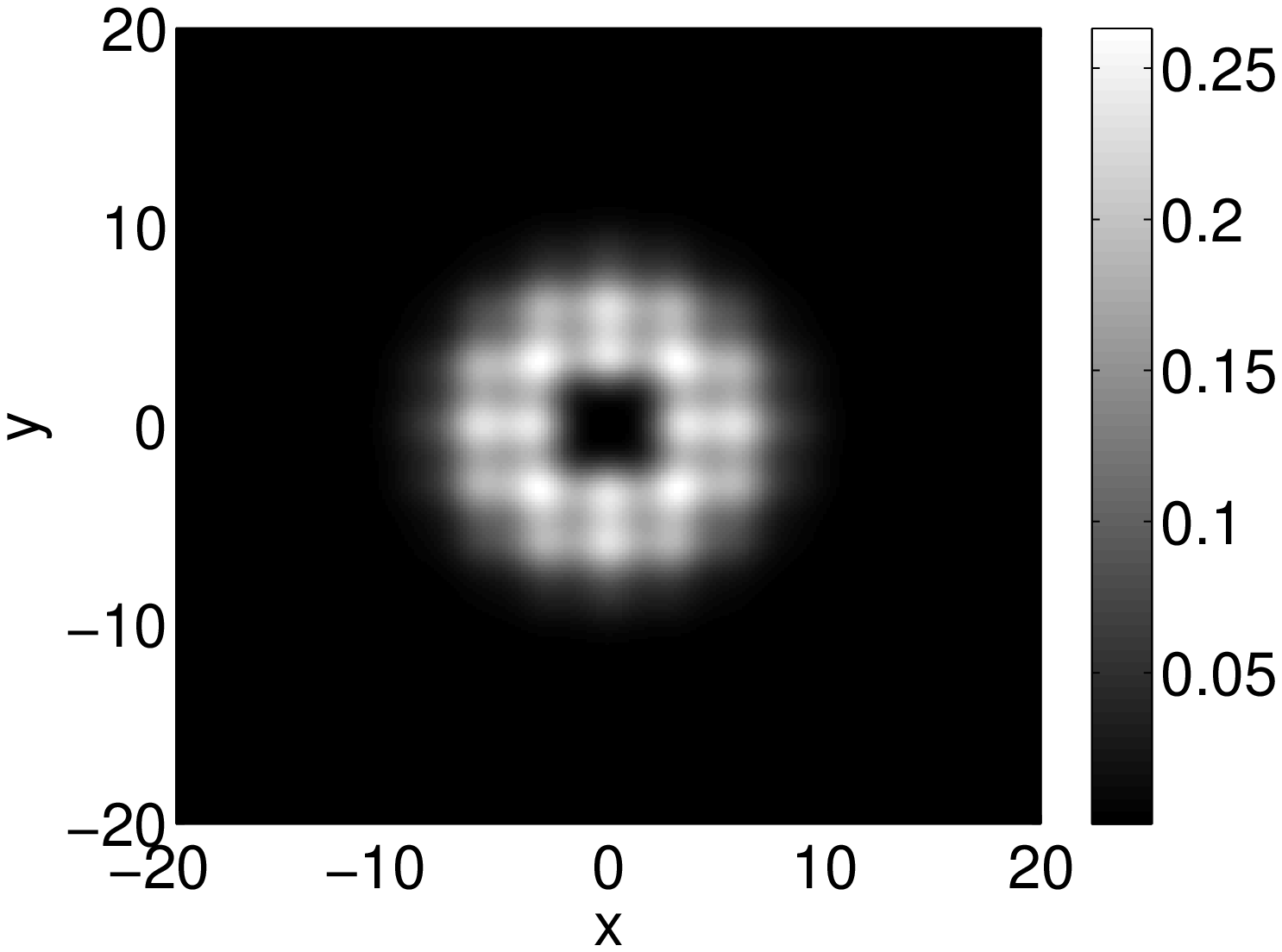}
\includegraphics[width=60mm]{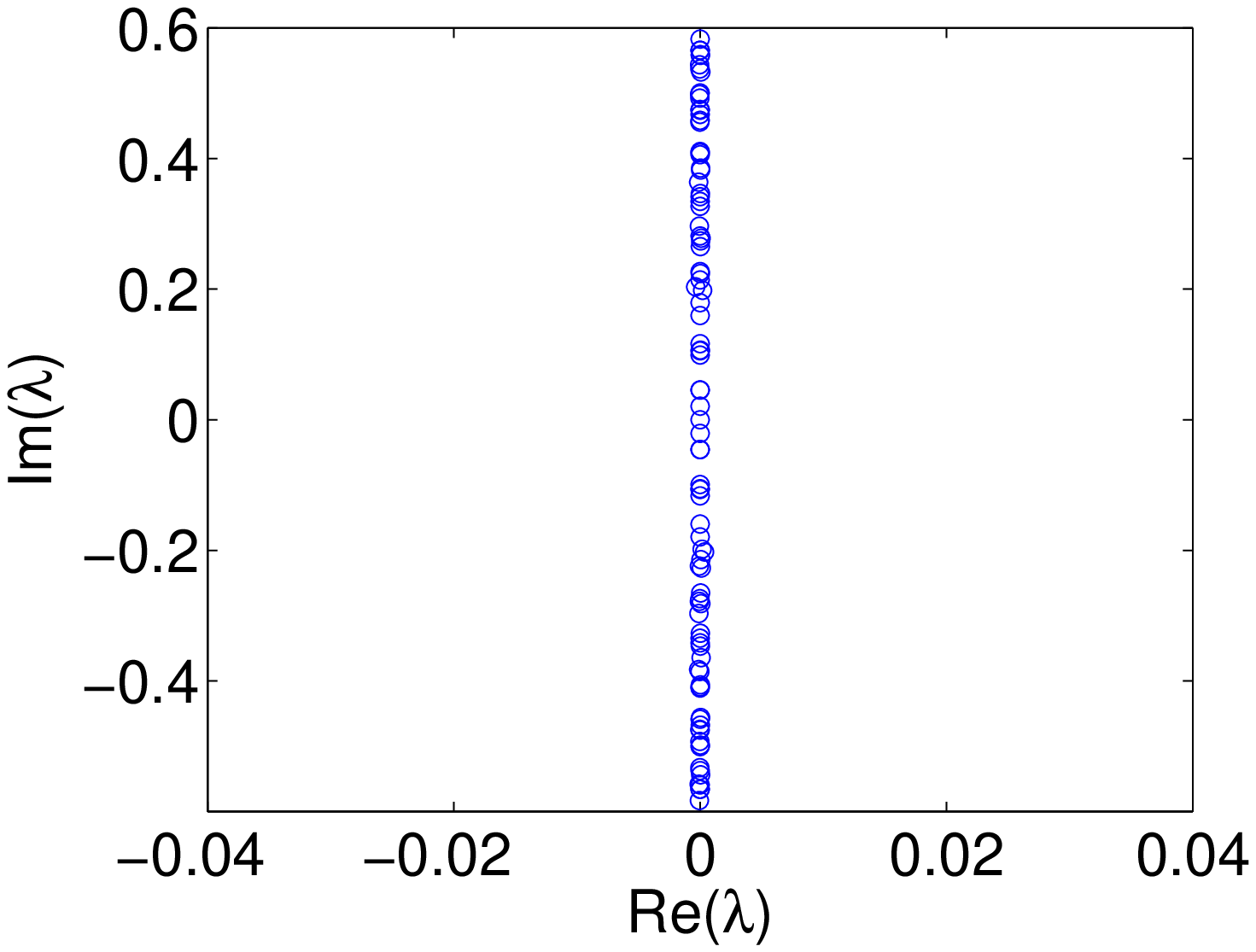}
\includegraphics[width=60mm]{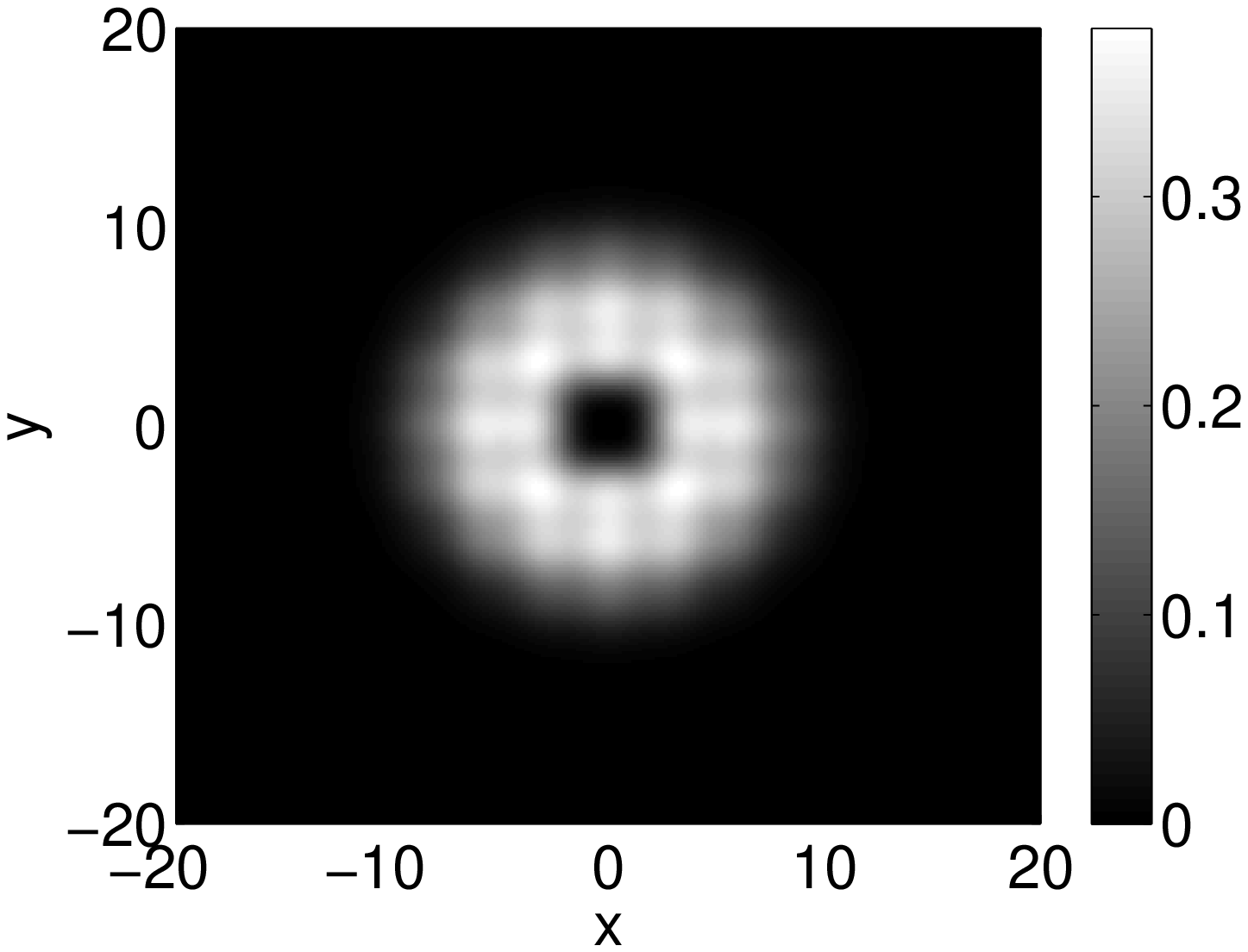}
\includegraphics[width=60mm]{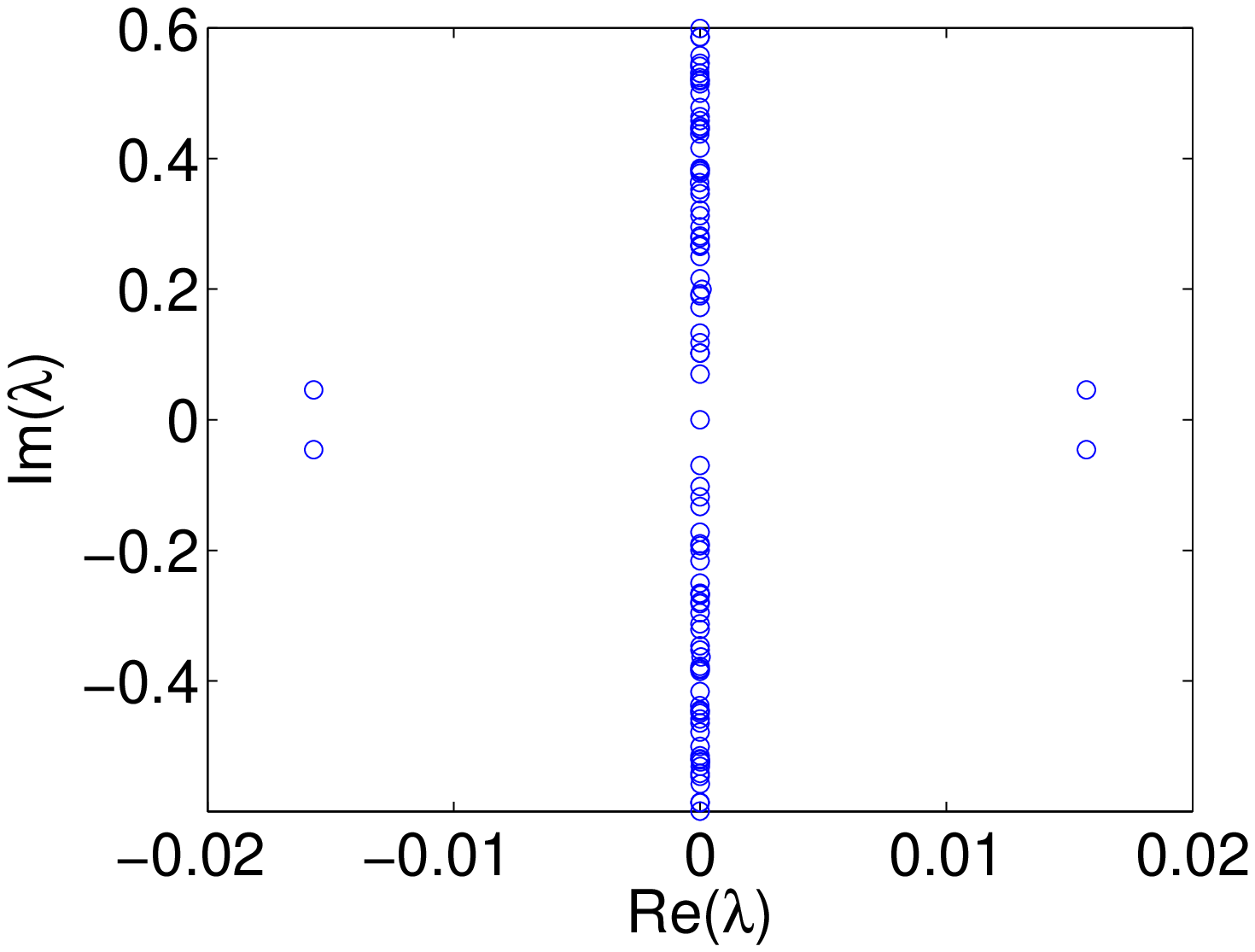}
\caption{(Color Online) Stable (top, $(0.25,0.71)$) 
and unstable (bottom, $(0.13,0.71)$) double charge vortices 
(both their profiles and their corresponding
spectral planes of eigenvalues) with $\phi=\pi/2$.}
\label{fig5}
\end{figure}

\begin{figure}
\includegraphics[width=100mm]{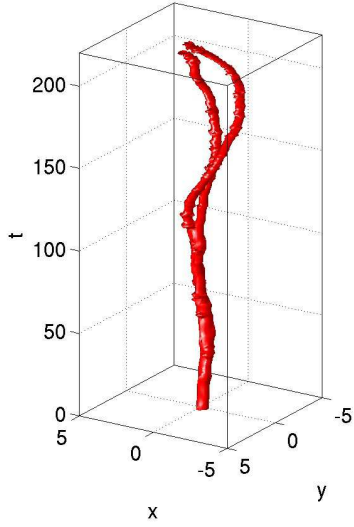}
\caption{(Color Online) The space-time dynamics, 
similarly to Fig. \ref{fig3}, 
of the unstable mode when $\phi=\pi/2$ for parameter values $(0.13,0.71)$.}
\label{fig6}
\end{figure}

\begin{figure}
\includegraphics[width=60mm]{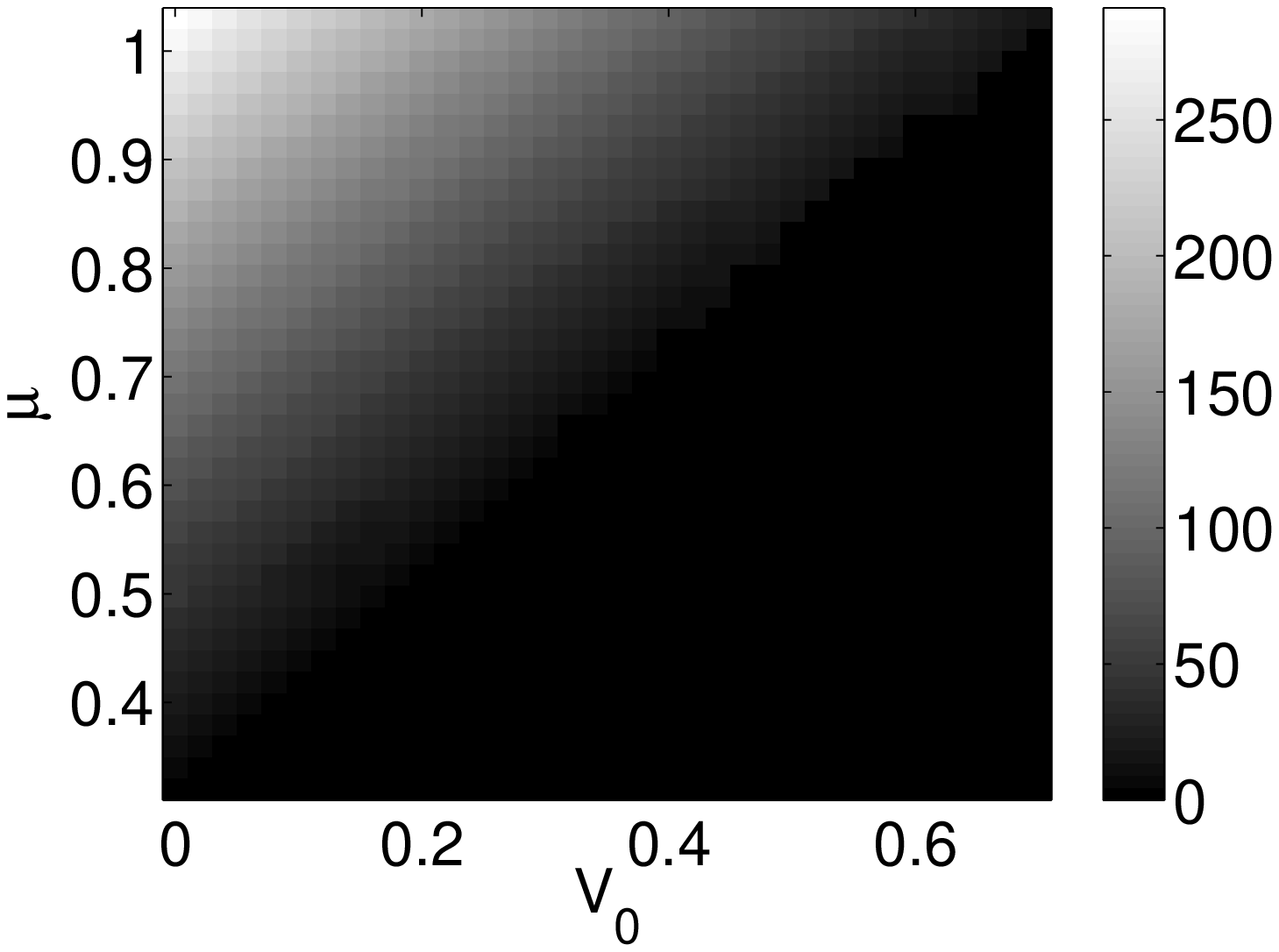}
\includegraphics[width=60mm]{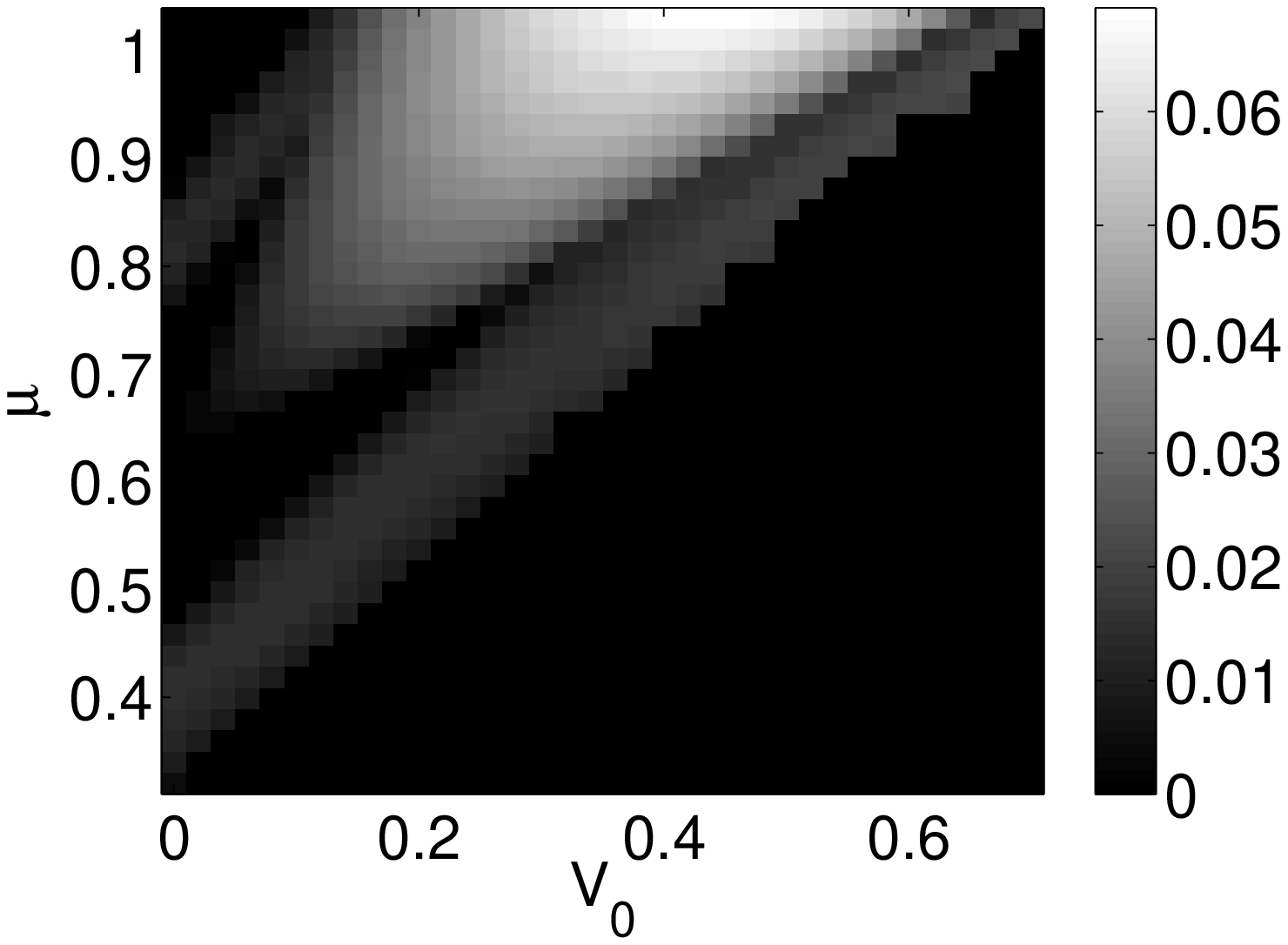}
\includegraphics[width=60mm]{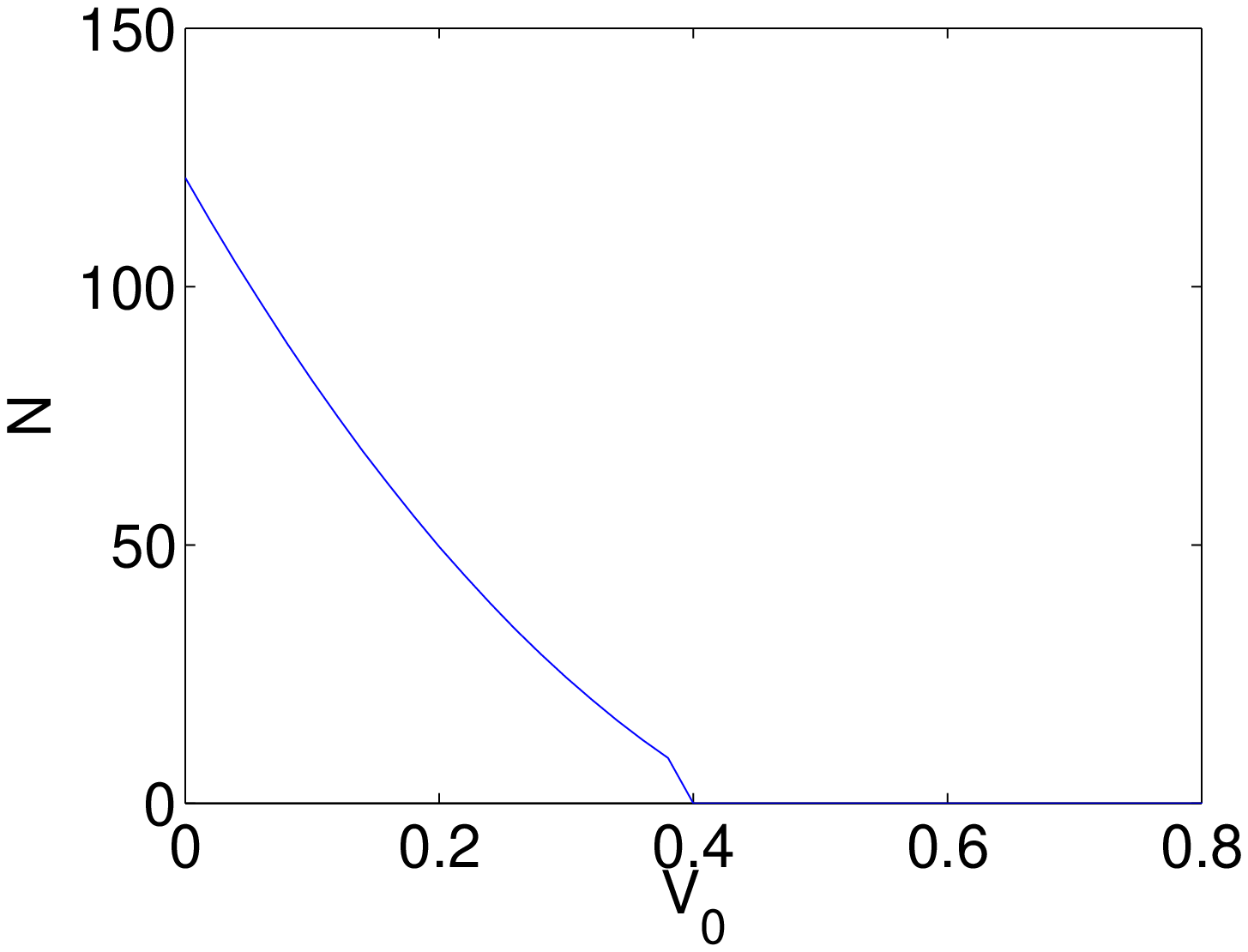}
\includegraphics[width=60mm]{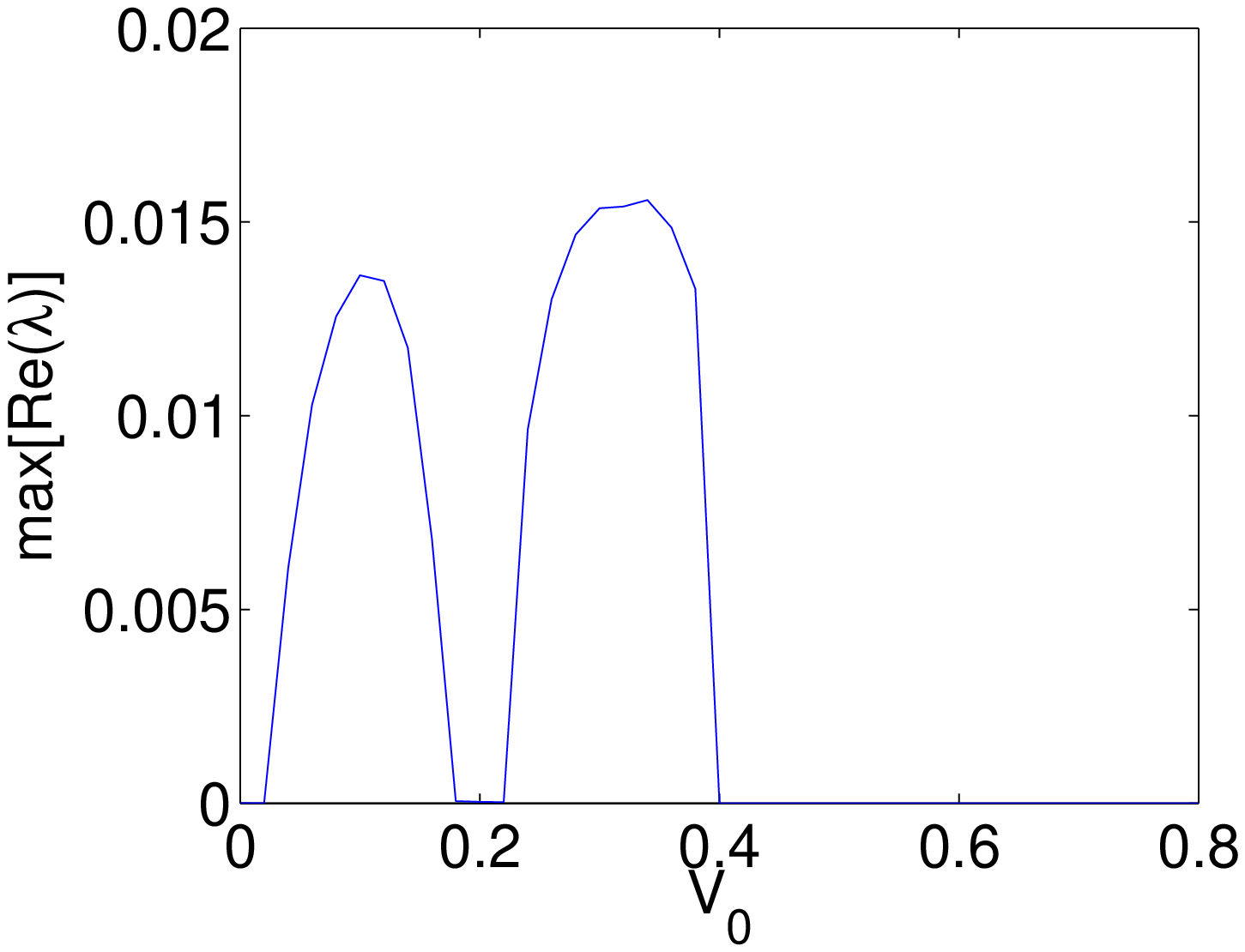}
\includegraphics[width=60mm]{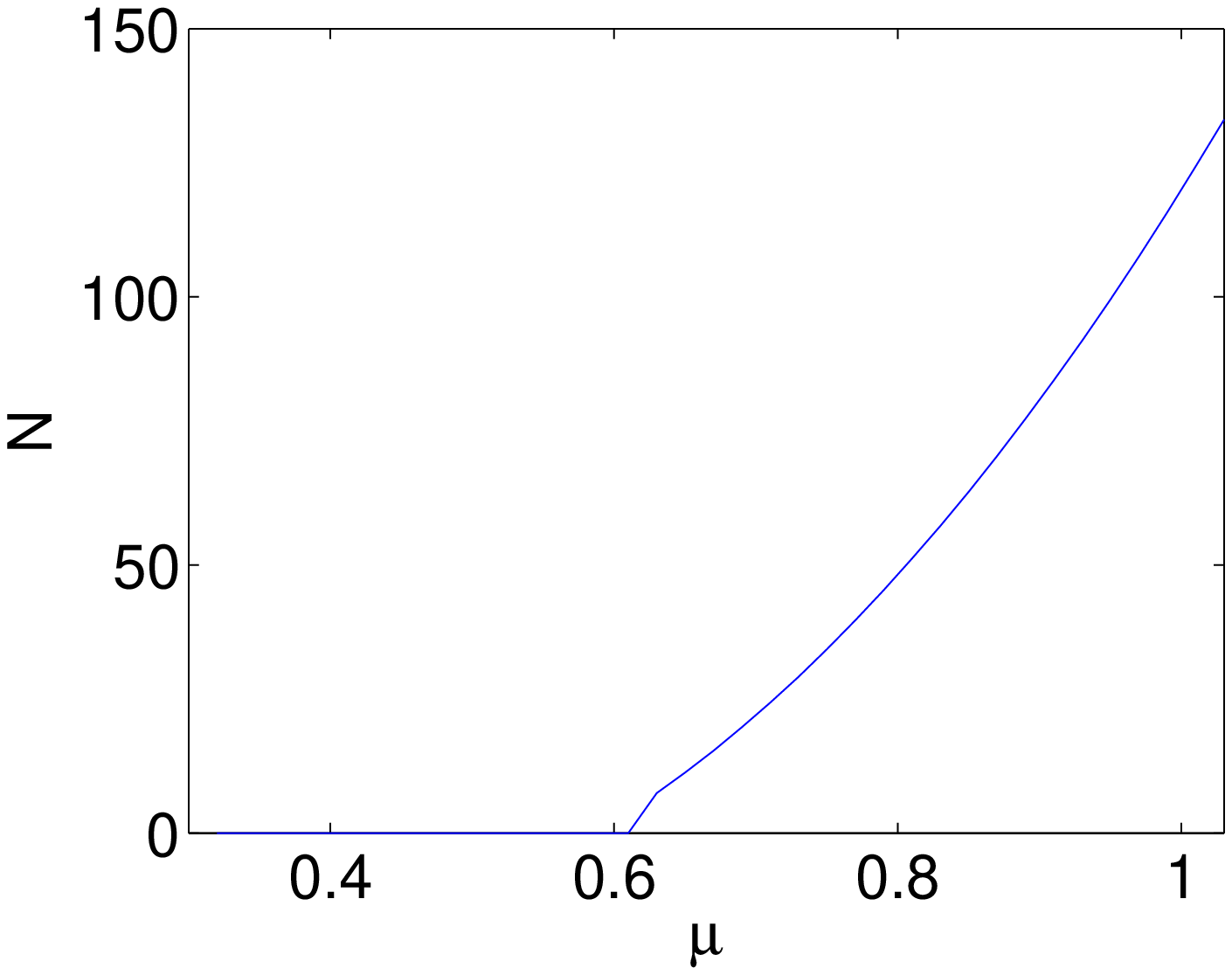}
\includegraphics[width=60mm]{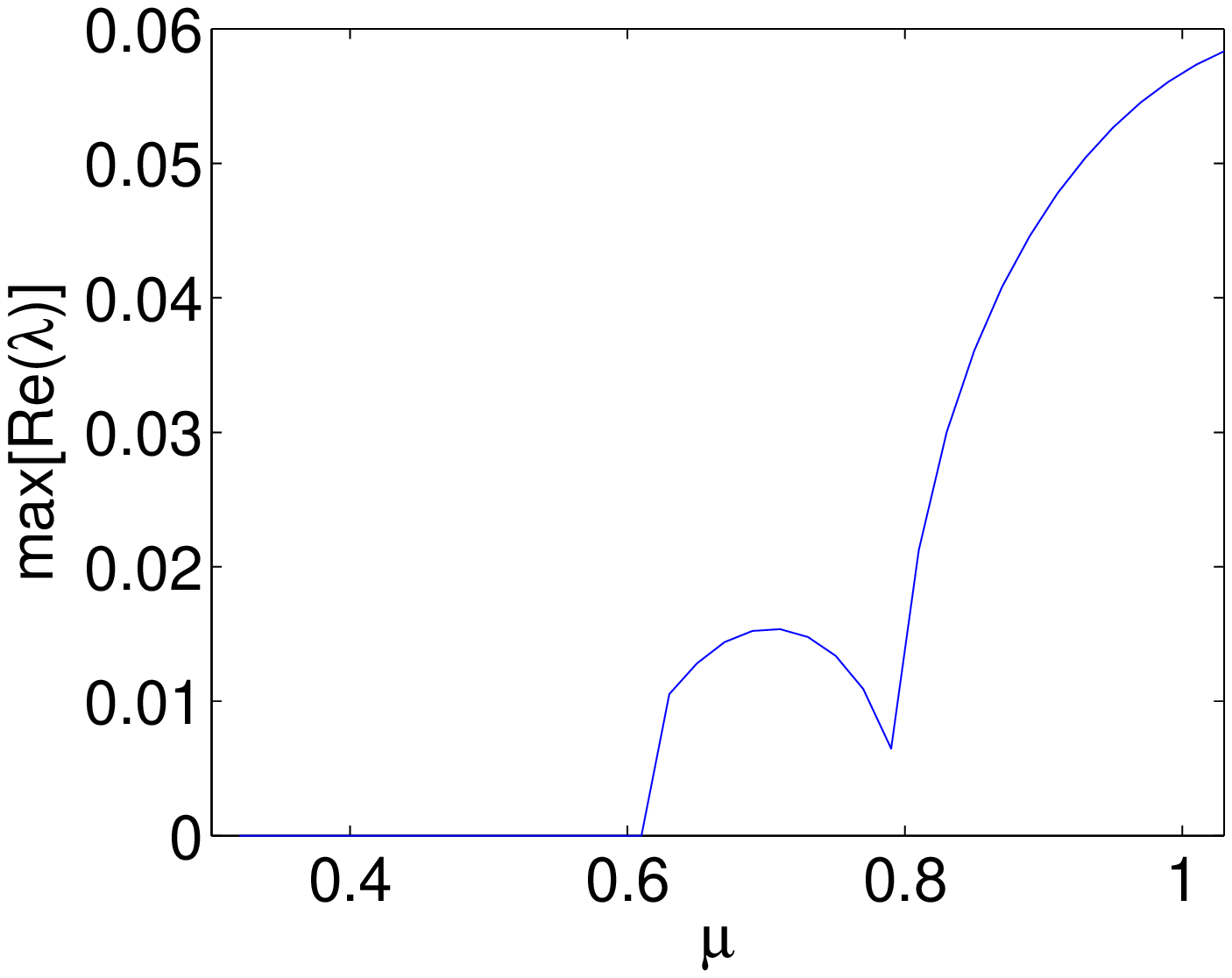}
\caption{(Color Online) Same as figure \ref{fig4} 
except with $\phi=0$. $V_0$ slices 
are for $\mu=0.71$ again and $\mu$ slices are for $V_0=0.3$.}
\label{fig7}
\end{figure}

\begin{figure}
\includegraphics[width=60mm]{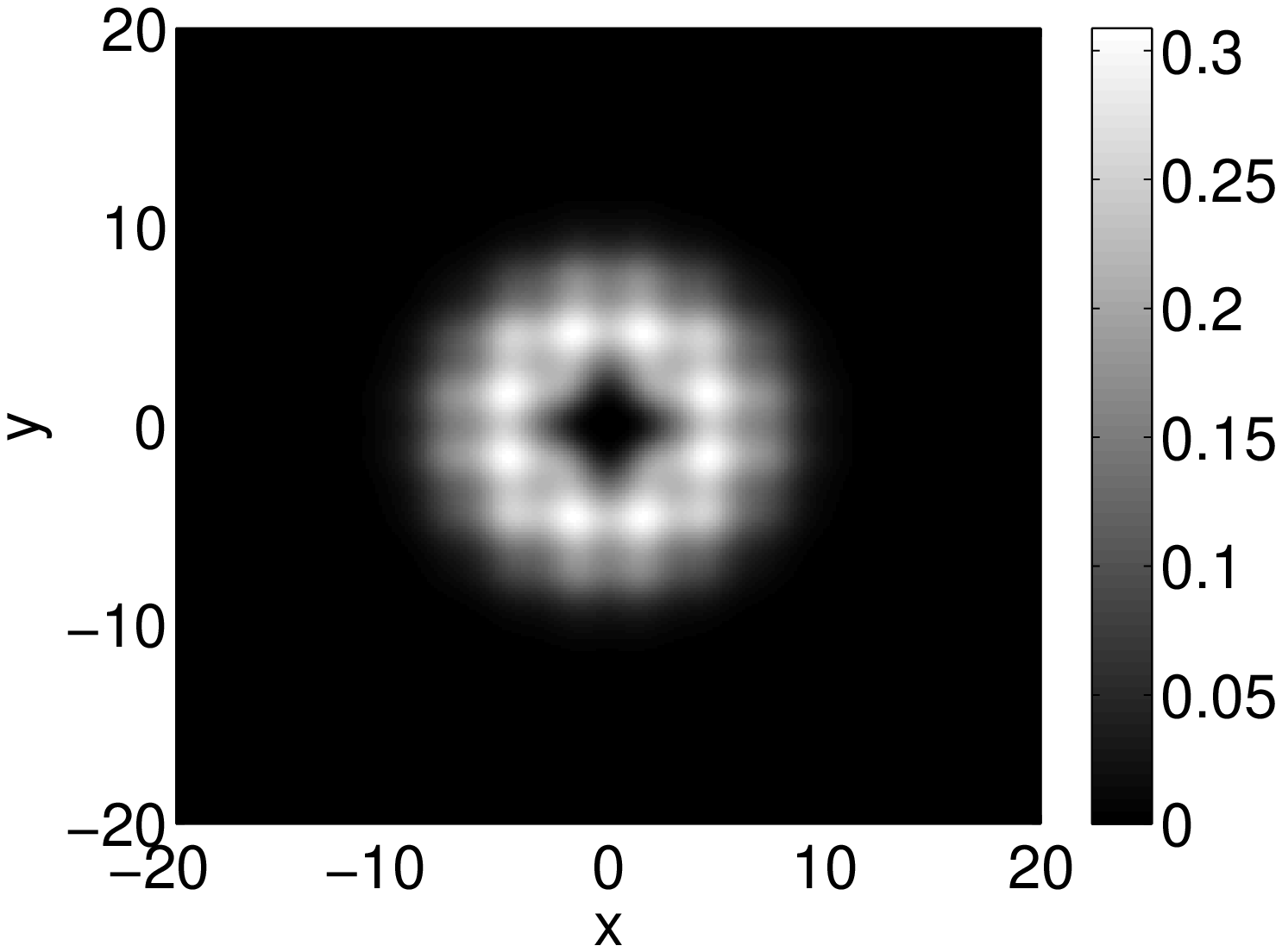}
\includegraphics[width=60mm]{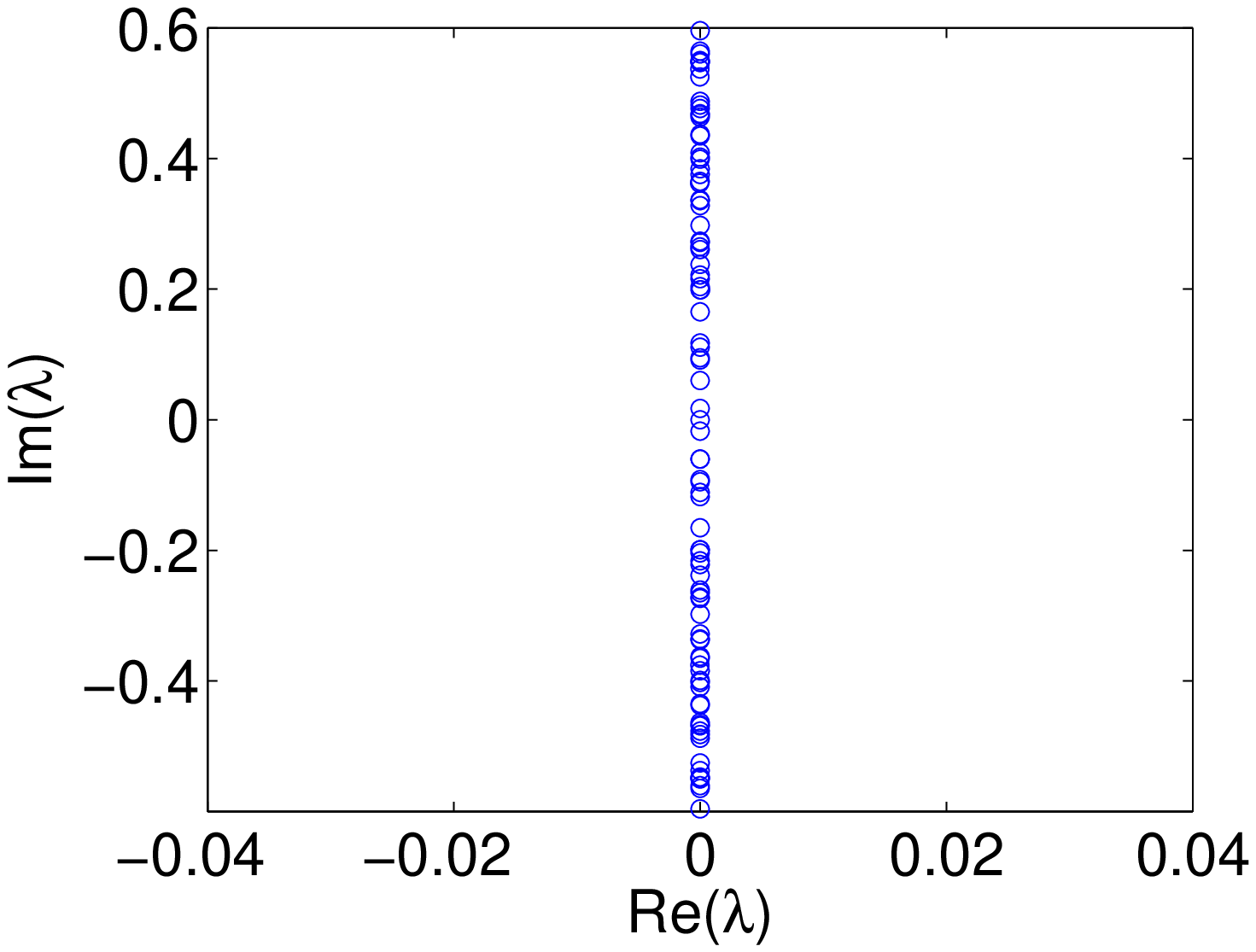}
\includegraphics[width=60mm]{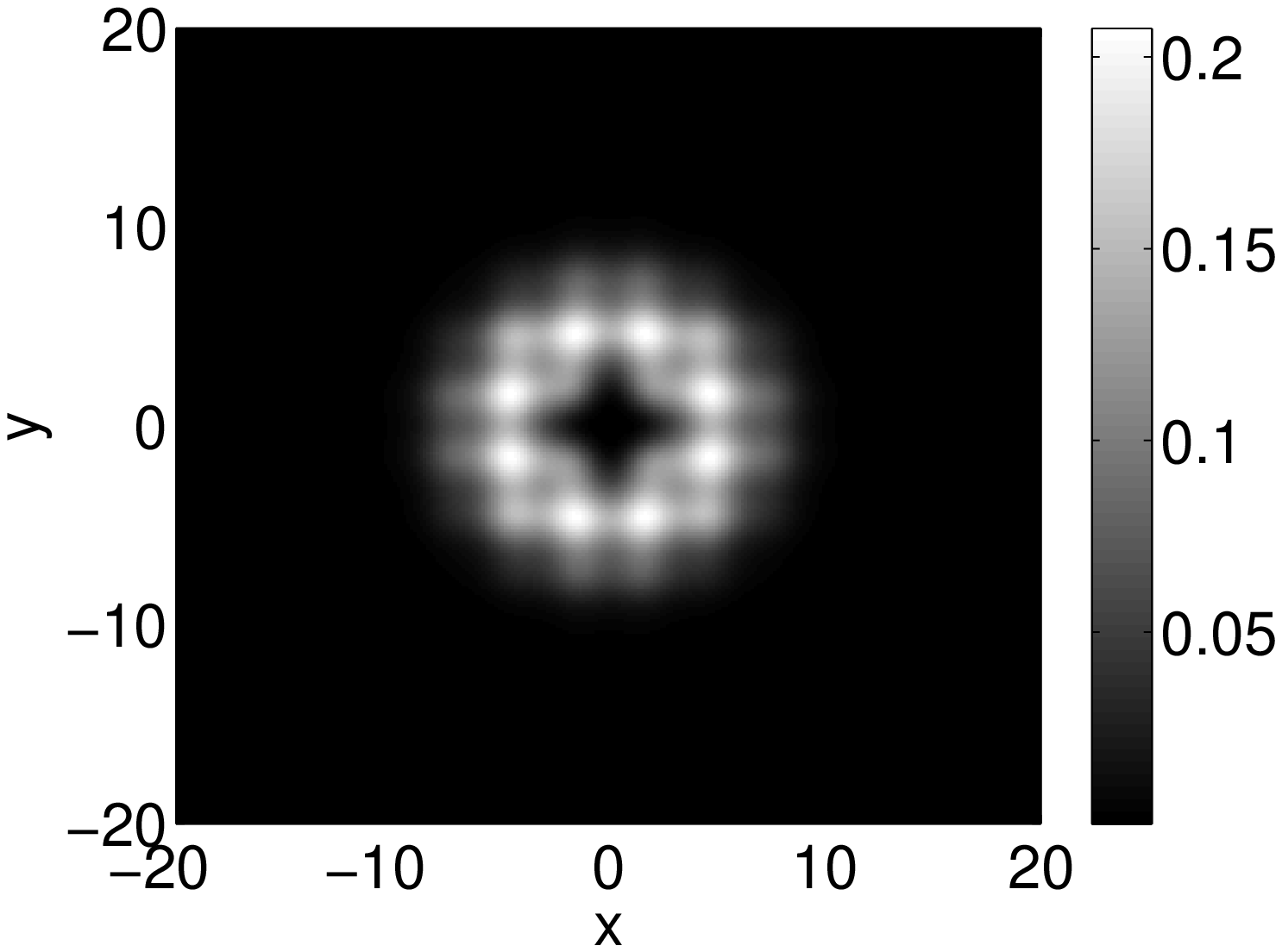}
\includegraphics[width=60mm]{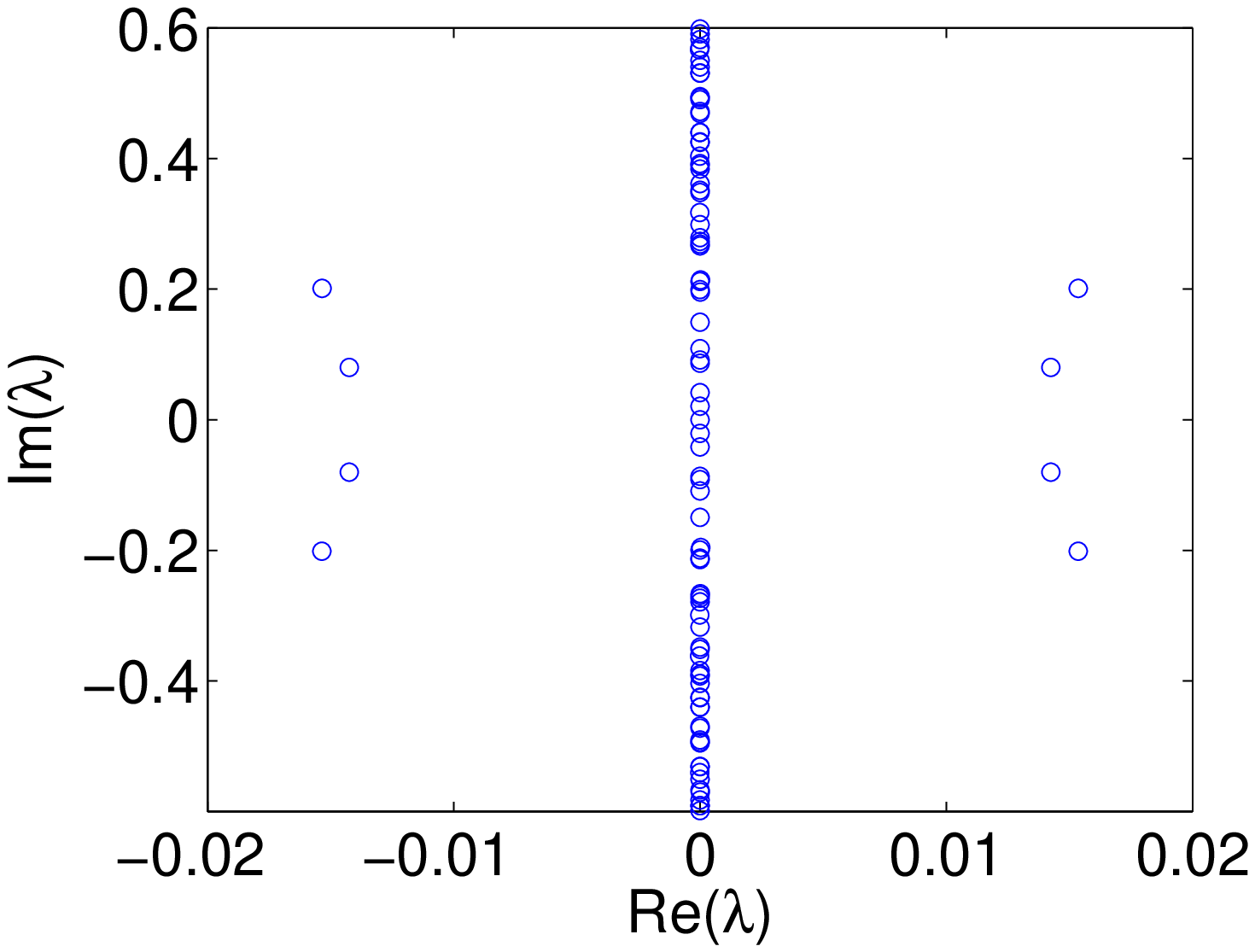}
\caption{(Color Online) Similarly to Fig. \ref{fig5}, 
stable (top, $(0.21,0.71)$) and 
unstable (bottom, $(0.3,0.71)$) double charge vortices but now for $\phi=0$.}
\label{fig8}
\end{figure}

\begin{figure}
\includegraphics[width=100mm]{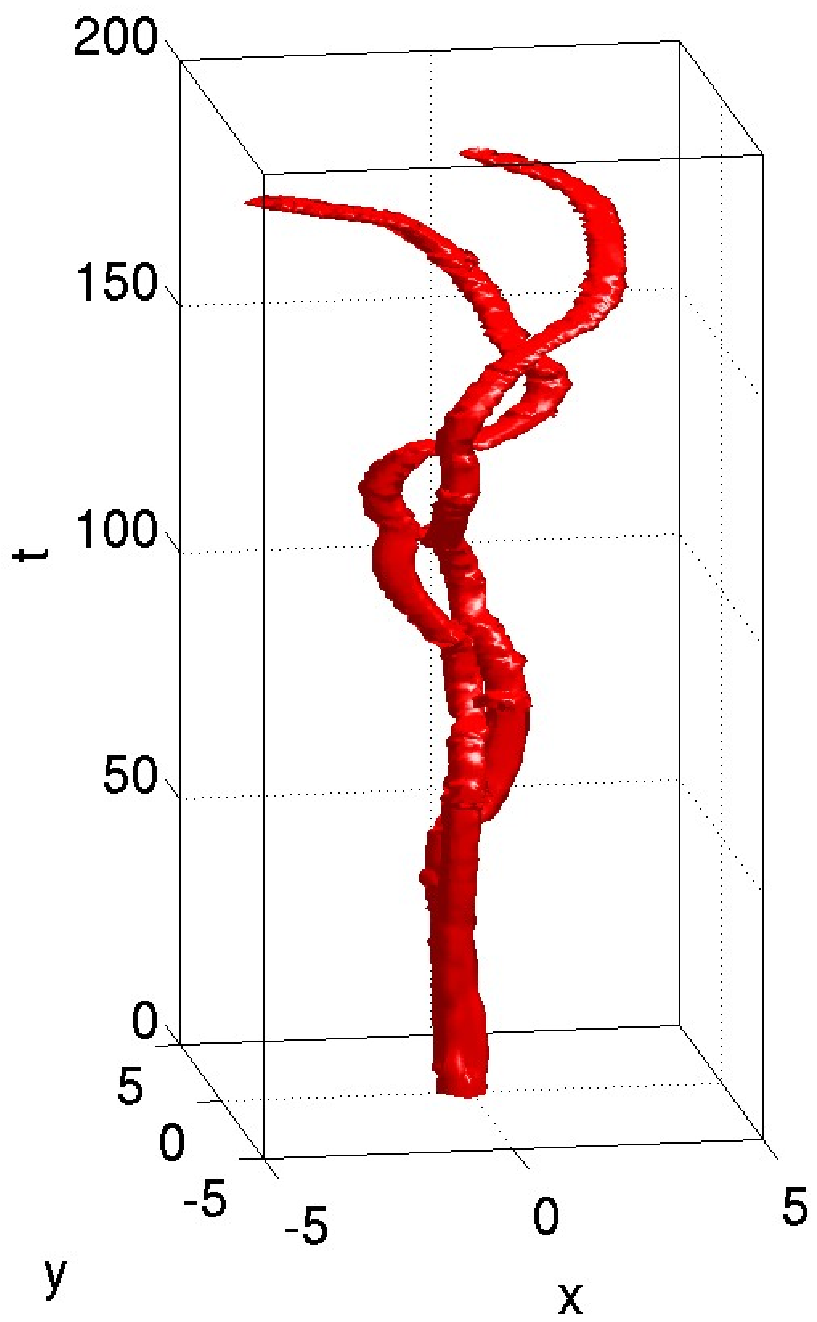}
\caption{(Color Online) The space-time dynamics 
of the unstable mode when $\phi=0$ for 
parameter values $(0.3,0.71)$ is illustrated.}
\label{fig9}
\end{figure}

The dependence of the structure of the solutions on the phase of the OL, 
$\phi$, becomes apparent in the case of these higher charged vortices. 
In the case $\phi=\pi/2$, 
the vortex is centered at a local minimum of the lattice 
and therefore, 
when it merges with the neighboring local minima of the condensate
background density, 
forms either an 'X' or a square comprised of four local maxima of the
lattice (minima of the condensate background), 
depending on the relative size of the condensate ($\mu$) 
and the lattice ($V_0$), as can be seen in figure \ref{fig5}.  
In the case of $\phi=0$, where the vortex merges with the neighboring 
local minima, it assumes the form of a cross comprised of five local
maxima of the lattice, as is shown in
Fig. \ref{fig8}. 

Finally, we examine the dynamical evolution of vortices in the
case of $\phi=\pi/2$ (Fig. \ref{fig6}) and $\phi=0$ (Fig. \ref{fig9}).
Over time, both  double charge vortices 
split into a pair of single charge vortices
which follow one another around in an oscillatory 
spiral motion. 
Also, it is clear 
that the vortex located at the maximum (Fig. \ref{fig8}) has typically 
somewhat larger instability growth rates and in fact may possess  
\textit{two} unstable eigenvalue quartets in its spectrum. This can
also be observed to have a bearing  
in the evolution images, particularly as is highlighted in
the contrast between figure \ref{fig6}, for parameter values for which the 
solution has a small instability growth rate, 
and figure \ref{fig9}, for parameter values for which the solution
has a strong growth rate. 
The spiraling in Fig. \ref{fig6} 
is slower and less violent than that of Fig. \ref{fig9}.

\begin{figure}
\includegraphics[width=50mm]{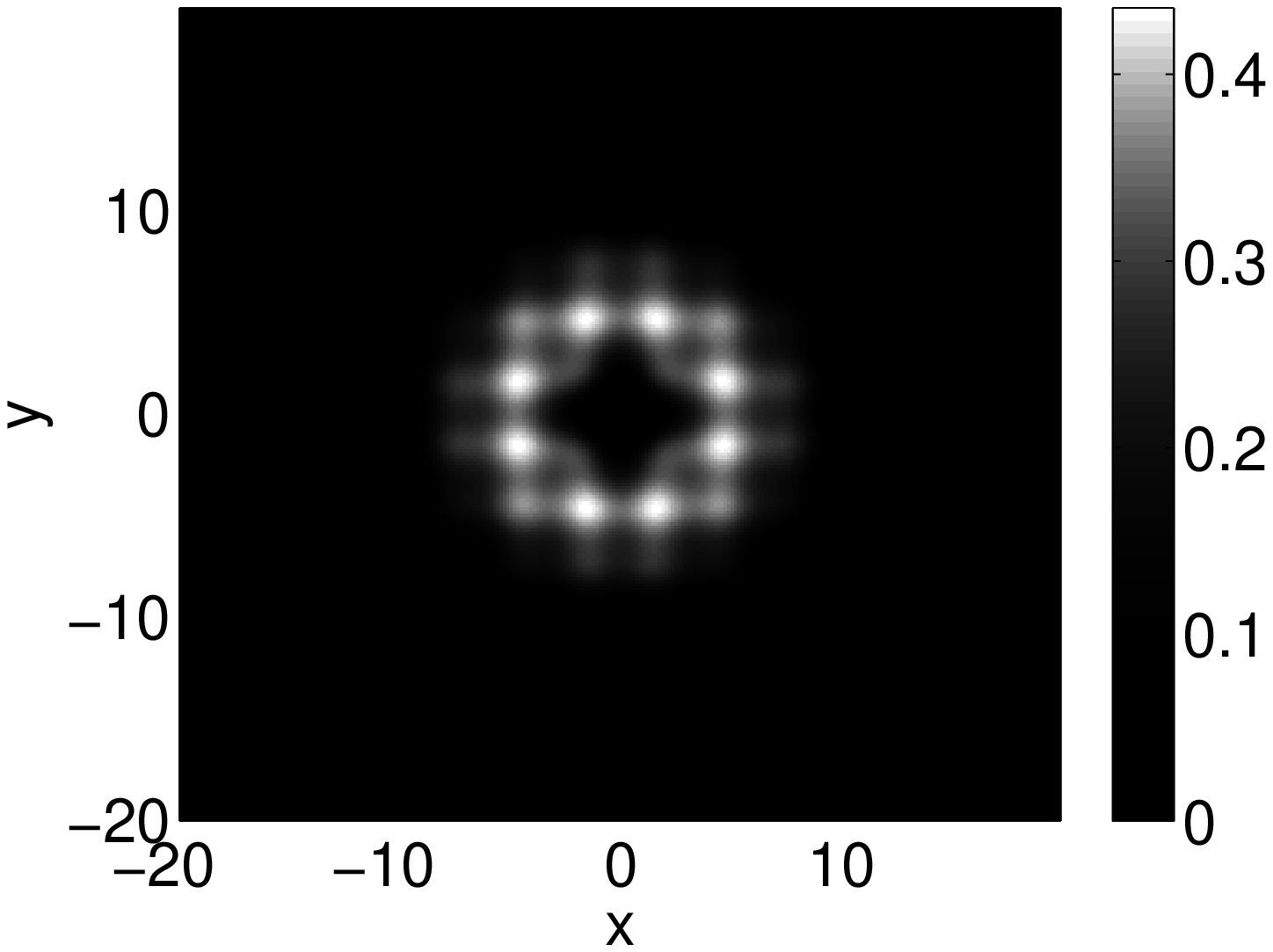}
\includegraphics[width=50mm]{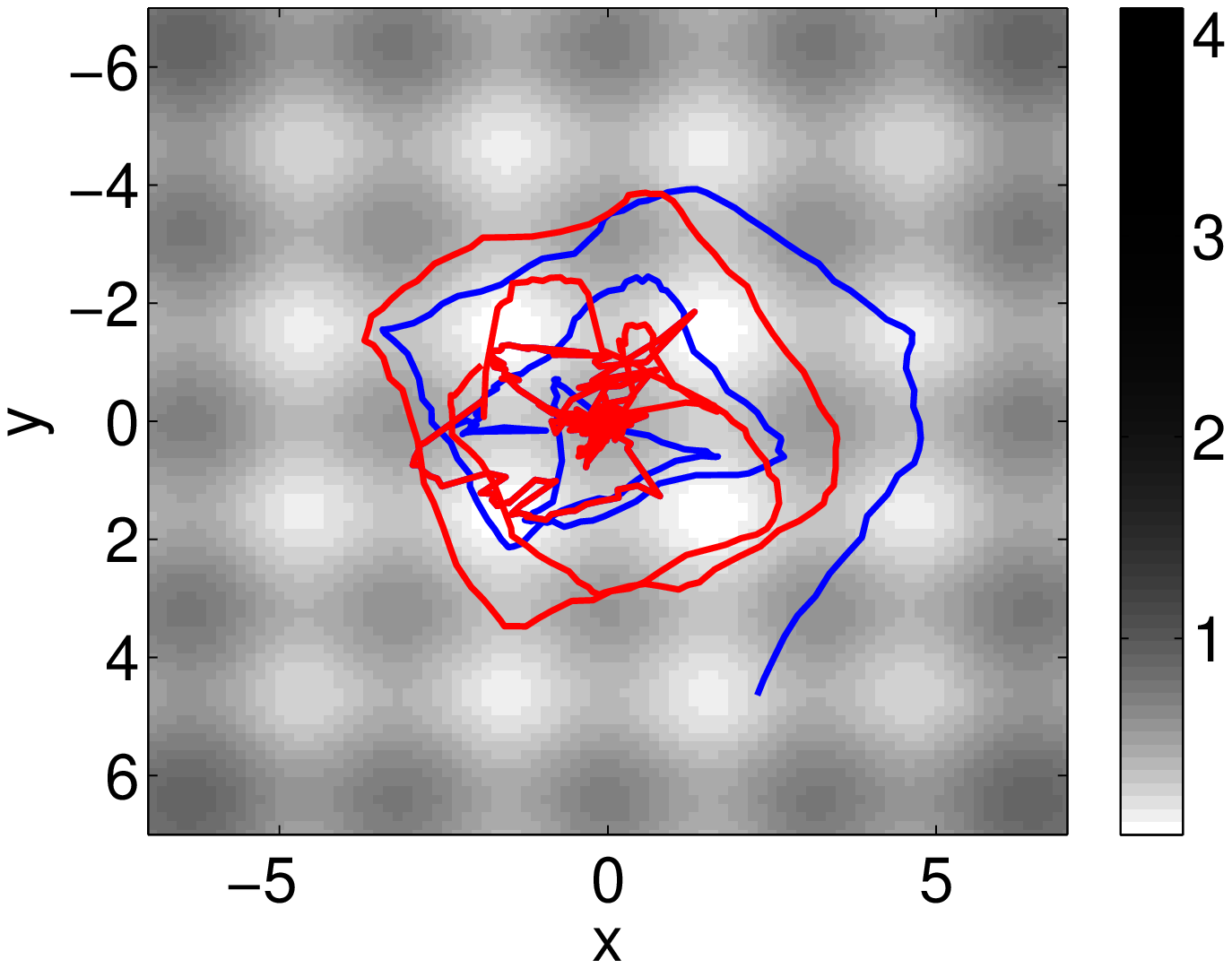}
\includegraphics[width=50mm]{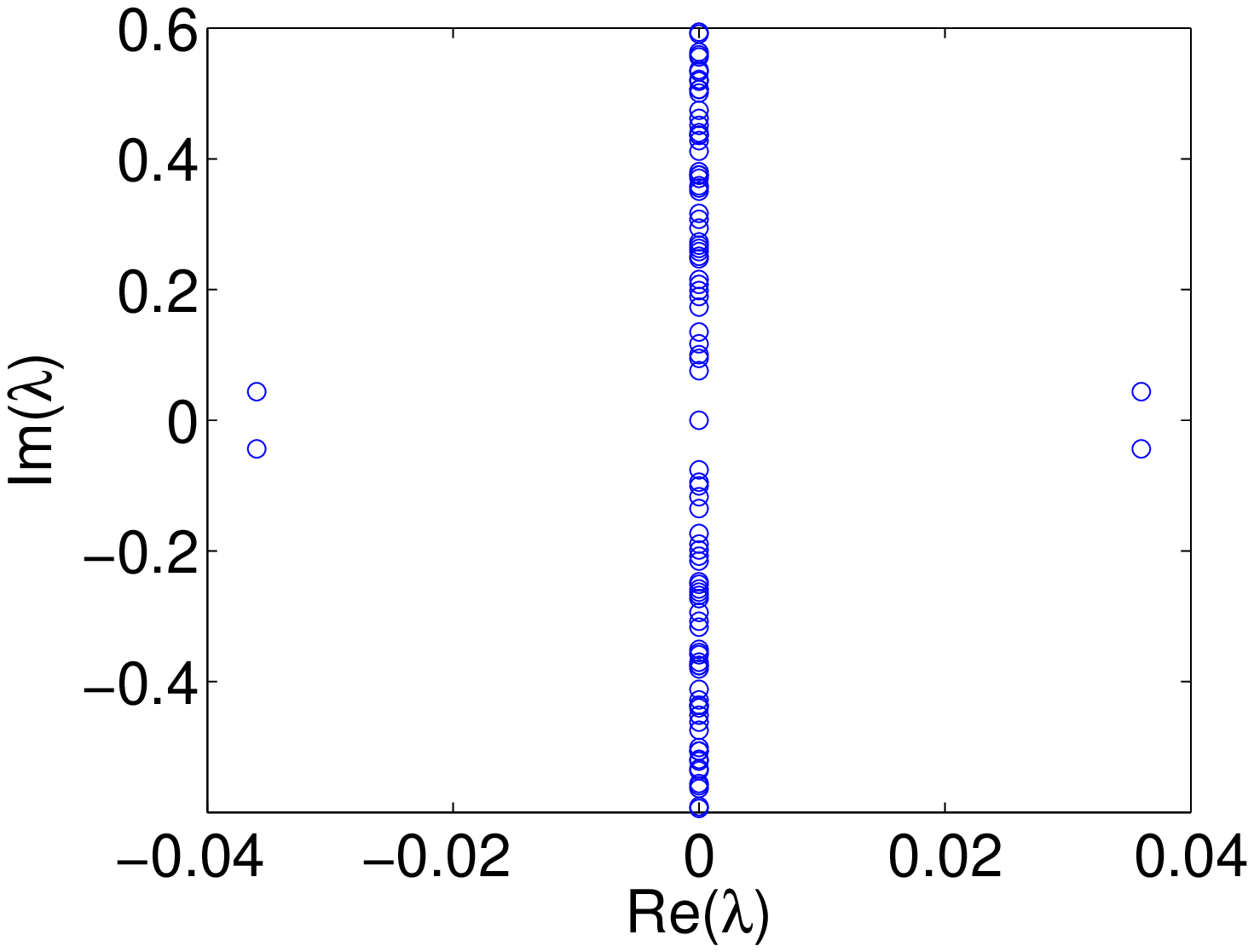}
\includegraphics[width=50mm]{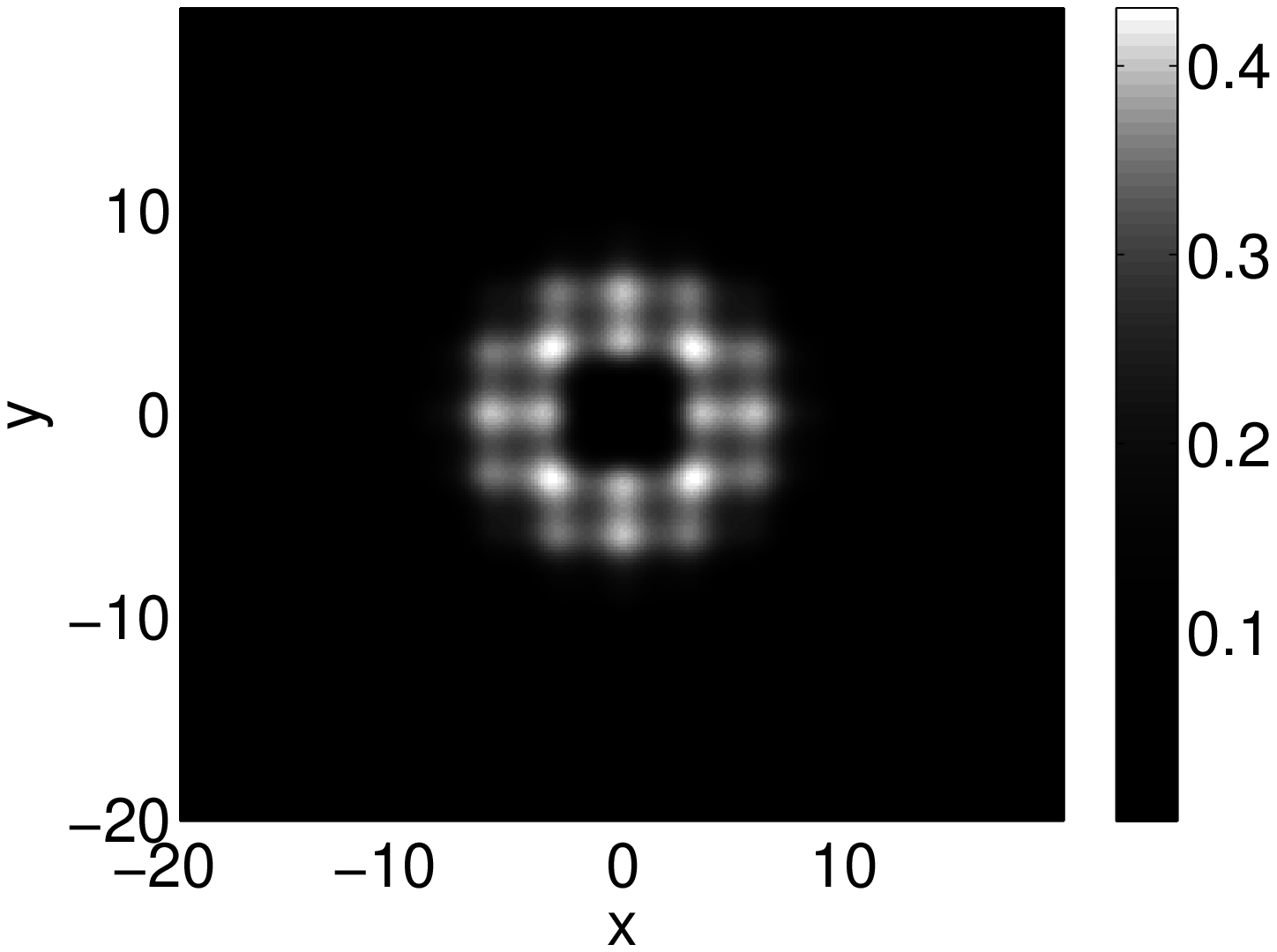}
\includegraphics[width=50mm]{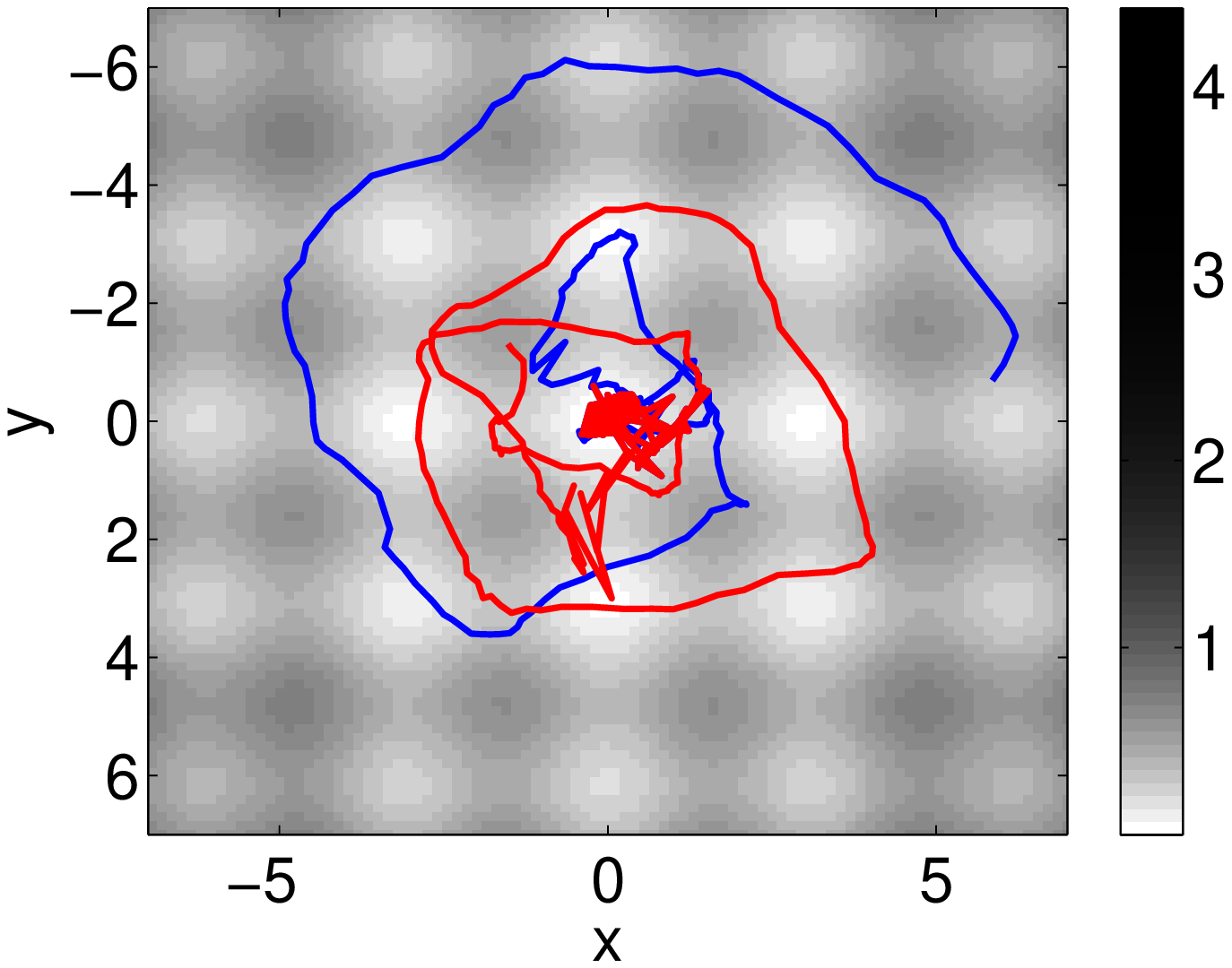}
\includegraphics[width=50mm]{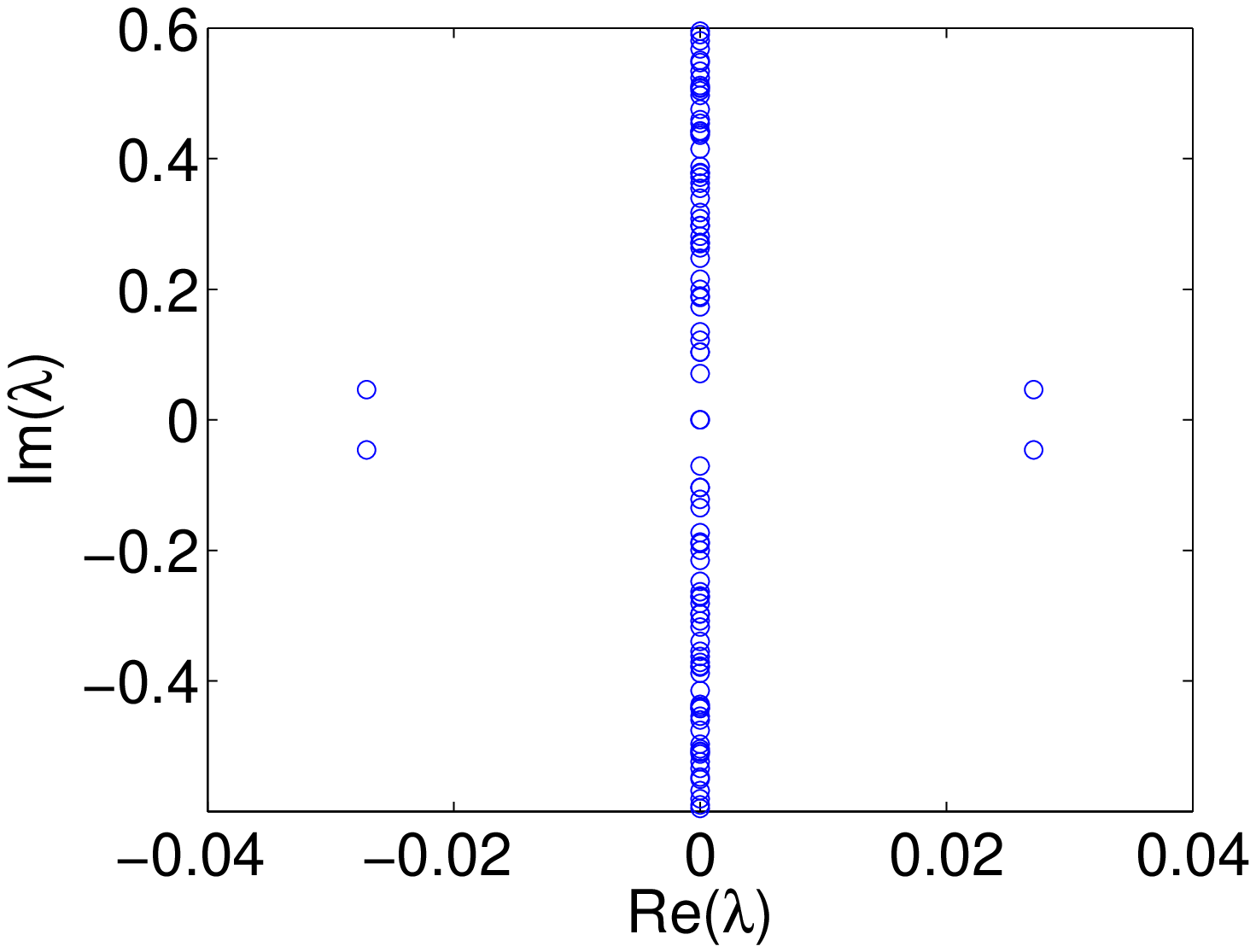}
\caption{(Color Online) 
The top row contains the profile (left) spectrum (right) and temporal
evolution for 450 time units of the charge 2 vortex with $\phi=0$,
where one can visualize the splitting into tow charge 1 vortices
in blue and red. The bottom row is the same for the case $\phi=\pi/2$.
Parameter values are equal in both cases, $(0.25,0.8)$.}
\label{figd2}
\end{figure}

Dynamics were also
performed for equal parameter values of $(0.25,0.8)$ and the 
vortex trajectory for 450 time units along with the profiles 
and spectra are presented in figure \ref{figd2}.  The magnitude 
of the real part of the single quartet in each case is 
comparable, as are the 
corresponding dynamics.  For $\phi=0$, ${\rm max [Re} (\lambda)]=0.036$,
while for $\phi=\pi/2$, ${\rm max [Re} (\lambda)]=0.027$. While 
this slight discrepency is hard to detect from the trajectories shown
in Fig. \ref{figd2}, we can define a similar diagnostic as in the 
single charge case.  Here we define $\textbf{p}_1$ and $\textbf{p}_2$
as in equation \ref{vtx_traj} and 
$t^{**}={\rm min} \{t; {\rm max_k}(|{p}_1^k-p_2^k|)>1.5 , k\in\{1,2\}\}$, 
where $\textbf{p}=(p^1,p^2)$,
in order to investigate this fundamentally different type of instability.
As we expect from the larger real part of its unstable quartet,
$t^{**}=152$ when $\phi=0$ is smaller than $t^{**}=216$ when $\phi=\pi/2$.

It is interesting to remark here the apparent difference
of this result with the case of the $S=1$ vortex, where
the potential with $\phi=0$ induced stability, while
that of $\phi=\pi/2$ could potentially lead to instability.
However, an important clarification should be made here.
The instability
of the vortices with $S=1$ and that of the vortices with $S=2$ are
{\it fundamentally} different. In particular, the former concerns the
spiraling of the vortices away from the center of the trap
(see Figs. \ref{fig3} and \ref{more_dyno_1}). 
On the other hand, the latter predominantly involves
the effect of {\it splitting} of the vortex of charge 2 into two
vortices of charge $S=1$ (see Figs. \ref{fig6}, \ref{fig9} and \ref{figd2}),
As a result of this splitting, the two same charge vortices
(as is discussed e.g. in \cite{us}),
start rotating around each other (while, of course, spiraling away from the
center of the trap). This type of effect (the break-up of the $S=2$
into two $S=1$ vortices) may even happen in the magnetic trap alone
for solutions in one of the instability bands at $V_0=0$,
rendering vortices of $S=2$ potentially unstable in that setting as
discussed e.g. in \cite{pu}. 

\begin{figure}
\includegraphics[width=100mm]{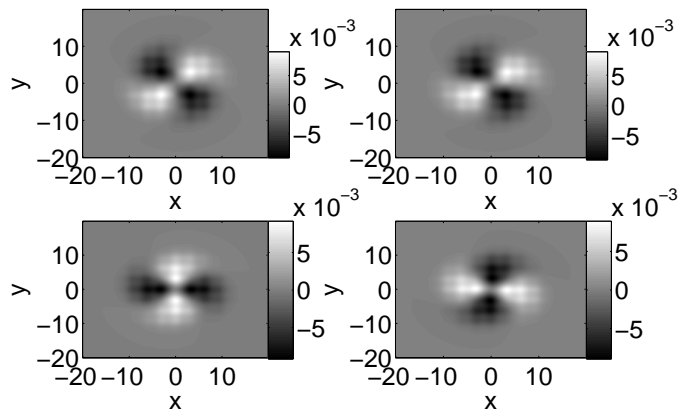}
\caption{(Color Online). The same set of images as in \ref{more_dyno_1_2} for the linearization around the charge 2 solution with $\phi=\pi/2$ for parameter values $(V_0,\mu)=(0.25,0.8)$ as in the evolution from the bottom row of Figure \ref{figd2}. We note that for the case of $\phi=0$, the corresponding eigenmodes
are qualitatively indiscernable from these shown here and are hence omitted.}
\label{more_dyno_2}
\end{figure}

In view of this, the two cases of $S=1$ and $S=2$ and their 
respective instability
growth rates (and trends) 
should not be directly compared as they physically refer
to different mechanisms. 
One of the two potentials (the $\phi=0$ case) can be thought of as
a ``local inhomogeneity''\footnote{An interesting reference examining the
effect of localized inhomogeneities in higher charge vortices in rotating
BECs can be found in the work of \cite{simula}.} 
within the central well of the optical lattice,
while the other one (the $\phi=\pi/2$) can be thought of, within the same
region, as an approximate parabolic potential (modulated by an appropriately
weak Gaussian past the first maximum of the optical lattice). 
While this picture can be fairly successful in capturing the 
stability characteristics of the optical lattice for small values
of the chemical potential (results not shown), for strong nonlinearities
(large chemical potentials) 
this picture is no longer valid. It is only
in the latter strongly nonlinear regime where the growth rates of the
instabilities in the cases of the two different phases of the optical
lattice potential are substantially different. Hence it is a challenging
task whether this difference can be intuitively understood, a question
that is outside the scope of the present work. Interestingly,
the unstable eigenmodes in this case have strong similarities in their
spatial profile between the $\phi=0$ and $\phi=\pi/2$ cases and
so only those for $\phi=\pi/2$ are shown in Fig. \ref{more_dyno_2}.



\section{$S=3$ Vortices}

\begin{figure}
\includegraphics[width=60mm]{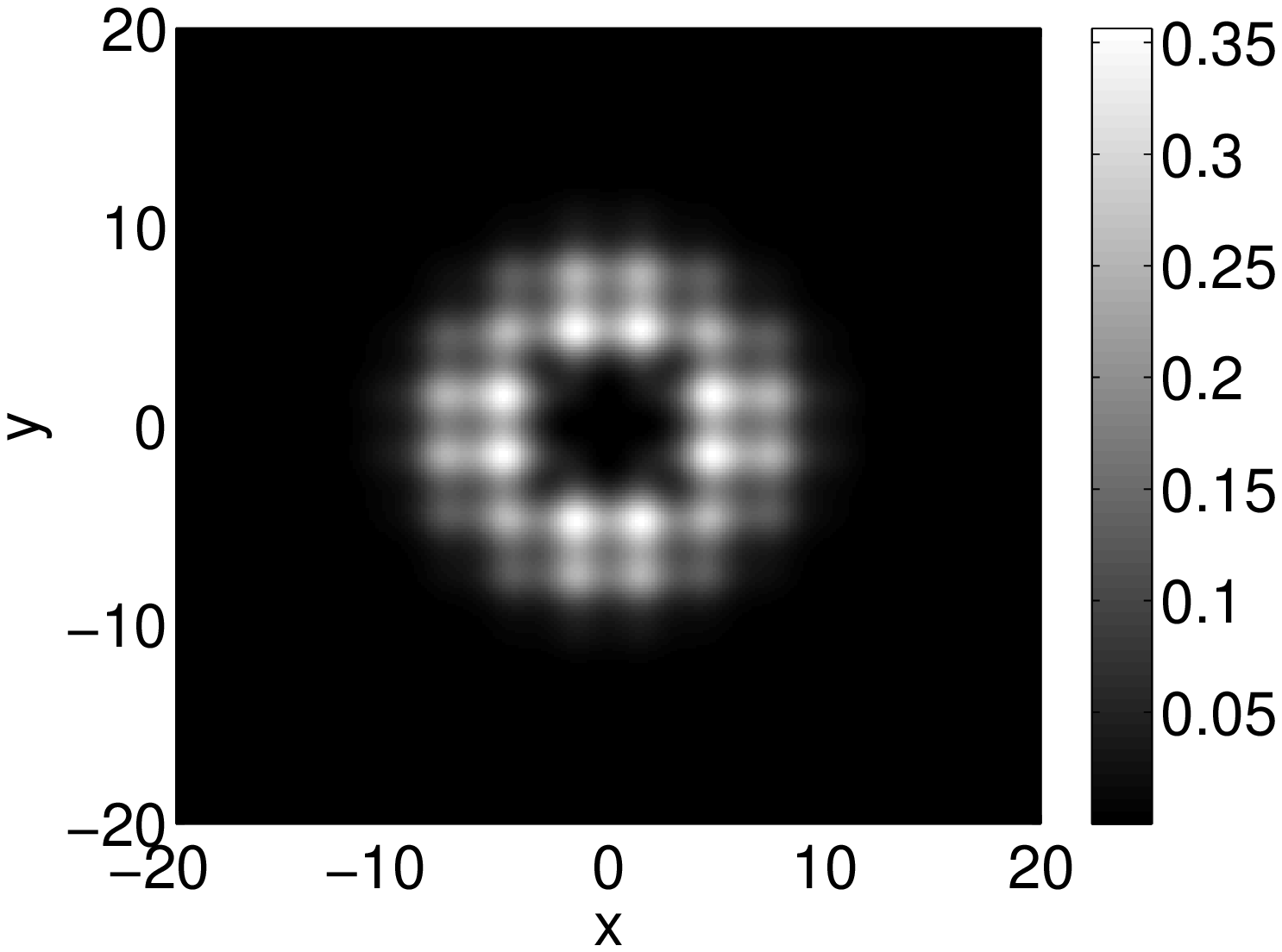}
\includegraphics[width=60mm]{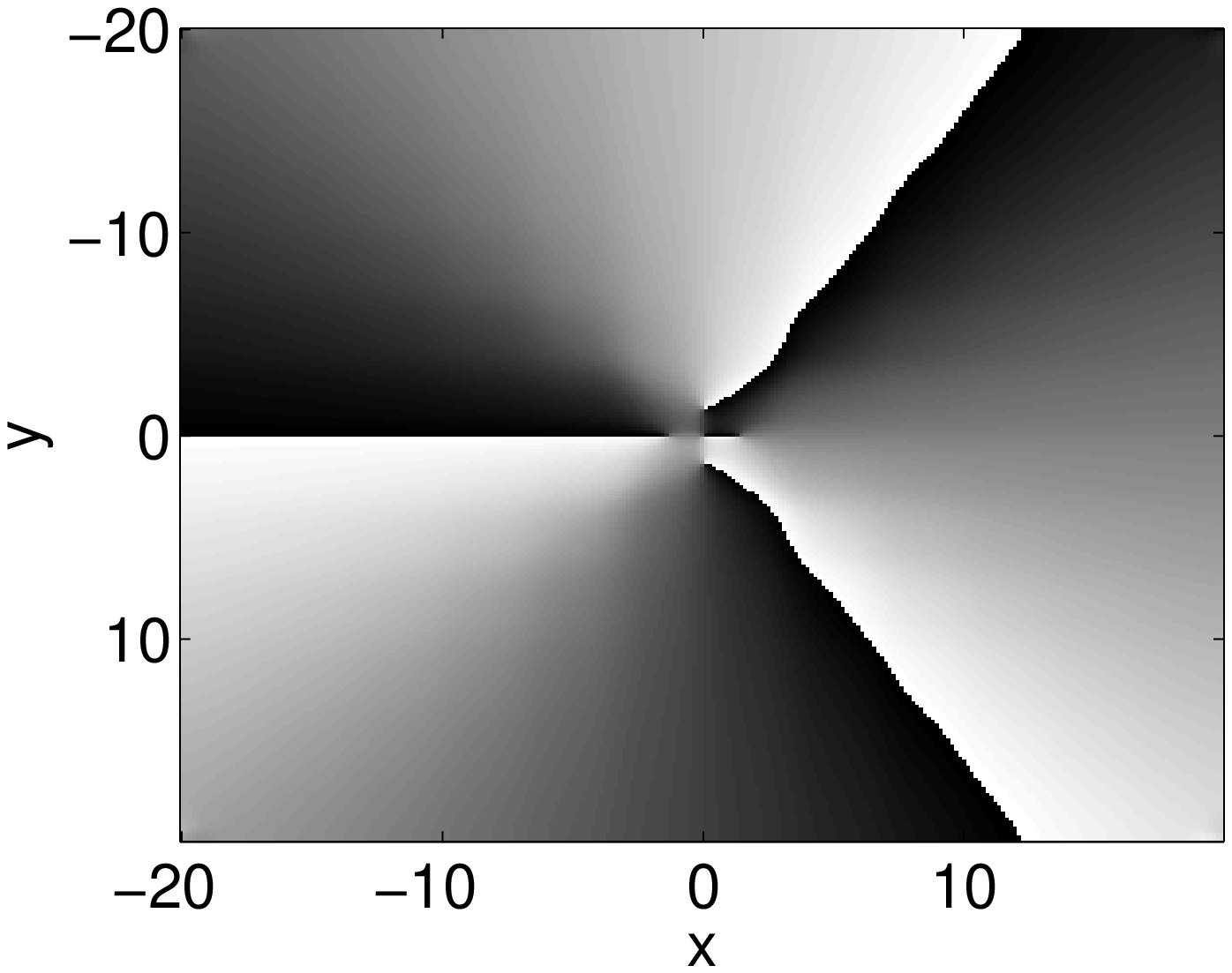}
\includegraphics[width=60mm]{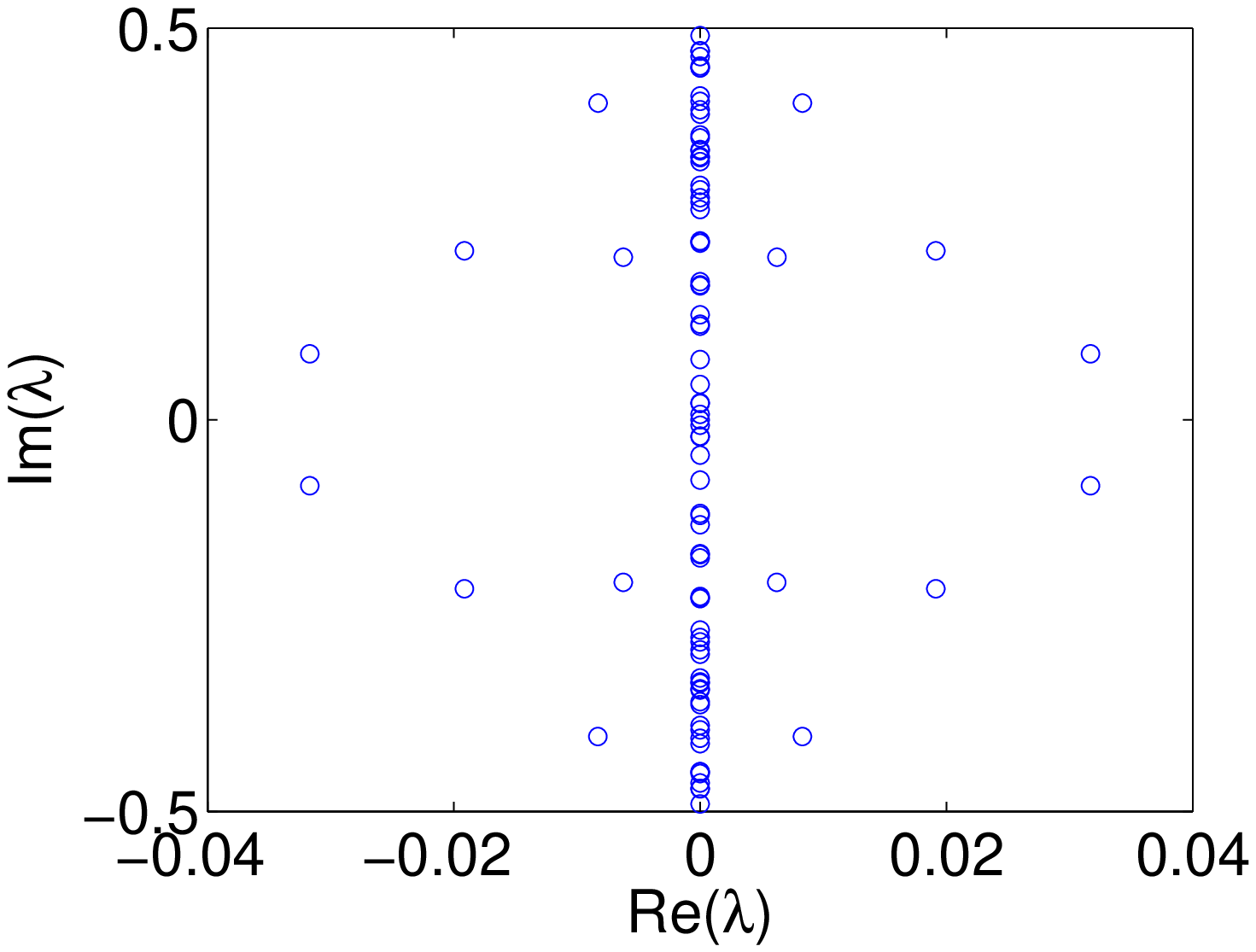}
\includegraphics[width=60mm]{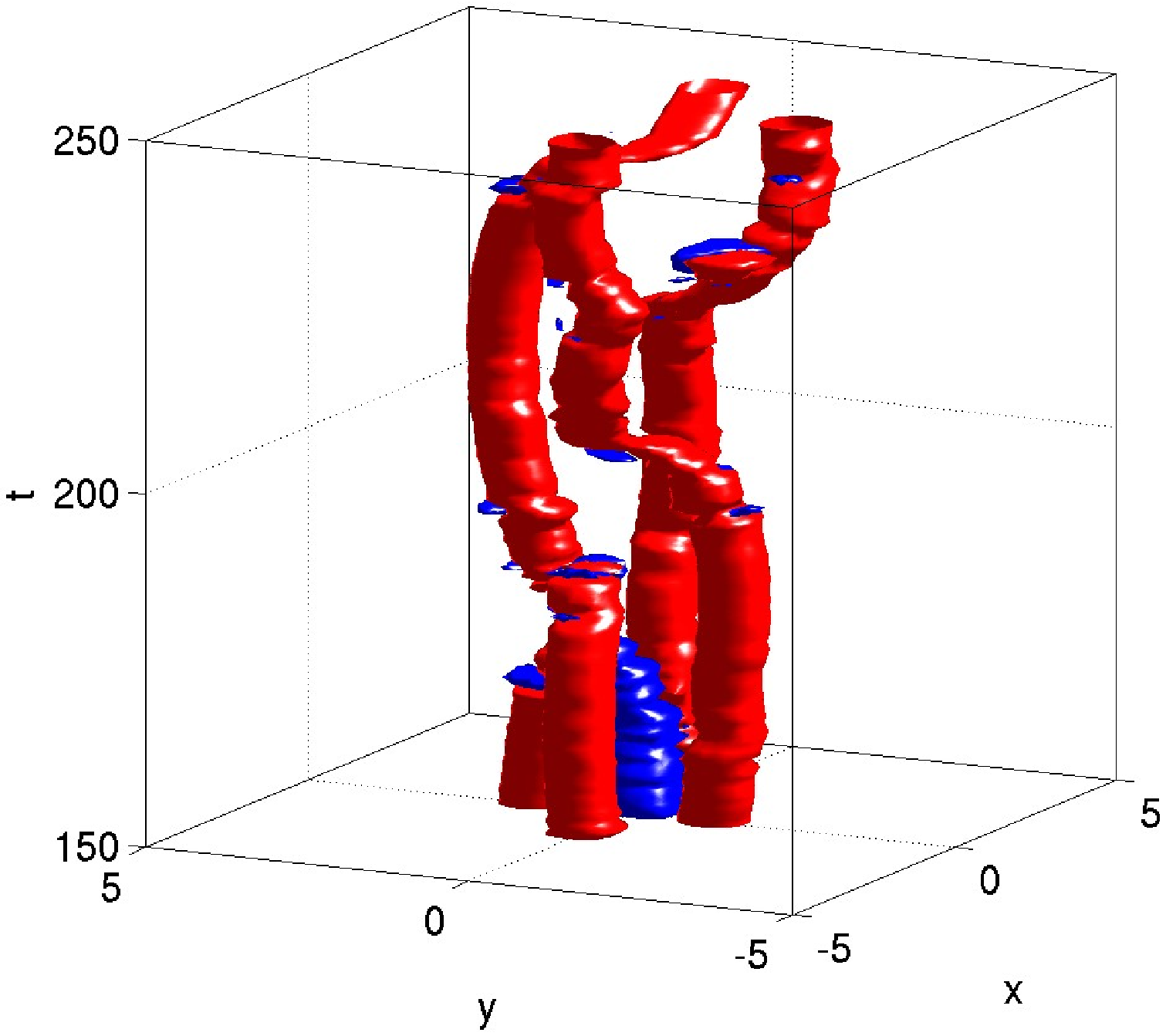}
\caption{(Color Online) Top row:
unstable charge $3$ vortex for $(0.44,1)$ for $\phi=0$; 
density profile (left) and corresponding phase diagram (right).  
Bottom row: spectrum of the solution (left) and dynamics of the
unstable mode for parameter values $(0.44,1)$, where red is a positive 
isosurface and blue is a negative surface of equal magnitude.}
\label{fig10}
\end{figure}

\begin{figure}
\includegraphics[width=60mm]{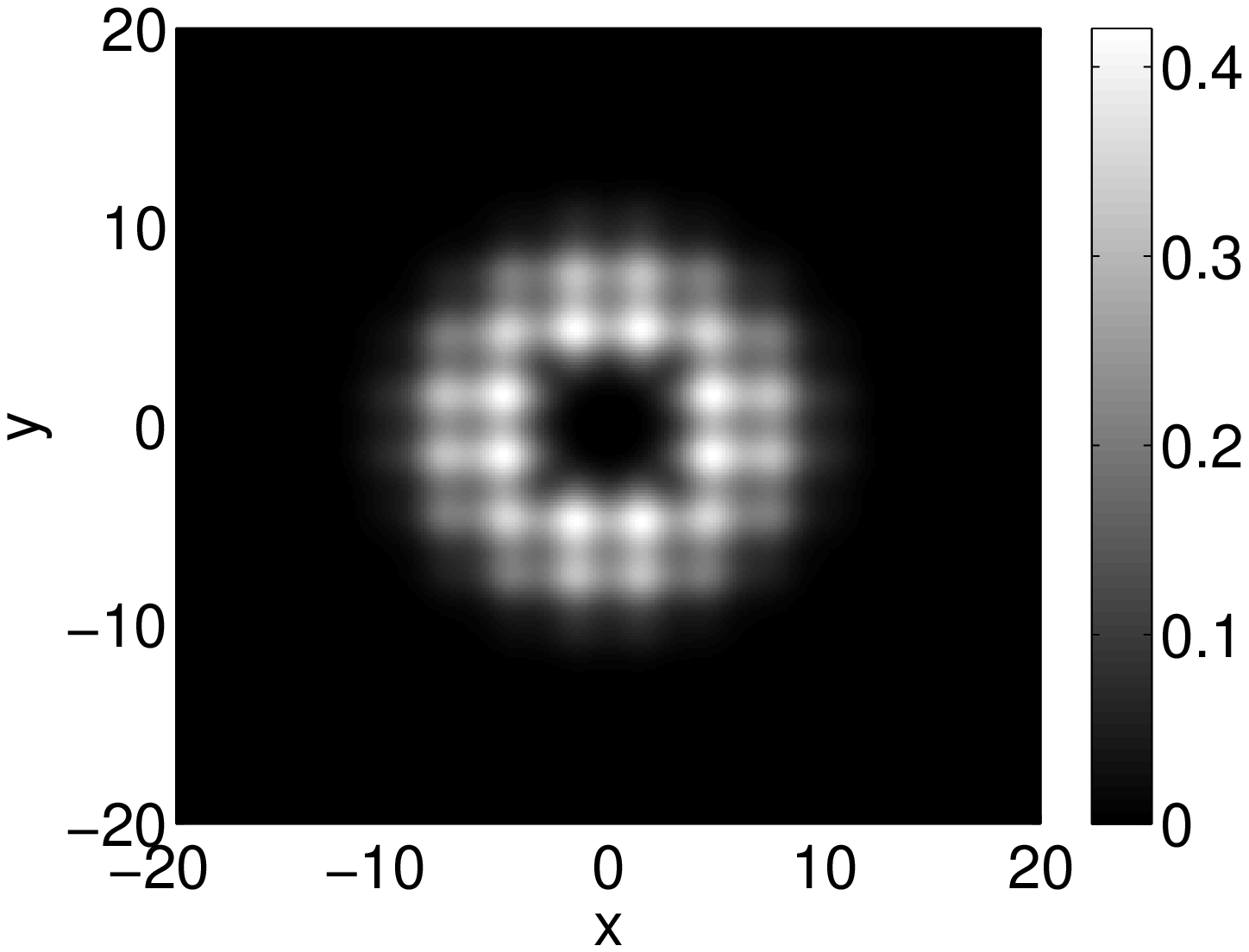}
\includegraphics[width=60mm]{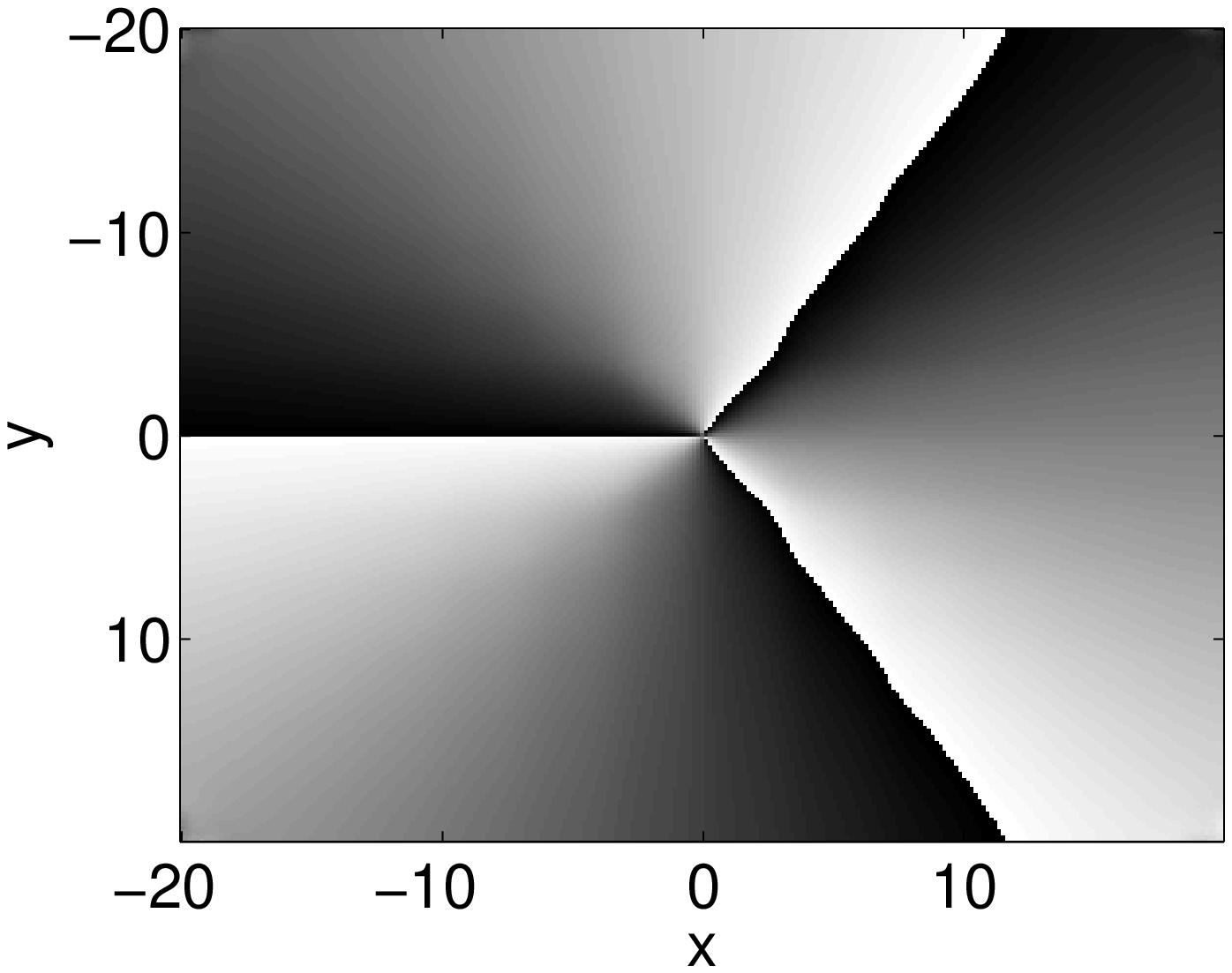}
\includegraphics[width=60mm]{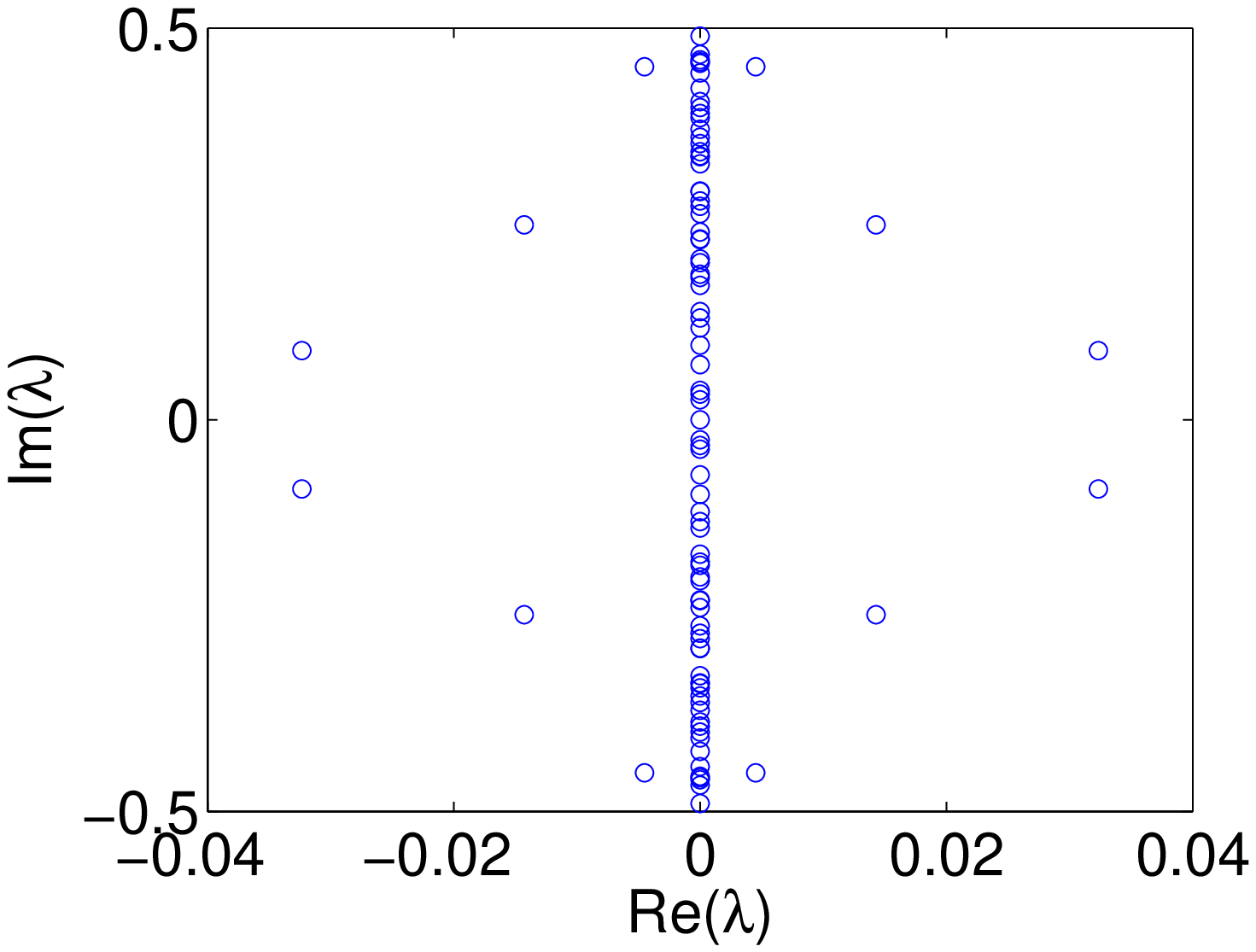}
\includegraphics[width=60mm]{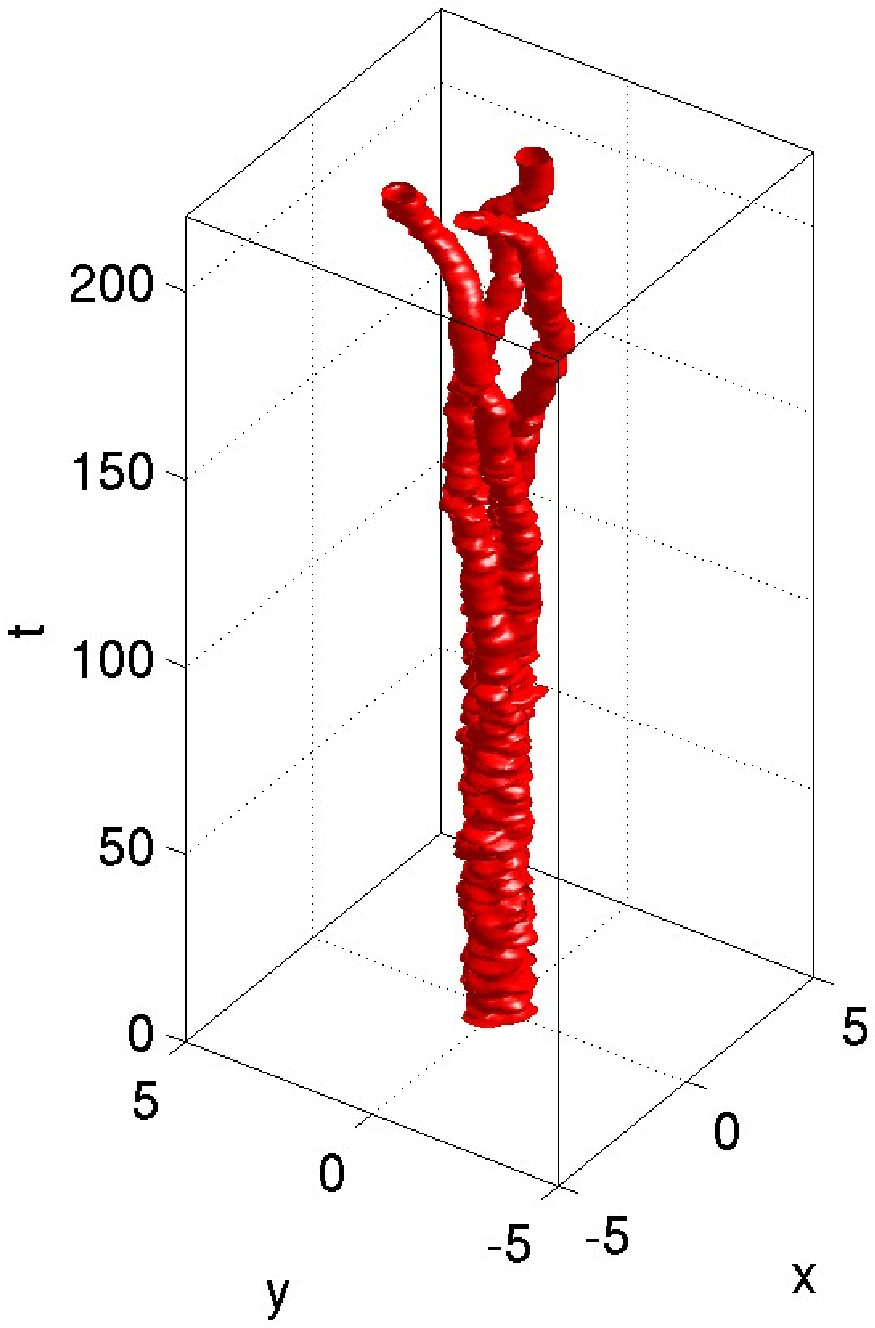}
\caption{(Color Online) Top row: 
Unstable charge $3$ vortex for $(0.37,1)$ for $\phi=0$; 
density profile (left) and corresponding phase diagram (right).  
Bottom row: spectrum of the solution (left) and its dynamical evolution
(right).}
\label{fig11}
\end{figure}

Finally, we briefly touch upon the case of triple quantized vortices. 
In this setting, and depending on parameter values, we were able to
identify two different fundamental types of solutions with this 
vorticity. One of them
consists of four positively charged and one negatively charged vortex,
forming a bound state with $S=3$, as is shown in Fig. \ref{fig10}.
However, we were also able to identify settings where the entire
structure appeared to be a single, point-like vortex as 
is shown in Fig. \ref{fig11}.

The dynamics over time of these solutions were consistent with the previous
cases, as can also be seen in figures \ref{fig10} and \ref{fig11}.  
The configuration with the $5$ single-charge vortices (of total
$S=3$) exhibits an interesting 
type of instability development. Around $t=150$ the negative charge 
vortex in the middle
collides with one of the $4$ positively charged vortices and they both 
dissappear,
while the remaining $3$ $S=1$ structures 
begin to meander around the domain. 
On the other hand, the triple charge vortex in figure \ref{fig11} 
splits into 
$3$ single charge vortices, which also spiral towards the periphery of
the BEC cloud.
All the cases shown for charge 3 are for $\phi=0$ , but we 
found similar results for $\phi=\pi/2$.

\section{Conclusion}

In this work, we have examined in some detail the existence regime, 
stability regions and dynamical evolution of vortices of single, double
and briefly also triple charge in a nonlinear Schr{\"o}dinger equation
involving both a parabolic and a periodic potential. This setting
is particularly relevant to pancake-shaped BECs in the presence of 
magnetic trapping and optical lattices, but similar applications may
arise in optics among other settings.
We constructed two-parameter diagrams of the modes and their
stability as a function of the solution's chemical potential and
the strength of the optical lattice. These continuations led to
the insight that single-charged vortices are typically stable 
for cosinusoidal potentials, while they are typically unstable
beyond a critical periodic potential strength for sinusoidal 
potentials. The instability is manifested through the outward
spiraling of the vortex. In the case of $S=2$ vortices, the
regions of instability widen (and, in fact, reach through all
the way to the limit of no optical lattice, a feature that did
not exist in the $S=1$ case) and now become fairly similar for
the cosinusoidal and sinusoidal cases. The instability dynamics
features the split up of the vortices and their subsequent rotation
around each other, while they are both spiraling towards the outskirts
of the BEC. Finally, both more complex scenaria of instability
(involving many quartets of eigenvalues) and more complicated
structures (including bound states of five vortices with total $S=3$)
were found to arise in the $S=3$ case.

It would be of particular interest to examine the same type of
solutions in the case of a three-dimensional magnetic and optical
trapping. Adapting the numerical technology of \cite{Huepe},
it should be possible to perform relevant three-dimensional
existence and stability computations, by efficiently solving the
linear system in the Newton method and identifying the dominant 
eigenvalues to infer how the stability is affected by introducing 
the $z$-dependence. In that problem, one can also use $\Omega$
as a parameter starting from the pancake-shaped limit of $\Omega \ll 1$,
and going towards the spherical trap limit of $\Omega \rightarrow 1$. 
It would be especially interesting to observe how the stability
properties are modified in the latter setting and how the unstable
dynamics may develop. Such studies are currently in progress and
will be reported in future publications.

\end{document}